\documentclass[
reprint,
superscriptaddress,
amsmath,amssymb,
aps,
pra,
floatfix
]{revtex4-2}

\usepackage{graphicx}
\usepackage{dcolumn}
\usepackage{bm}
\usepackage[colorlinks,linkcolor=magenta,citecolor=magenta,urlcolor=magenta]{hyperref}

\begin{document}

\title{Nonlocal Exciton-Photon Interactions in Hybrid High-Q Beam Nanocavities with Encapsulated MoS$_2$ Monolayers}

\author{Chenjiang Qian}
\email{chenjiang.qian@wsi.tum.de}
\author{Viviana Villafañe}
\author{Pedro Soubelet}
\author{Alexander Hötger}
\affiliation{Walter Schottky Institut and Physik Department, Technische Universit{\" a}t M{\" u}nchen, Am Coulombwall 4, 85748 Garching, Germany}
\author{Takashi Taniguchi}
\affiliation{International Center for Materials Nanoarchitectonics, National Institute for Materials Science, 1-1 Namiki, Tsukuba 305-0044, Japan}
\author{Kenji Watanabe}
\affiliation{Research Center for Functional Materials, National Institute for Materials Science, 1-1 Namiki, Tsukuba 305-0044, Japan}
\author{Nathan P. Wilson}
\author{Andreas V. Stier}
\author{Alexander W. Holleitner}
\author{Jonathan J. Finley}
\email{finley@wsi.tum.de}
\affiliation{Walter Schottky Institut and Physik Department, Technische Universit{\" a}t M{\" u}nchen, Am Coulombwall 4, 85748 Garching, Germany}
\begin{abstract}
	Atomically thin semiconductors can be readily integrated into a wide range of nanophotonic architectures for applications in quantum photonics and novel optoelectronic devices.
	We report the observation of nonlocal interactions of \textit{free} trions
	in pristine hBN/MoS$_2$/hBN heterostructures coupled to single mode (Q $>10^4$) quasi 0D nanocavities.
	The high excitonic and photonic quality of the interaction system stems from our integrated nanofabrication approach simultaneously with the hBN encapsulation and the maximized local cavity field amplitude within the MoS$_2$ monolayer.
	We observe a nonmonotonic temperature dependence of the cavity-trion interaction strength, consistent with the nonlocal light-matter interactions in which the extent of the center-of-mass wavefunction is comparable to the cavity mode volume in space.
	Our approach can be generalized to other optically active 2D materials, opening the way towards harnessing novel light-matter interaction regimes for applications in quantum photonics.
\end{abstract}
\maketitle


Monolayer crystals of transition metal dichalcogenides (TMDs) are ideally suited as the active material for solid-state cavity quantum electrodynamics (cQED) investigations \cite{RevModPhys.90.021001, Ardizzone_2019}.
They have very large exciton binding energies $\geq 100$ meV \cite{PhysRevLett.120.057405,Goryca2019}, linewidths close to the homogeneous limit when suitably encapsulated by hexagonal boron nitride (hBN) \cite{Wierzbowski_2017,PhysRevX.7.021026,Raja2019}, and, strong optical absorption strengths close to the excitonic transitions exceeding $\sim10\%$ per atomically thin layer \cite{Li_PRB2014}.
In addition, 2D materials can be readily attached to a wide range of substrates \cite{Castellanos-Gomez_2014}, making them ideally suited for hybrid solid-state cQED experiments \cite{Schneider_2018}.
Indeed, recent work has reported strong light-matter coupling for monolayer TMDs using diverse photonic resonator geometries including planar open-fiber cavities \cite{Dufferwiel_2015}, photonic crystals \cite{Zhang2018,Ma_2020,2107.00078} and nanoplasmonic TAMM resonators \cite{Lundt_2016, Hu_2016, Flatten_2016}.
However, the direct coupling of 2D semiconductors to quasi 0D nanophotonic modes whilst preserving excellent excitonic properties and high cavity quality (Q) factor has remained a challenge.
Most commonly, nonencapsulated TMD monolayers are stacked directly on top of pre-fabricated photonic structures using pick-and-place assembly \cite{Wu2015,Ye2015,Li2017,2010.05458}.
In this case, local strain arising from the non-planar substrate, and spatially varying local dielectric screening result in a disordered energy landscape that perturbs the excitonic properties of 2D semiconductors \cite{Raja2019,Rhodes2019}.
Whilst dielectric disorders can be partially mitigated by full hBN encapsulation \cite{Wierzbowski_2017,PhysRevX.7.021026,Raja2019}, this approach results in a trade-off between the strength of the disorder potential and the cavity-TMD coupling strength, by moving the TMD monolayer away from the antinode of the cavity field \cite{2010.05458}.

Besides improving the optical properties of TMDs, full hBN encapsulation also enhances the transport properties of excitons \cite{Ju2014,Wierzbowski_2017,PhysRevX.7.021026,Raja2019}.
Indeed, markedly contrasting temperature dependencies of the exciton transport properties have been observed in hBN-encapsulated and bare TMD monolayers \cite{C9NR07056G,doi:10.1021/acsnano.6b05580}.
In quasi 0D nanocavities, exciton motion is expected to play an important role; light-matter couplings can enter the \textit{nonlocal} regime \cite{PhysRevB.86.085304} in which excitons sample different spatial regions of the cavity mode having dissimilar electric field amplitudes, during their lifetime.
This is in strong contrast to the typical situation such as the III-V quantum dots, where the emitter size is much smaller than the emission wavelength and excitons are stationary, such that electric field is constant across the exciton wavefunction.
In typical case, light-matter interaction is governed by the dipole approximation with a strength $\mathbf{d}\cdot\mathbf{E}\left( \mathbf{r}\right)$ determined by the electron-hole dipole moment $\mathbf{d}$ and the local electric field $\mathbf{E}\left( \mathbf{r}\right)$ at the emitter position $\mathbf{r}$.
In 2D materials, the center-of-mass motion of free excitons spatially extends over $\mu$m lengthscales \cite{doi:10.1021/acsnano.0c05305}, much larger than the emission wavelength.
Therefore, $\mathbf{E}\left( \mathbf{r}\right)$ is no longer constant during the exciton lifetime and the exciton-photon interaction enters the \textit{nonlocal} regime \cite{PhysRevB.86.085304,doi:10.1002/qute.201900024}.
Here, the light-matter interaction is determined by an interplay in spatial between the cavity mode field and the center-of-mass wavefunction of the exciton \cite{PhysRevB.86.085304}.

\begin{figure*}
	\includegraphics[width=\linewidth]{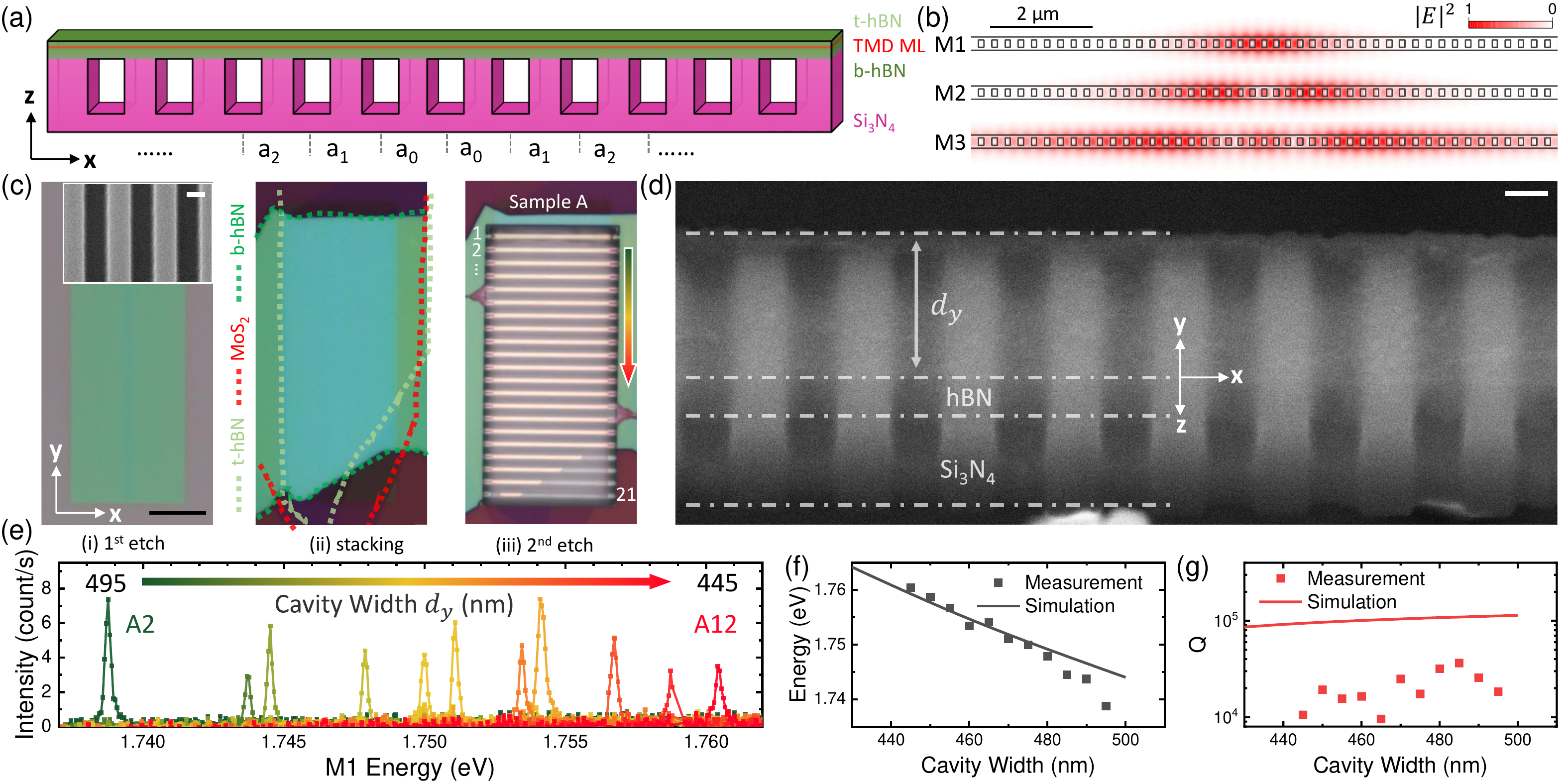}
	\caption{\label{f1}
		(a) Schematic of cavity structures.
		(b) Maps of $\vert E \vert^2$ distribution of three modes M1-3.
		(c) Images of sample A with cavities A1-A21, recorded at different stages of the fabrication process: (i) after first etch (close up in inset), (ii) after stacking the TMD heterostructure and (iii) following the second etch to define nanobeams. Here the thickness of top (bottom) hBN is 15 (55) nm.
		(d) Image of one typical cavity (tilted 45$^\circ$).
		(e) Room temperature PL spectra of cavities A2-A12 (arrow in (c)iii).
		(f)(g) Experimental and calculated cavity mode energy (f) and Q-factors (g). The refractive indexes used in the calculation are $n_{\mathrm{Si_3N_4}}=2.00$ and $n_{\mathrm{hBN}}=1.82$.
		The black scale bar in (c) is 10 $\mathrm{\mu m}$ and the white scale bar in (c)(d) is 100 nm.
	}
\end{figure*}

Here, we observe nonlocal light-matter interactions in a hBN/MoS$_2$/hBN/Si$_3$N$_4$ hybrid photonic crystal nanocavities.
Our optimized cavity structure solves the trade-off problem by integrating the hBN/MoS$_2$/hBN as a functional dielectric part of the cavity structure rather than an attachment.
Therefore, the high excitonic quality and the strong cavity-MoS$_2$ overlap are achieved simultaneously.
Our nanofabrication approach with encapsulated 2D-materials provide cavity mode Q-factors $\geq 10^4$, comparable to the best active III-V and silicon nanocavities explored for solid-state cQED experiments \cite{Yoshie2004,Hennessy2007,doi:10.1002/lpor.201200043}.
Optical spectroscopy performed as a function of lattice temperature shows that light-matter interactions between the cavity modes and spatially extended free trions (TXs) operate in the nonlocal regime \cite{PhysRevB.86.085304, doi:10.1002/qute.201900024}.
The interaction strength is shown to be consistent with the shrinking center-of-mass wavefunction of TX as temperature increases.
Our work demonstrates the significance of both optical and transport properties in hybrid 2D-material-cQED systems, and the sensitivity of nonlocal effects to the environment.
These results provide ways to the efficient control of exciton-photon interactions and enable hybrid 2D-material-cQED systems for novel optoelectronic and quantum photonic devices.


The cavity structure is presented in Fig.~\ref{f1}(a) and the typical electric field distribution calculated using finite-difference time-domain methods for the first three modes (M1-M3) are presented in Fig.~\ref{f1}(b).
Usually both the diameter of air holes and the periodicity decrease linearly in the cavity center for optimized nanobeam cavities \cite{Li2017,2010.05458}.
Here, to simplify fabrication processes all nanoscale trenches were chosen to have the same width of $140\ \mathrm{nm}$, but their separation $a_i$ ($i\in {0,1,2..}$) follows a Gaussian function $a_i/a=1-A\cdot \mathrm{exp}(-i^2/(2\sigma^2))$, where $a$ is the lattice constant and $A=0.1,\sigma=4$ define a smoothly varying photon confinement for high Q-factors \cite{Akahane2003}.

The key fabrication steps are outlined in Fig.~\ref{f1}(c), where the images of sample A ($a=270\ \mathrm{nm}$) are presented with cavities marked by A$i$ ($i\in {1,2..}$).
Fabrication began by etching nanoscale trenches into Si$_3$N$_4$.
Subsequently, large ($\geq 10^4\ \mathrm{\mu m^2}$) 2D flakes were exfoliated \cite{Huang2020} before being assembled into a hBN/MoS$_2$/hBN heterostructure on top of the etched trenches using a viscoelastic dry transfer process \cite{Pizzocchero2016}.
The samples were completed in a second etching step that divides the heterostructure into multiple parallel nanocavity beams with a width $d_y$ (Fig.~\ref{f1}(d)) that tunes the frequency of the cavity modes, followed by a final wet under-etch to produce freestanding nanobeams.
The hBN/MoS$_2$/hBN heterostructure was only etched in the second etching and is not perforated, which retains pristine excitonic properties of the MoS$_2$ monolayer and greatly reduces disorder-induced optical losses.
Detailed design and fabrication is in supplement \cite{supplement}.

Figure~\ref{f1}(e) shows typical photoluminescence (PL) spectra of the fundamental mode M1 from cavities A2-A12 with varying $d_y$.
Luminescence from the MoS$_2$ filtered through the cavity mode is observed when exciting using a 532 nm cw-laser.
The laser spot area is $1\ \mathrm{\mu m}$ and the power density is $28.8\ \mathrm{kW/cm^2}$.
We observe cavity linewidths $\hbar\gamma_C\sim\ 200\ \mathrm{\mu eV}$, close to our resolution limit $\hbar\gamma_{res}=130\ \mathrm{\mu eV}$.
Thus the measurement of exact Q-factors is limited by the spectral resolution.
The Q-factor estimated by deconvolution $\mathrm{Q}=\omega_C/\left( \gamma_C-\gamma_{res}\right)$ \cite{JIMENEZMIER1994741}, where $\omega_C$ is the cavity mode frequency, are universally $\geq10^4$ (for details see supplement \cite{supplement}).
These values are more than one order of magnitude larger than hBN cavities hitherto reported \cite{Kim2018,doi:10.1002/adom.201801344,doi:10.1021/acsnano.0c01818} and comparable to state-of-the-art III-V photonic crystal structures \cite{Yoshie2004,Hennessy2007}.
The observed functional dependence of the cavity energies and Q-factors on the extensive geometry are in good accord with simulated results as shown in Fig.~\ref{f1}(f)(g).
The experimental Q-factors are smaller than theoretical predictions due to the disorder in fabrication.

\begin{figure}
	\includegraphics[width=\linewidth]{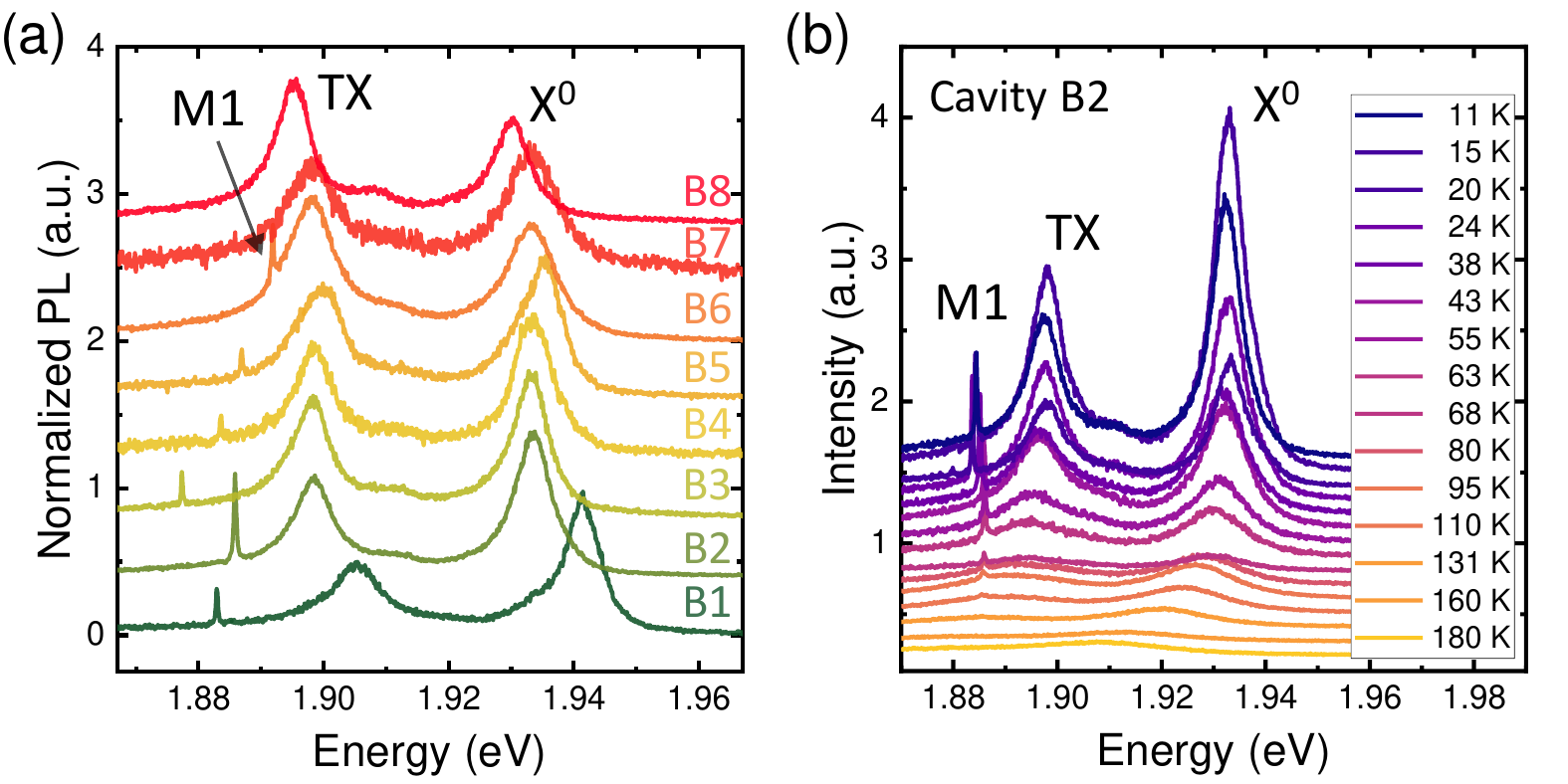}
	\caption{\label{f2}
		(a) PL spectra recorded from sample B ($a=250\ \mathrm{nm}$) with cavities B1-B8 ($d_y=530-460\ \mathrm{nm}$) at 11 K. Excitation power is $1.4\ \mathrm{kW/cm^2}$.
		Emission centered at 1.93 (1.90) eV is X$^0$ (TX).
		X$^0$ has a linewidth $\sim 5\ \mathrm{meV}$, indicative of the high excitonic quality \cite{PhysRevX.7.021026}.
		The sharp peaks on cavities B1-B6 at the red side of TX are the M1 modes.
		The nonmonotonic shift of M1 in B1-B2 reflects inhomogeneity of the hBN thickness.
		(b) Temperature-dependent PL spectra recorded from cavity B2 with excitation power $2.1\ \mathrm{kW/cm^2}$.
		As $T$ increases, the M1 peak intensity first continuously decreases as the exciton emission decreases, but then suddenly disappears after reaching the blue side of TX ($T>131\ \mathrm{K}$).
	}
\end{figure}

We continue by exploring the coupling between the MoS$_2$ monolayer and the fundamental M1-mode in our high-Q nanocavities.
In all the cavities investigated, the emission from cavity mode could only be observed when M1 is \textit{red} detuned from TX ($\Delta\omega=\omega_{TX}-\omega_{M1}\geq0$).
This can be clearly seen in Fig.~\ref{f2}, where the PL spectra is measured from cavities B1-B8 on another sample B (Fig.~\ref{f2}(a)), and from the cavity B2 where $\Delta\omega$ is mainly tuned via the temperature-dependent $\omega_{TX}$ (Fig.~\ref{f2}(b)).
The bare cavity mode is also slightly tuned by the temperature but much less compared to $\omega_{TX}$ (for full details see supplement \cite{supplement}).
In both sets of data, emission from the M1 mode quenches as $\Delta\omega<0$.
We note that this observation is distinct from previous studies of nonencapsulated monolayers attached to pre-fabricated nanocavities, where emission from cavity modes could be readily observed even when its energy is higher than the neutral exciton (X$^0$) \cite{Li2017}.
We traced this phenomenon to the combined impact of reabsorption of cavity photons by the MoS$_2$ monolayer for long photon lifetimes (high-Q) and the continuous energy spectrum of free excitons \cite{10.1002/9783527628155.nanotech004}.
This differs significantly from the situation with discrete quantum emitters \cite{10.1002/9783527628155.nanotech004}, such as QDs, due to the increased phase space of exciton states with non-zero momentum that can be accessed via inelastic scattering for blue detunings ($\Delta\omega<0$).
We estimate the impact of enhancing the photon lifetime in the high-Q cavity.
Taking a typical absorption coefficient of $\alpha=2.8\times{10}^4\ \mathrm{m^{-1}}$ \cite{KWAK2019102202}, we estimate that reabsorption of cavity photons by the MoS$_2$ becomes significant for $\alpha(\mathrm{Q}/\omega_C)(c/n_r)\gg 1$ where $\mathrm{Q}/\omega_C$ is the cavity photon lifetime, $c$ is the speed of light and $n_r$ is the refractive index.
For the maximum cavity-TMD overlap, we estimate $Q\gg900$ to denote the limit beyond which reabsorption begins to limit the exciton-photon interaction.
Our cavity Q exceeds this estimated threshold and, moreover, the MoS$_2$ monolayer is inserted close to the antinode of the cavity mode.
Thus, the cavity mode quenches for blue detunings $\Delta\omega<0$.

\begin{figure}
	\includegraphics[width=\linewidth]{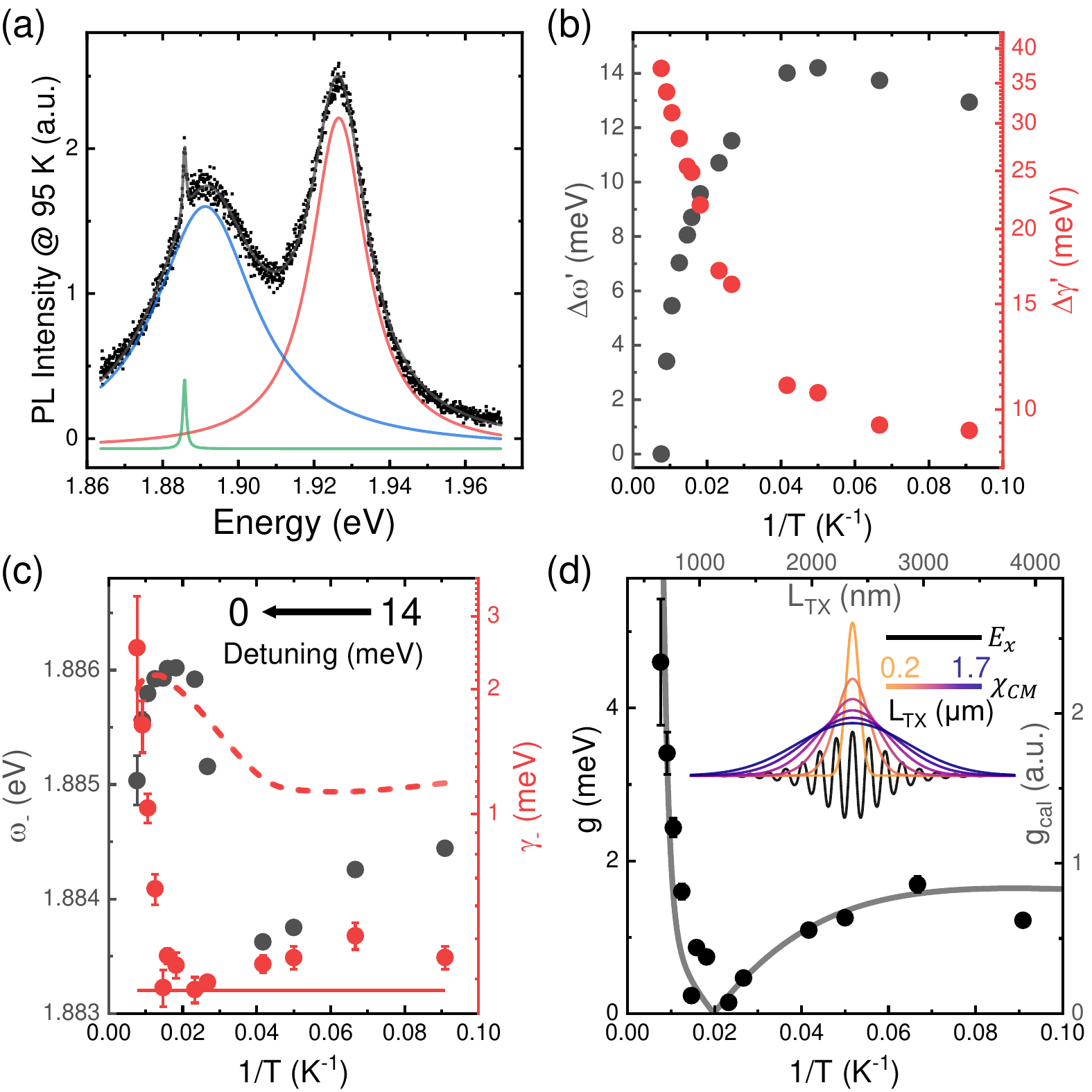}
	\caption{\label{f3}
		(a) PL spectra at 95 K and the multi-Lorentz fitting.
		(b) $\Delta\omega'$ and $\Delta\gamma'$ extracted from experimental data.
		(c) $\omega_-$ (energy of the cavity-branch peak) and $\gamma_-$. Red solid line is the bare cavity linewidth $\gamma_C$. Red dashed line is the $\gamma_-$ predicted with a constant interaction strength $g'$.
		(d) Results of interaction strength $g$. Points are experimentally extracted at different temperatures. The solid line shows theoretical results calculated with the cavity electric field $E_x$ from mode M1 and the center-of-mass wavepacket $\chi_{CM}$ with different spatial extent $L_{TX}$ presented in the inset.
	}
\end{figure}

To analyze the coupling of TXs to the high-Q cavity mode, we extracted the peak energy and linewidth of the cavity mode and TX from the spectra in Fig.~\ref{f2}(b) by multi-Lorentz fitting, e.g., the spectra at 95 K presented in Fig.~\ref{f3}(a).
Since the interaction is at low excitation level without non-linearity (for full details see supplement \cite{supplement}), the Jaynes-Cummings model is used to model the cavity-TX interaction \cite{2010.05458}.
Hereby, when the cavity-TX interaction is weak, the interaction strength $g$ can be calculated from the linewidth of cavity-like polariton $\gamma_-$, the energy detuning ($\Delta\omega'$) and linewidth difference ($\Delta\gamma'$) between the exciton-like and cavity-like polariton branch by $\gamma_-=\gamma_C+g^2\Delta\gamma'/\left( \Delta\omega'^2+1/4\Delta\gamma'^2 \right)$.
The experimental extracted parameters are presented in Fig.~\ref{f3}(b)(c), and the results of $g$ are summarized in Fig.~\ref{f3}(d) as the central results of this work.
Fist, the values of $g$ are far smaller than that expected of a strong coupling thus the small $g$ approximation in the calculation is right.
Furthermore, the experimentally determined $g$ (black dots) is found to be highly nonmonotonic as the temperature $T$ increases (detuning reduces):
$g$ first \textit{reduces} with increasing $T$ before \textit{increasing} rapidly - a behaviour that can be accounted for by nonlocal light-matter interactions of mobile excitons in the hybrid nanocavity \cite{PhysRevB.86.085304,doi:10.1002/qute.201900024}.
This situation can occur, e.g., for plasmonic systems with tightly confined electromagnetic modes \cite{Andersen2011,PhysRevB.97.245405,Iranzo291} or in situations where the exciton wavefunction has a large spatial extent, e.g., large QDs \cite{PhysRevLett.122.087401}.
Additionally, it can be observed that the $\gamma_-$ predicted by a constant $g'$ of 4 meV (red dashed line in Fig.~\ref{f3}(c)) is clearly different to the experimental observation thus a constant $g'$ is obviously not our case.

We continue to explain how this behaviour is expected in the nonlocal regime of light-matter interactions.
In the local regime, $\mathbf{E}\left( \mathbf{r}\right)$ which usually involves $e^{i\mathbf{k}\cdot\mathbf{r}}$ ($\mathbf{k}$ is the wave vector) is approximately constant within the exciton wavefunction.
This translates to $\mathbf{k}\cdot\mathbf{r}\ll 1$ for all $\mathbf{r}$ within the electron and hole wavefunctions, whereas nonlocal regime occurs when $\mathbf{k}\cdot\mathbf{r}$ is non-negligible \cite{PhysRevB.86.085304,doi:10.1002/qute.201900024,PhysRevLett.122.087401}.
The free excitons in the hBN encapsulated MoS$_2$ are generated locally but their center-of-mass samples positions having different local cavity field ($\mathbf{k}\cdot\mathbf{r}\gg 1$) before recombining.
Hereby, TXs were modelled in the weak confinement regime (effective Bohr radius $\ll$ the spatial extent of the center-of-mass wavefunction) for which the total wavefunction is separable into components arising from the center-of-mass motion $\chi_{CM}\left(\mathbf{R}\right)$ and internal dynamics of the e-h pair $\chi_{rel}\left(\mathbf{r'}\right)$, respectively.
$\mathbf{R}=\left(m_e\mathbf{r_e}+m_h\mathbf{r_h}\right)/\left(m_e+m_h\right)$ corresponds to the center-of-mass motion and $\mathbf{r'}=\mathbf{r_e}-\mathbf{r_h}$ accounts for the internal relative motion of the e-h pair \cite{PhysRevB.86.085304,Goryca2019}.
Due to the large exciton binding energy in hBN-encapsulated TMDs \cite{doi:10.1021/acs.nanolett.6b03276,PhysRevLett.120.057405,Goryca2019}, $\chi_{rel}\left(\mathbf{r'}\right)$ only extends over a few nanometers \cite{RevModPhys.90.021001,Goryca2019}.
Moreover, it is expected to be fully independent of temperature since the exciton binding energy is $\gg k_B\cdot T$.
In contrast, due to exciton-phonon coupling, the spatial extent of the center-of-mass wavefunction has a negative temperature relation in hBN-encapsulated TMDs \cite{Wang2016,C9NR07056G,doi:10.1021/acsnano.6b05580}.
Generally acoustic photons result in $L_{TX}$ that varies sub-linearly with $1/T$, while optical phonons result in $L_{TX}$ that varies super-linearly \cite{Zinovev1983}.
We performed calculations using a Gaussian wavepacket to describe the center-of-mass motion $\chi_{CM}\left(x\right)=\left(1/\pi\right)^{1/4}(1/L_{TX})^{1/2}e^{-x^2/2L_{TX}^2}$ \cite{PhysRevB.86.085304}, and considered the spatial extent $L_{TX}$ following a $1/T$ dependence (colored lines in Fig.~\ref{f3}(d) inset).
As the unit cell in MoS$_2$ ($\sim 0.3\ \mathrm{nm}$) is much smaller than the TX emission wavelength, the interaction strength is $g_{cal}=\vert\int \chi_{CM}(x)E_x(x)dx\vert$ \cite{PhysRevB.86.085304} where $E_x(x)$ is the cavity electric field of mode M1 (black line in Fig.~\ref{f3}(d) inset).
The resulting $L$-dependence of $g_{cal}$ is presented by the grey line in Fig.~\ref{f3}(d) and reproduces our experimental findings (dots) remarkably well, despite the simplicity of our model.
Therefore, the non-trivial temperature dependence of cavity-TX interaction are best reproduced by the nonlocal light-matter interaction physics.
At $T=300\ \mathrm{K}$, the corresponding $L_{TX}$ is $480\pm 30\ \mathrm{nm}$.
Considering the non-degenerate single optical mode in the cavity is spatially coherent, and the enhancement by hBN encapsulation and Si$_3$N$_4$ stress \cite{C9NR07056G,Chai2017,John_2020}, this value of $L_{TX}$ close to the recently reported diffusion length of $300\ \mathrm{nm}$ is a reasonable result \cite{doi:10.1021/acsnano.0c05305}.
Here, we note that the extent of the center-of-mass wavefunction and the exciton diffusion length both describe the spatial distribution of excitons.
However, they are not strictly equivalent since the exciton diffusion length also includes inelastic scattering processes.
The theoretical analysis used here is based on the center-of-mass wavefunction that is not necessarily equal to the diffusion length.
Thus, any distinction does not have any substantive impact on the conclusions drawn from Fig.~\ref{f3}.

Other factors such as gas condensation and localized excitons (LXs) are unlikely to explain the nonmonotonic $g$ since they have very little affect on the cavity polariton linewidth (for details see supplement \cite{supplement}).
In addition, we note that the center-of-mass wavefunction $\chi_{CM}\left(\mathbf{R}\right)$ is \textit{not} equivalent to a Gaussian spatial distribution of excitons $\chi^2_{CM}(x)$.
In the latter case, the interaction strength would be $g_{cal}^2=\int \chi^2_{CM}(x)E_x^2(x)dx$ corresponding to a monotonic decreasing $g$ as $L$ increases (for details see supplement \cite{supplement}).
Conversely, here $E_x(x)$ has both positive and negative values depending on the position $x$, and thus the integral $\int \chi_{CM}(x)E_x(x)dx$ has both positive and negative contributions, resulting in the minimum interaction strength for specific constellations, such as when $L_{TX}=1.1\ \mathrm{\mu m}$ around twice the wavelength of the cavity electric field in Fig.~\ref{f3}(d).


In summary, we explored novel light-matter interaction regimes of free trions in a hybrid high-Q photonic crystal nanocavity embedded with pristine hBN/TMD/hBN heterostructures.
The optimized structure provided quasi 0D modes with $Q>10^4$, exciton linewidths of MoS$_2$ approaching homogeneous limit and large cavity-MoS$_2$ overlap.
These advances facilitated the demonstration of nonlocal interaction between the cavity and free trions.
Since our approaches can be applied to any 2D materials, our work provides an ideal platform to investigate cQED and quantum photonics using 2D materials.
Therefore, additional interesting phenomena could be expected in future work, e.g., the interaction with site-selectively generated defects \cite{Klein2019} or Moiré exciton lattices \cite{Seyler2019,Baek2020}, towards highly scalable quantum photonic devices.

\begin{acknowledgments}
	All authors gratefully acknowledge the German Science Foundation (DFG) for financial support via grants FI 947/8-1, DI 2013/5-1 and SPP-2244, as well as the clusters of excellence MCQST (EXS-2111) and e-conversion (EXS-2089). C. Q. and V. V. gratefully acknowledge the Alexander v. Humboldt foundation for financial support in the framework of their fellowship programme.
	K. W. and T. T. acknowledge support from the Elemental Strategy Initiative conducted by the MEXT, Japan (Grant Number JPMXP0112101001) and JSPS KAKENHI (Grant Numbers 19H05790 and JP20H00354).

\end{acknowledgments}

\newpage

\section*{Supplementary Information}
\setcounter{figure}{0}
\renewcommand{\figurename}{SFig.}

\section{\label{sec1}Introduction}

Here we present the supplementary information for investigations on cavity quantum electrodynamics (cQED) with high-Q nanocavity and transition metal dichalcogenide (TMD) monolayers full encapsulated with hexagonal boron nitride (hBN), including the nanocavity design, the fabrication method and the additional experiment results.
In Sec.~\ref{sec2}, we present the nanocavity design.
3D finite difference time domain (FDTD) method is used to simulate cavity properties including mode frequency, quality factor (Q) and field distribution.
For the interaction between quasi 0D cavity and TMDs, a problem is the trade-off between excitonic properties and the cavity-TMD overlap \cite{2010.05458}.
As discussed in the main paper, hBN-encapsulation will move the TMD monolayer away from the nanophotonic structure thus weaken the cavity-TMD coupling.
Therefore, directly fabricating nanophotonic structure on hBN \cite{Kim2018,doi:10.1002/adom.201801344}, or integrating hBN as a functional dielectric part of the cavity \cite{doi:10.1021/acsnano.0c01818} are potential ways to solve this problem.
However, till now hBN is hard to etch perfectly \cite{doi:10.1116/1.4826363,Caldwell2014}, which limits the Q-factor of hBN or hBN/Si$_3$N$_4$ cavities to a few thousands \cite{Kim2018,doi:10.1002/adom.201801344,doi:10.1021/acsnano.0c01818}.
Moreover, etching nanostructures through hBN will also perforate the TMD monolayer, affecting the pristine optical and transport properties of excitons. 
Here we propose a transverse nanobeam photonic crystal cavity with hBN/TMD/hBN/Si$_3$N$_4$ hybrid structures to solve these problems.
Compared to the conventional nanobeam cavity, the anti-rough property of our cavity is one order of magnitude better.
Furthermore, nanoscale etching in the TMD monolayer is avoided, retaining the pristine optical and transport properties of excitons as much as possible.
These advantages provide the basis for the observation of new exciton-photon interaction phenomena as discussed in the main paper.

In Sec.~\ref{sec3}, we introduce detailed fabrication methods consisting of the nanofabrication and the transfer of 2D materials.
The nanofabrication mainly contains ebeam lithography (EBL) and inductively coupled plasma reactive ion etching (ICPRIE), which are common technologies widely applied in nanophotonics.
From scanning electron microscope (SEM) images, we observe an ultra-smooth etching of the periodic nanoscale trenches in Si$_3$N$_4$ and a rough etching of hBN, consistent to the theoretical prediction discussed in Sec.~\ref{sec2}.
The 2D flakes are exfoliated on polyvinyl alcohol (PVA) films \cite{Huang2020} which helps achieving huge flakes ($>\mathrm{40\ \mu m}$), and then stacked with the dry transfer method \cite{Pizzocchero2016}.
During these processes, the contamination is generated.
Therefore a vacuum annealing \cite{Jain_2018} is implemented to clean the cavities, with which a great improvement in the lifetime and emission intensity of TMD excitons are observed.

In Sec.~\ref{sec4}, we introduce measurement setups and present additional experimental results, along with further discussions which strengthen conclusions in the main paper and develop interesting topics for future investigations.
Firstly, properties of bare cavity mode and bare TMD excitons are presented, providing comparisons to the exciton-photon polaritons.
The linewidth of cavity modes measured from PL spectra (45 pm minimum) is limited by the setup resolution ($\sim$ 50 pm).
Thereby, the deconvolution is used to estimate actual Q values as discussed in the main paper.
Here we present the high-resolution reflectivity measurement to further demonstrate the setup resolution.
At low temperature, TMD excitonic properties in the cavity, such as the energy, linewidth and power-dependence of neutral exciton (X$^0$), trion (TX) and localized exciton (LX), generally agree well to previous literatures \cite{PhysRevX.7.021026,Shree_2019}.
Secondly, additional data for the cavity-TX interaction is presented and discussed.
The power-dependent photoluminescence (PL) spectra demonstrate the interaction in the linear regime at low excitation levels.
Finally, we present some further discussions on the non-local interaction.
The transport properties of TMD excitons are a non-trivial problem and not fully investigated yet \cite{Zinovev1983,doi:10.1021/acsnano.6b05580,C9NR07056G}, depending on the intrinsic properties, exciton-phonon scatterings, excitation conditions and many other factors.
In addition to the linear shrinking of the center-of-mass wavefunction used in the main paper, we also show the result with a super-linear shrinking which can also well fit the experiment results.
Besides the lattice temperature used in this work, we discuss some other potential methods to control the non-local effects such as the external electromagnetic field.
These discussions indicate potential further investigations on the hybrid 2D-material cQED system.

\section{\label{sec2}Nanocavity Design}

In this section, FDTD simulations are calculated with Lumerical FDTD Solutions.
The decay of cavity Q with rough hBN etching sidewall is simulated.
Results show the cavity Q decay one order of magnitude smaller in our transverse cavity compared to the conventional cavity.
Additionally, simulations with other possible fabrication errors are also presented.
Only a few cavity Q decay is observed, demonstrating the robustness of our optimized cavity.
Near field extension and far field profile are also calculated and discussed.

\subsection{\label{sec2a}General Design}

\begin{figure*}
	\includegraphics[width=\linewidth]{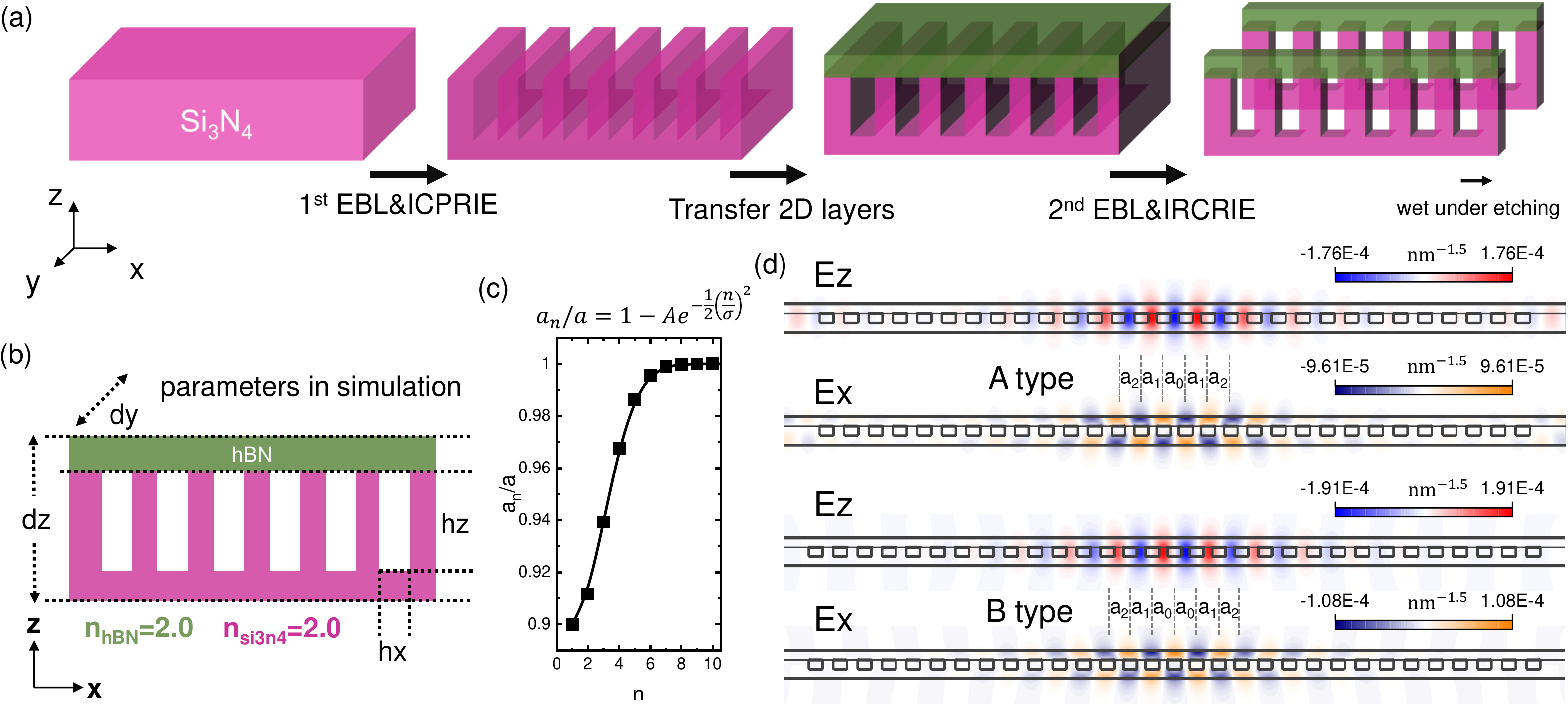}
	\caption{\label{des}
		General design of cavities.
		(a) Schematic of structures and fabrication processes. First EBL and ICPRIE to fabricate nanoscale trenches in Si$_3$N$_4$, followed by stacking 2D flakes on top. Then second EBL and ICPRIE to divide nanobeams. Final wet under etching to remove Si below Si$_3$N$_4$.
		(b) Parameters used in FDTD simulations. For brevity we use 2.0 for the refractive index of both hBN and Si$_3$N$_4$.
		(c) The separation between nanoscale trenches follows a Gaussian distribution.
		(d) The cavity mode function of two types of cavities. The main difference is that, for type A the 2D flakes in cavity center stands on Si$_3$N$_4$, with strong $E_z$ field. Whilst for type B the 2D flakes in cavity center suspends on trenches, with strong $E_x$ field.
	}
\end{figure*}

To integrate the hBN or hBN-encapsulated TMD into nanocavities, we propose the transverse 1D nanobeam photonic crystal cavity, due to the obvious advantages of pristine unetched TMD monolayer and smooth periodic nanoscale trenches.
SFigure~\ref{des}(a) shows the designed fabrication processes, including two steps of nanofabrication (EBL + ICPRIE) and one step of stacking 2D flakes between the two nanofabrications.
These technologies are all common for nanophotonics or 2D materials, thus the design has a high feasibility.
If we neglect the difference between hBN and Si$_3$N$_4$, the structure is almost same to conventional cavities by just exchanging y- and z- direction, thus here we call it transverse nanobeam cavity.
SFigure~\ref{des}(b) shows the parameters in FDTD simulations, where $h_{x,z}$ is the width and depth of nanoscale trenches, $d_z$ is the whole nanobeam height in z-direction and $d_y$ is the whole nanobeam width in y-direction.
$\left(d_z+h_z\right)/2$ must be 200 nm which is the value of Si$_3$N$_4$ thickness on the Si substrate.
The separation between nanoscale trenches $a_n$ follows the Gaussian distribution $a_n/a=1-A\cdot \mathrm{exp}(-n^2/(2\sigma^2))$ (SFig.~\ref{des}(c)) to form a smoothly varying photon confinement in the nanobeam center \cite{Akahane2003,Kim2018}.
$a$ is the lattice constant for the photonic crystal. $A$ and $\sigma$ describe the Gaussian profile.
We note that the refractive index of hBN $n_{hBN}$ is not well known yet.
It is reported anisotropic and the value varies between $1.8-2.4$ in different experiments \cite{doi:10.1002/pssb.201800417,Rah:19}.
For simplify we set 2.0 for both the refractive index of hBN and Si$_3$N$_4$ in most simulations, which results in the energy difference between simulation and experiment.
E.g., $n_{hBN}$ is specifically corrected to 1.82 in simulations to fit the experimental results in Fig.~\ref{f1} in the main paper.

For the cavity-TMD interaction, we mainly focus on the cavity electric field at the position of the TMD monolayer.
SFigure~\ref{des}(d) shows the cavity electric field distribution of two types of cavities A and B.
Type A has one $a_0$ in the center.
Within the hBN region where TMD monolayer could be encapsulated, $E_z$ is symmetric in x direction while $E_x$ is antisymmetric.
The 2D material in the cavity center ($x=0$) stands on Si$_3$N$_4$.
Type B has two $a_0$ in the center. Oppositely, $E_z$ is antisymmetric in x direction while $E_x$ is symmetric. The 2D material in the cavity center is suspended.
For a quantum emitter placed at the cavity center such as quantum dot, type A is usually used in conventional cavities (center point in type B is etched) \cite{Li2017,Kim2018}.
In this work, the TMD monolayer is in-plane to xy-plane, and we mainly focus on intralayer excitons.
The in-plan exciton dipole interacts with $E_x$, thus the x-polarized cavity field ($E_x$ both symmetric in x and y direction) in type B cavity improves the interaction.
As $E_x$ is strongest around the top (SFig.~\ref{des}(d)), the position of TMD monolayer is optimized by a thin top hBN and a thick bottom hBN for the strong $E_x$.
In contrast, type A cavity the out-plane $E_z$ could potentially couple to the spin-forbidden dark excitons or interlayer excitons with out-of-plane dipoles \cite{Zhou2017,Rivera2018}.

Considering the Si$_3$N$_4$ thickness and target cavity energy around $1.7-1.9$ eV, optimized parameters are achieved as $a=\mathrm{270\ nm}$, $h_x=\mathrm{120\ nm}$, $h_z=\mathrm{150\ nm}$ (thus total hBN thickness is 50 nm), $d_z=\mathrm{500\ nm}$, $A=0.1$ and $\sigma=4$ for the type B cavity after series of simulations.
Three modes M1-3 are observed in the simulation with cavity frequencies at 429, 415 and 406 THz, corresponding to the three cavity field distribution in Fig.~\ref{f1} in the main paper.
Following theoretical works are generally based on these parameters.
The quantized electric field in the cavity is
\begin{eqnarray}
	\mathbf{E}\left(\mathbf{r}\right)=i\left(\frac{\hbar\omega_{C}}{2\epsilon_r\epsilon_0}\right)^{\frac{1}{2}}\lbrack a\bm{\alpha}\left(\mathbf{r}\right)-a^{+}\bm{\alpha^\ast}\left(\mathbf{r}\right) \rbrack \nonumber
\end{eqnarray}
where $a^{+}$ and $a$ are the creation and destruction operators of a photon in the cavity mode and $\bm{\alpha}\left(\mathbf{r}\right)$ is the normalized cavity mode function. $\hbar\omega_C$ is the cavity energy and $\epsilon_r$ ($\epsilon_0$) is the relative (vacuum) permittivity.
The cavity mode function $\bm{\alpha}\left(\mathbf{r}\right)$ is calculated by normalizing the electric field
\begin{eqnarray}
	\bm{\alpha}\left(\mathbf{r}\right)=\frac{\epsilon_r\left(\mathbf{r}\right)\cdot\mathbf{E_{cal}}\left(\mathbf{r}\right)}{\sqrt{\Delta V \sum_{r'} \vert\epsilon_r\left(\mathbf{r'}\right)\cdot\mathbf{E_{cal}}\left(\mathbf{r'}\right)\vert^2}} \nonumber
\end{eqnarray}
where $\mathbf{E_{cal}}\left(\mathbf{r}\right)$ is the electric field extracted from the FDTD simulation, and $\Delta V$ is the volume of one mesh unit in the simulation (therefore $\sum{\vert\bm{\alpha}\left(\mathbf{r}\right)\vert^2}\Delta V=1$).
For brevity we use the cavity mode function $\bm{\alpha}\left(\mathbf{r}\right)$ to describe the electric field in this work, e.g., $E_x$ means $\alpha_x$ and $E_z$ means $\alpha_z$.

\subsection{\label{sec2b}Roughness of hBN}

\begin{figure}
	\includegraphics[width=\linewidth]{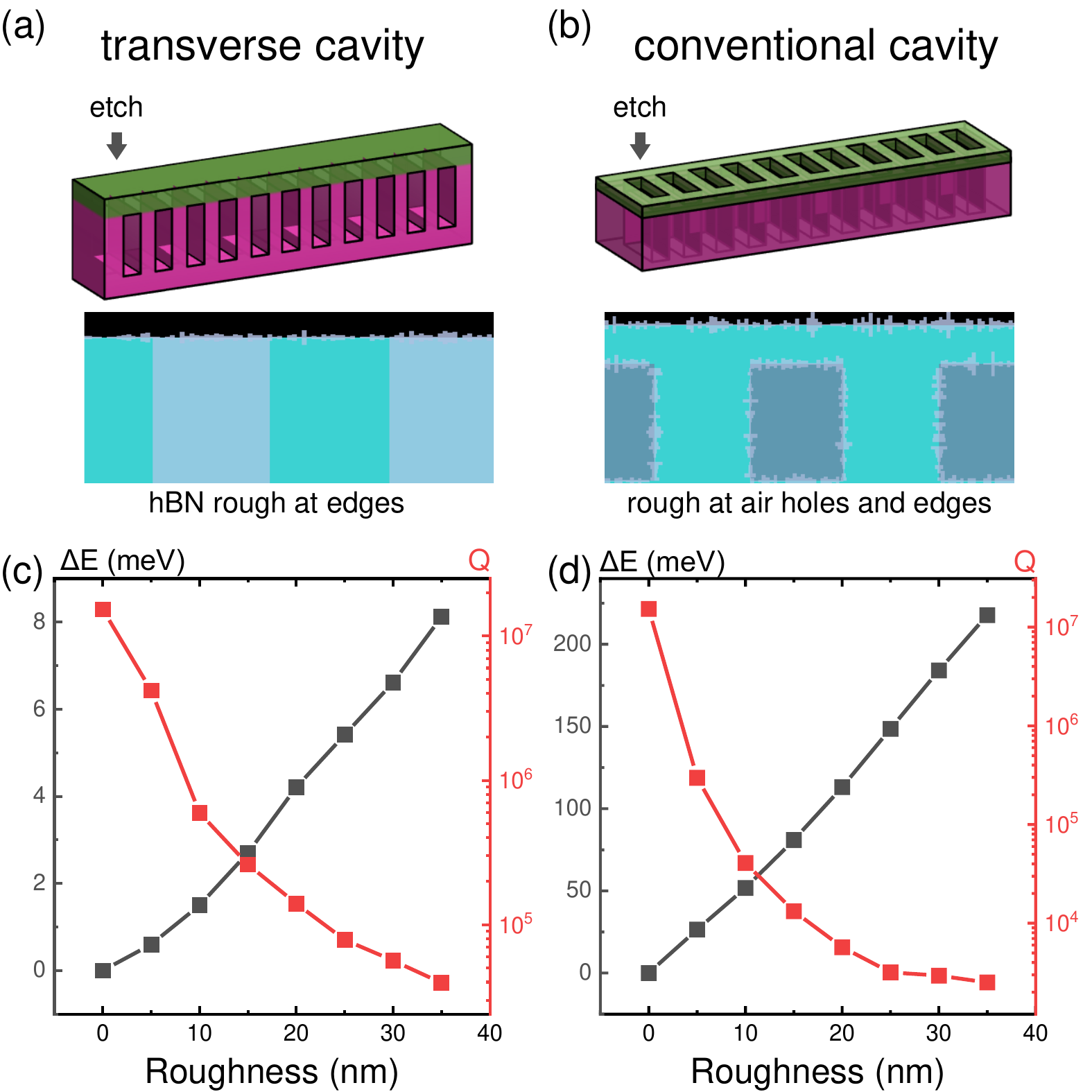}
	\caption{\label{rou}
		Comparison between the transverse and conventional cavity.
		(a) Schematic of the transverse structure. Due to the etching direction, hBN roughness will exist at 2 side surfaces.
		(b) Schematic of the conventional structure. Due to the etching direction, hBN roughness will exist both in nanoscale rectangle holes and side surfaces.
		(c)(d) Simulation results for (c) the transverse and (d) conventional cavity vs. etching roughness.
	}
\end{figure}

Here small etching cubes are added to simulate the rough hBN etching sidewalls as shown in SFig.~\ref{rou}(a)(b), with the size $\vert\Delta_{rough}\cdot Rn\vert$ where $\Delta_{rough}$ is the roughness and $Rn$ is a random number following standard normal distribution.
SFigure~\ref{rou}(a) shows the situation for the transverse cavity.
Etching of hBN only occurs in the 2nd nanofabrication, thus the roughness only exists in two boundaries.
SFigure~\ref{rou}(b) shows the conventional cavity.
As nanoscale rectangle holes are also etched in hBN, the roughness exists in both rectangle holes and boundaries.
The cavity energy shift and Q decrease of M1 mode are shown in SFig.~\ref{rou}(c)(d), and results in the transverse cavity are both one order of magnitude better than those in the conventional cavity.
This is not surprise, because in photonic crystals the periodic holes form the photonic band gap which confines the photons, thus roughness in the periodic holes is more fatal.

\subsection{\label{sec2c}Other Fabrication Errors}

\begin{figure}
	\includegraphics[width=\linewidth]{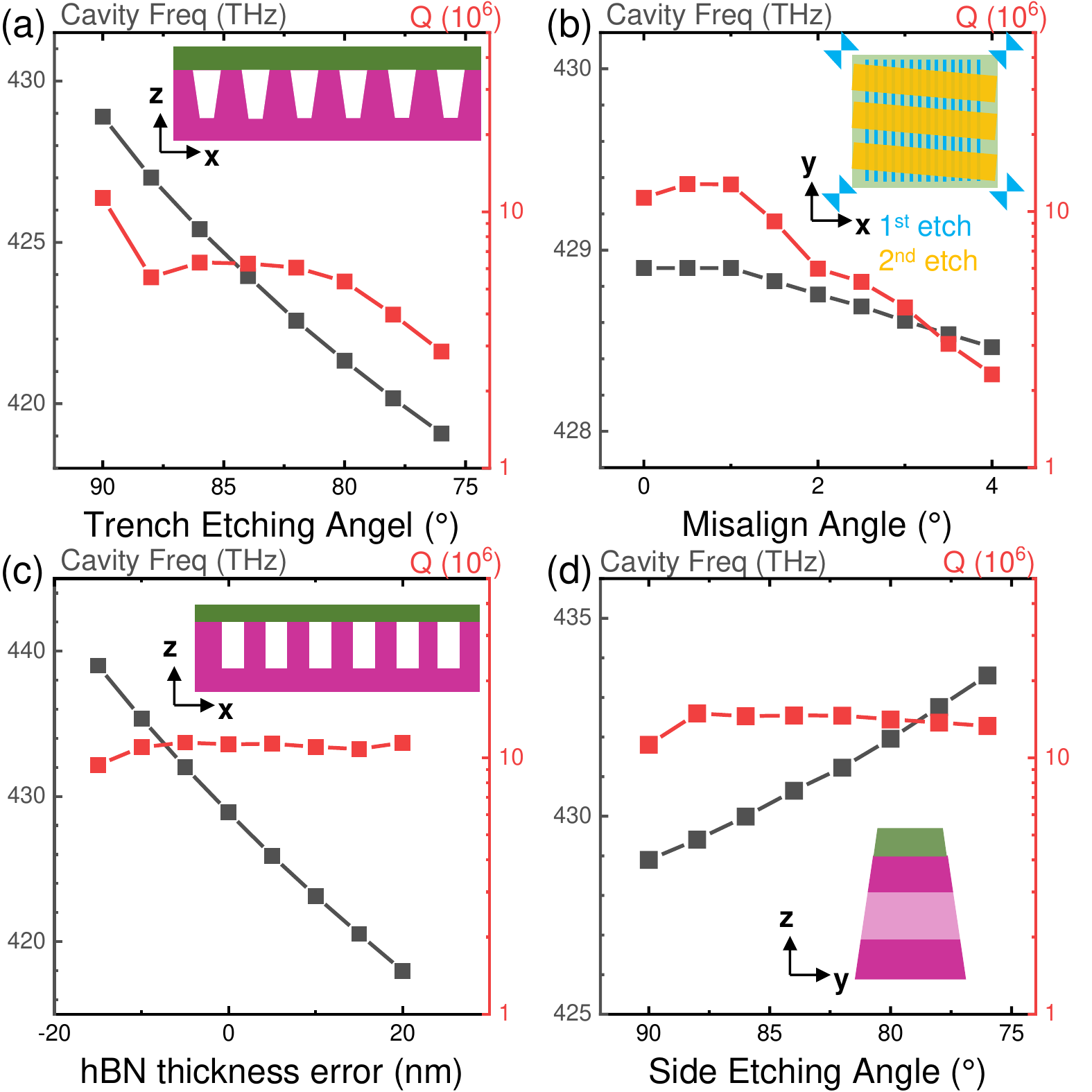}
	\caption{\label{rob}
		Robustness with other fabrication errors.
		(a) With etching angle in nanotrenches.
		(b) With misalignment between 1st and 2nd nanofabrication. (inset) From bottom to top: blue patterns are nanotrenches (1st etching), green patterns are the 2D flakes, and yellow patterns are the 2nd etching to divide nanobeams.
		(c) With hBN thickness error.
		(d) With etching angle in 2nd nanofabrication.
	}
\end{figure}

Besides the roughness, other possible fabrication errors are also considered.
SFigure~\ref{rob}(a) shows the simulation with etching angle in nanotrenches.
The etching angle of Si$_3$N$_4$ in the recipe we use is $\sim$ 83$^\circ$ as shown latter in Sec.~\ref{sec3b}.
Here for brevity we set a constant etching angle as shown in SFig.~\ref{rob}(a) inset.
With etching angle of $76^\circ$, the decrease of Q is smaller than one order of magnitude.
SFigure~\ref{rob}(b) shows the simulation with alignment error between the 1st and 2nd nanofabrication.
As shown in SFig.~\ref{rob}(b) inset, obviously a small misalignment of x-shift or y-shift in the 2nd etching (yellow part) to the 1st etching (blue part) will not affect the final result of cavities.
The rotation misalignment (between the yellow and blue part) is the only factor which makes difference.
With the misaligned angle of $4^\circ$, which is already a huge value for a normal EBL machine, the decrease of Q is also smaller than one order of magnitude.
We also considered other possible errors such as hBN thickness and etching angle in the 2nd etching, as shown in SFig.~\ref{rob}(c)(d).
Within reasonable errors, little Q decrease is observed.
In addition to the results with roughness in Sec.~\ref{sec2a}, the FDTD simulations demonstrate the high robustness of the transverse cavity.

\subsection{\label{sec2d}Extension}

\begin{figure}
	\includegraphics[width=\linewidth]{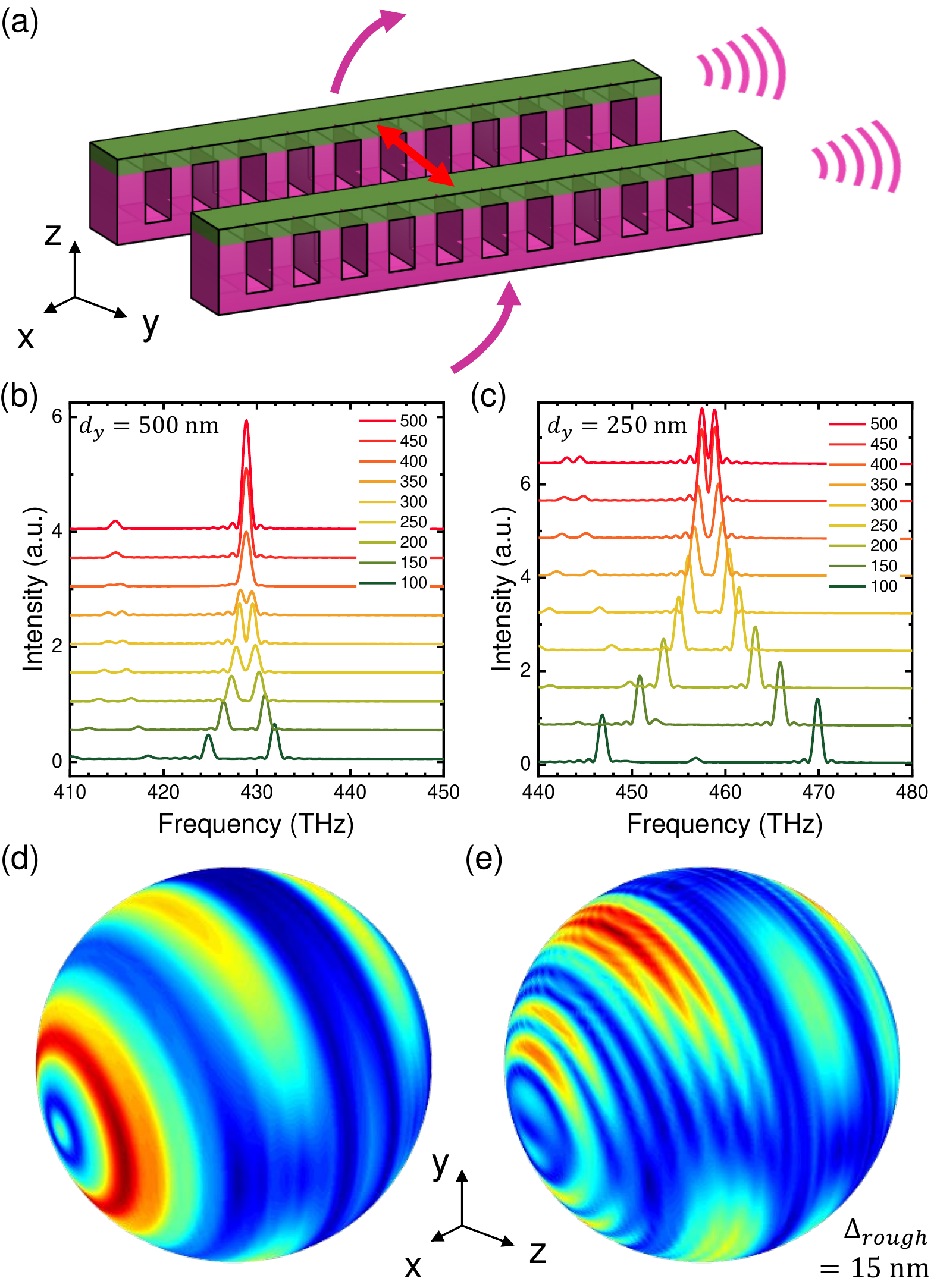}
	\caption{\label{far}
		Near-field and far-field extension.
		(a) Schematic of the on-chip integration. Cavities, waveguides and grating couples could couple to each other and be integrated on one chip.
		(b) Simulated spectra of two coupled cavities, with the gap distance between two cavities from 100 to 500 nm.
		The cavity-cavity coupling disappears with gap larger than 500 nm.
		(c) Same simulation in (b) but with cavity width $d_y=\mathrm{250\ nm}$. The coupling is $\sim$3 times stronger than that in (b) with $d_y=\mathrm{500\ nm}$.
		(d) Far-field profile of the cavity (coordinate attention).
		(e) Far-field profile of the cavity with a roughness (Sec.~\ref{sec2b}) of 15 nm.
	}
\end{figure}

On-chip integrability is one feature of photonic crystal nanocavities.
For the nanobeam cavity, the easiest way to couple cavities is to control the gap distance between them, as shown in SFig.~\ref{far}(a).
The results with cavity width $d_y=\mathrm{500\ nm}$ are presented in SFig.~\ref{far}(b), where almost no coupling is observed with the gap of 500 nm.
Therefore, in the real sample the gap between two cavities 1.5 $\mathrm{\mu m}$ is enough to avoid the cavity-cavity coupling.
The mode splitting (coupling strength between two nanocavities) is $\sim$ 7 THz with the gap distance of 100 nm as shown in SFig.~\ref{far}(b).
Smaller $d_y$ would improve the cavity-cavity coupling strength.
As shown in SFig.~\ref{far}(c), with $d_y=\mathrm{250\ nm}$, the mode splitting with the gap distance of 100 nm is $\sim$ 23 THz which increases over 3 times compared to that with $d_y=\mathrm{500\ nm}$.

SFigure~\ref{far}(d) shows the far-field profile of cavity mode M1, calculated by collecting electromagnetic field at a cubic box ($16\times 1.6\times 1.6\ \mathrm{\mu m}$ size in xyz) which envelope the nanocavity ($18\times 0.5\times 0.3\ \mathrm{\mu m}$ size) in the center.
The radiation of cavity mode in y direction is quite strong.
This is why in PL spectra we usually observe cavity peaks from neighbour cavities.
We note that in reality, rough surfaces should play the main role in far field profile because they dominate the photon losses (Q decay).
The rough-related radiation direction is somehow random, so the real far-field profile might be different to the simulation results, such as the simulation results with a roughness of 15 nm (Sec.~\ref{sec2b}) shown in SFig.~\ref{far}(e).

\section{\label{sec3}Fabrication Methods}

\begin{figure*}
	\includegraphics[width=\linewidth]{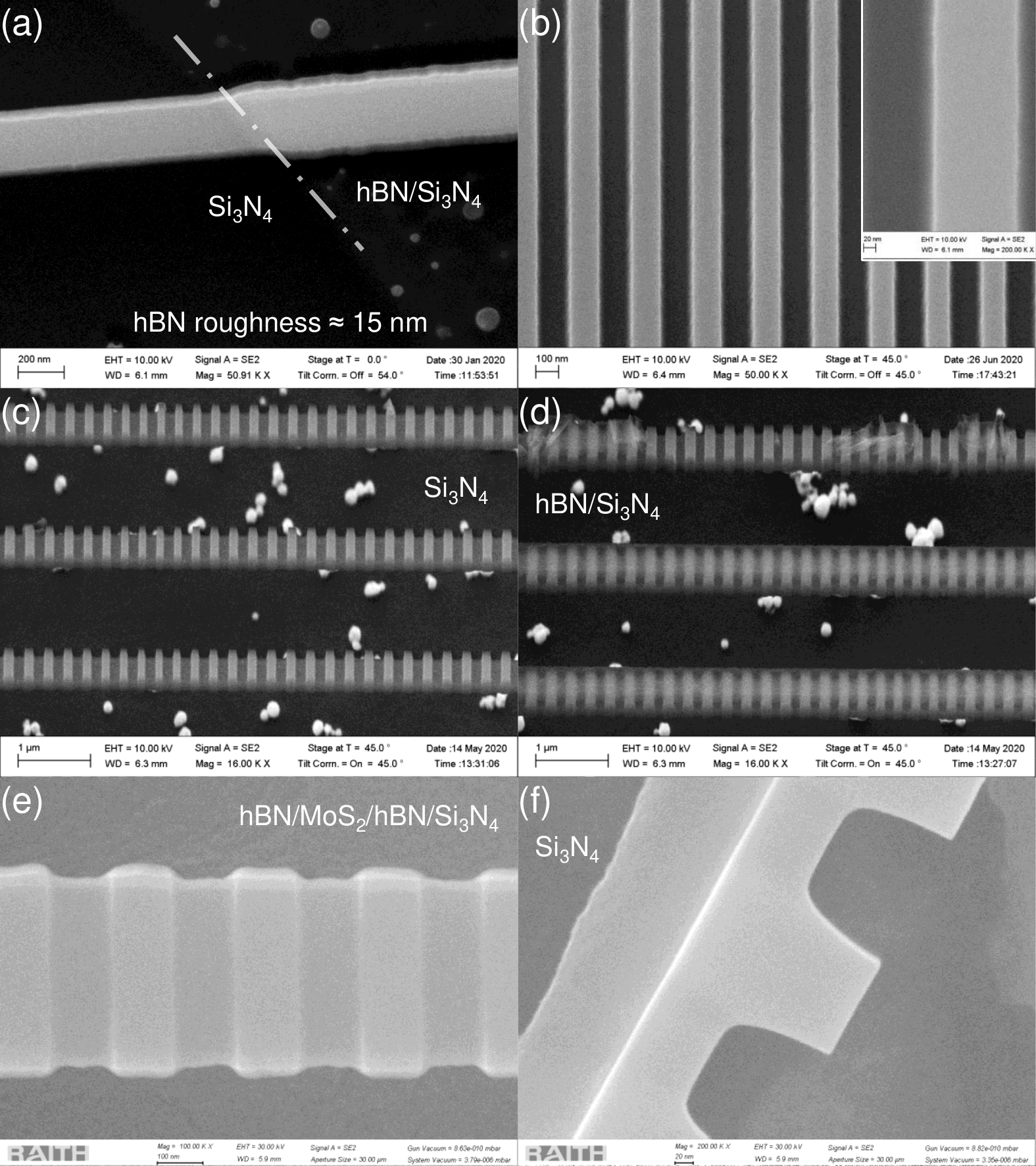}
	\caption{\label{fab}
		SEM images during nanofabrication processes.
		(a) An etching test sample.
		(b) Nanoscale trenches in Si$_3$N$_4$ with width $h_x=140\ \mathrm{nm}$ and depth $h_z=120\ \mathrm{nm}$ after the 1st nanofabrication. Even with 200k magnification (inset) few roughness is observed.
		(c)(d) Cavity structures without and with hBN on top. Titled 45$^\circ$.
		(e) Complete hBN/MoS$_2$/hBN/Si$_3$N$_4$ hybrid cavity.
		(f) A broken cavity providing the side view of 1st nanofabrication in Si$_3$N$_4$.
	}
\end{figure*}

As shown in Sec.~\ref{sec2a} SFig.~\ref{des}, the fabrication mainly contains 2 steps of nanofabrication and 1 step of stacking 2D materials.
The Si$_3$N$_4$ wafer in this work is from Active Business Company GmbH, consisting of $\mathrm{200\ nm}$ thick LPCVD grown Si$_3$N$_4$ on $\mathrm{525\ \mu m}$ thick Si substrate.
In nanofabrication, the EBL machine is eLINE from Raith GmbH, and the ebeam resist is AR-P 6200 from Allresist GmbH.
The ICPRIE machine is PlasmaPro 80 from Oxford Instruments.
The SEM images are taken with the eLINE or NVision 40 from ETH Zürich.
The hBN for etching optimization (SFig.~\ref{fab}(a)) is from Takashi Taniguchi's Group.
The hBN and MoS$_2$ for cavities are from HQ Graphene.
The stacking machine is home built.
Chemicals used in the whole fabrication processes are mainly from Merck KGaA.

In this work, we mainly present the results from two samples marked by A and B with the series of cavities A$i$ and B$i$.
Sample A is fabricated with the lattice constant $a=270\ \mathrm{nm}$ and the target energy of M1 around the LX band 1.75 eV, of which the results are shown in Fig.~\ref{f1} in the main paper.
Sample B is fabricated with $a=250\ \mathrm{nm}$ and the target energy of M1 around the free exciton band 1.9 eV, of which the results are shown in Fig.~\ref{f2}-\ref{3} in the main paper.
Sample A is not annealed, which means the TMD exciton quality is quite low with strong fluctuations and non-radiative recombination thus the exciton-photon interaction is quite weak.
Therefore, the TMD excitons only act as a light source, providing the platform to investigate the properties of bare cavity modes.
sample B is annealed and the significant improvements of TMD excitons are observed.

\subsection{\label{sec3a}2D Flakes Preparation}

\begin{figure}
	\includegraphics[width=\linewidth]{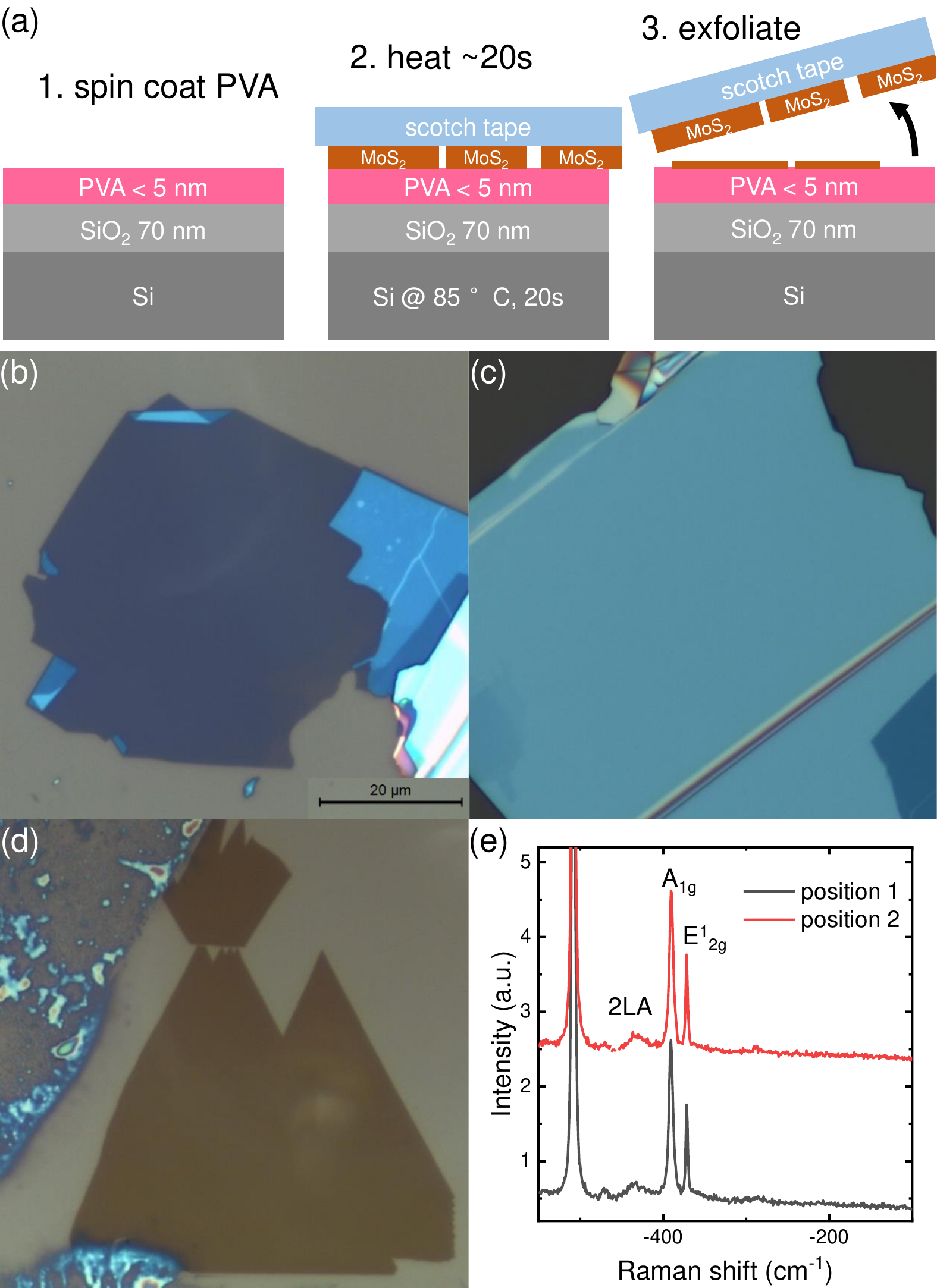}
	\caption{\label{fla}
		Preparation of 2D flakes.
		(a) Schematic of the PVA exfoliation \cite{Huang2020}.
		(b) Top and (c) bottom hBN flake.
		(d) MoS$_2$ monolayer (brown) and scotch tape glue (colourful).
		(e) Raman spectra of MoS$_2$ monolayer exfoliated on PVA-coated SiO$_2$.
		Note: the bright rings in (b)(d) are not contamination but due to light reflection in the microscope.
	}
\end{figure}

Large 2D flakes are demanded to cover all nanotrenches etched in the 1st nanofabrication.
In x-direction the nanobeam length is designed as $\mathrm{20\ \mu m}$ for 65 periodic nanotrenches inside (more trenches higher Q).
Thus the 2D flakes should have $\mathrm{20\ \mu m}$ length in x direction.
In y-direction, the 2D flake size determines the number of cavities fabricated on one sample.
To cover a wide range of cavity mode energies (by varying series $d_y$), the size of 2D flakes in y-direction is the larger the better.
Therefore, we exfoliate 2D materials on PVA coated SiO$_2$ substrate \cite{Huang2020}, a method which could easily get large flakes ($20-200\ \mathrm{\mu m}$) as schematically shown in SFig.~\ref{fla}(a).

SFigure~\ref{fla}(b)(c) are the microscope images of the top and bottom hBN.
The thickness is judged by the color contrast compared to the SiO$_2$ substrate \cite{doi:10.1063/1.4803041}.
E.g., the blue grey in SFig.~\ref{fla}(b) indicates a thickness of $\sim$ 15 nm and the dark blue in SFig.~\ref{fla}(c) indicates a thickness of $\sim$ 55 nm.
We note that a small error in hBN thickness is not a problem as discussed in Sec.~\ref{sec2c} SFig.~\ref{rob}(c).
Additionally, the thickness of a large hBN flake is usually inhomogeneous, especially at flake edges, reflected by the inhomogeneous color in SFig.~\ref{fla}(b)(c).
Thus the energy of first several cavities (at the flake edge) is usually nonmonotonic as shown in Fig.~\ref{f2}(a) in the main paper.

SFigure~\ref{fla}(d) shows the MoS$_2$ monolayer with a size of $50\ \mathrm{\mu m}$ which is enough to cover all cavity structures. The glue from the scotch tape could be observed around the monolayer flake.
As the glue exists only outside the monolayer flake, it will not affect quality of monolayer (Raman in SFig.~\ref{fla}(e)) as well as excitons in the hBN/MoS$_2$/hBN heterostructure.

\subsection{\label{sec3b}Nanofabrication}

The nanofabrication contains EBL and ICPRIE.
The electron voltage in EBL is 30 kV with current $\sim$ 15 pA.
In the 1st nanofabrication to etch nanotrenches, the resist thickness is 270 nm and the ebeam dose is $\mathrm{150\ \mu C/cm^2}$.
In the 2nd nanofabrication to divide nanobeam cavities, the resist thickness is 480 nm and the ebeam dose is $\mathrm{125\ \mu C/cm^2}$.
The resist in two steps are both developed in Amyl Acetate at 5 $^\circ$C for 1 min.
In both two steps, SF$_6$ and C$_4$F$_8$ with ratio 3:2 are used in ICPRIE etching, with the pressure of 13.5 mTorr, HF power of 15 W and ICP power of 220 W.
Some recent works report on the ebeam-induced etching (EBIE) which could improve the etching roughness of hBN \cite{doi:10.1002/adom.201801344}.
However, the EBIE etching is sensitive to secondary and backscattered electrons so the etching time need to be precisely matched to the hBN thickness \cite{doi:10.1002/adom.201801344}.
Additionally, the gas used in EBIE is usually H$_2$O for hBN and XeF$_2$ for TMDs \cite{doi:10.1002/adom.201801344,Stanford2018}.
Different gases would make the etching very complex for hBN/TMD/hBN heterostructures.
Therefore, ICPRIE with the recipe shown above is used in this work, which could etch both hBN, MoS$_2$ and Si$_3$N$_4$.

SFigure~\ref{fab}(a) shows the SEM image of an etching test sample.
The left smooth part is Si$_3$N$_4$, and the right rough part is hBN on Si$_3$N$_4$.
Meanwhile, in the hBN region we also observe many spots at bottom.
These spots should originate from the react between hBN and ions during ICPRIE etching, and cannot be pumped away by vacuum \cite{Caldwell2014,doi:10.1002/adom.201801344}.
Some of spots left on sidewalls resulting in the roughness, and others drop to the bottom of etched regions as shown in SFig.~\ref{fab}(a).
The roughness here approximately corresponds to $\sim$ 15 nm in Sec.~\ref{sec2b}, which theoretically decay the cavity Q by $\sim$ one or two order of magnitudes.
Compared to other possible fabrication errors in Sec.~\ref{sec2c}, the roughness plays the main role in the cavity Q decay and photon losses.

SFigure~\ref{fab}(b) shows the nanoscale trenches etched in Si$_3$N$_4$ after the 1st nanofabrication. Inset is the image with a 200k magnification.
The roughness of the etching boundary is $<$ 3 nm, which is a very good result close to the ebeam resolution.
Based on the consideration of mechanical stability, the etch depth here is chosen as $h_z=120\ \mathrm{nm}$, a value smaller than 150 nm in the simulation (Sec.~\ref{sec2a}).

SFigure~\ref{fab}(c)(d) shows hBN/Si$_3$N$_4$ cavities after the 2nd nanofabrication and the wet under etching, where (c) is a control group without hBN on top, and (d) is complete cavities with hBN on top.
The small bright balls at bottom is some residual KOH during the wet under etching, which could be removed by an additional cleaning in the HCl solution as discussed latter in Sec.~\ref{sec3d}.
In SFig.~\ref{fab}(d), hBN is almost transparent and Si$_3$N$_4$ is very bright, indicating the conductive of hBN much higher than that of Si$_3$N$_4$.
The hBN on one cavity (top in SFig.~\ref{fab}(d)) is broken during the thickness measurement with Dektek step profiler.

SFigure~\ref{fab}(e) shows a cavity with all hBN/MoS$_2$/hBN layers.
The corresponding PL spectra is shown in Fig.~\ref{f1} in the main paper.
SFigure~\ref{fab}(f) is one broken cavity.
The ICPRIE etching in the 2nd nanofabrication is not enough, so some residual Si$_3$N$_4$ extending out of cavity edge is observed.
This cavity provides the chance to check the first nanofabrication from side view.
The magnification is 200k and no roughness is observed.
The etching angle of Si$_3$N$_4$ is not very vertical, resulting in the round corner at the bottom of trenches.
The round corner should benefit the cavity Q, so we just keep this etching recipe.
Generally, SEM images in SFig.~\ref{fab} demonstrate the good fabrication of our sample and thus the high cavity Q we achieved is not surprise.

\subsection{\label{sec3c}Stacking of 2D Flakes}

\begin{figure}
	\includegraphics[width=\linewidth]{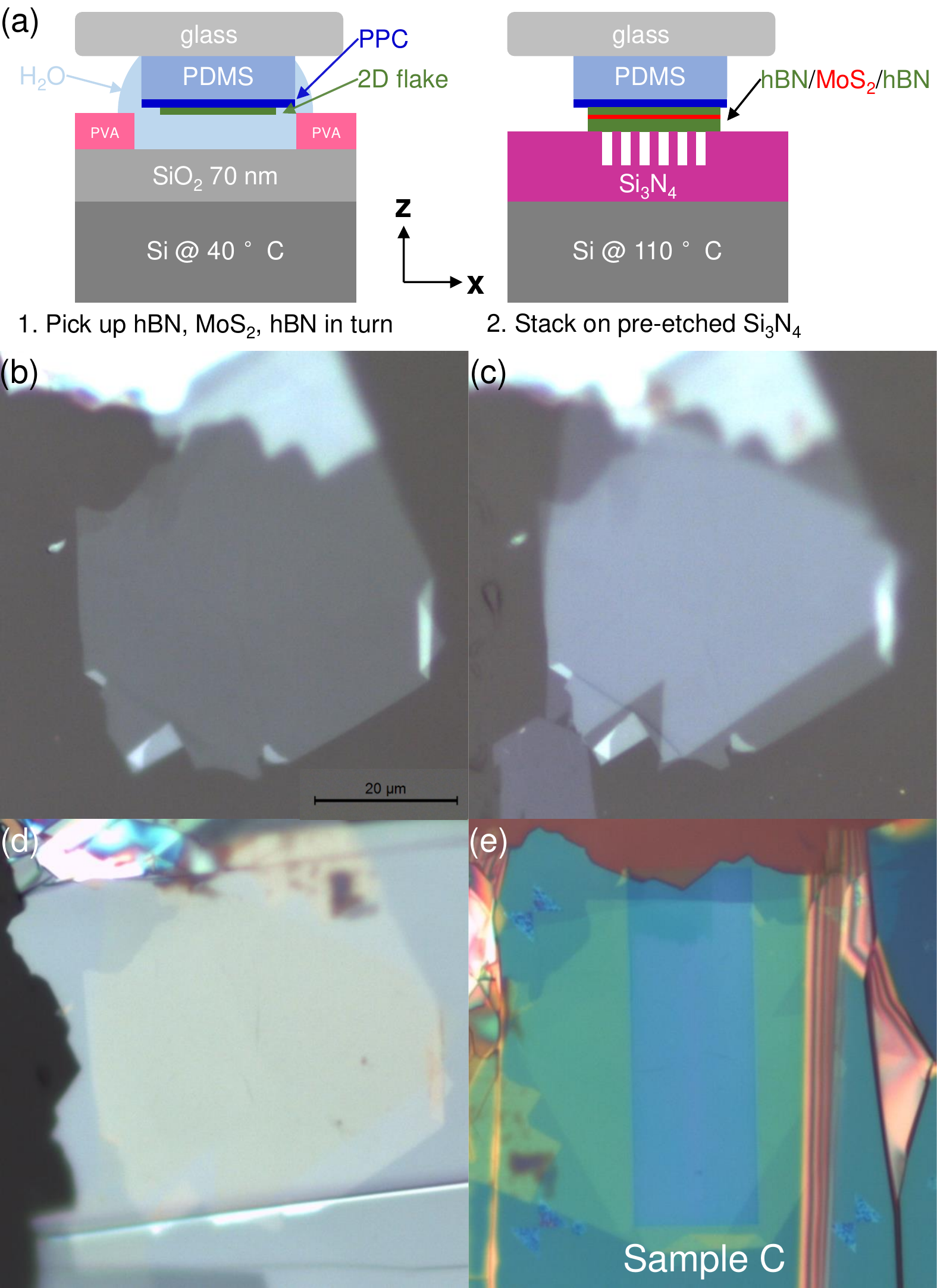}
	\caption{\label{sta}
		Stacking of 2D flakes.
		(a) Schematic of processes.
		(b)-(e) Microscope images during the stacking processes.
		(b) Stamp with top hBN.
		(c) Stamp with top hBN and MoS$_2$ monolayer.
		(d) Stamp with top hBN, MoS$_2$ monolayer and bottom hBN.
		(e) hBN/MoS$_2$/hBN layers together with PPC film released on the nanotrenches in Si$_3$N$_4$ substrate.
	}
\end{figure}

The dry transfer method with a Polypropylene Carbonate (PPC) stamp (SFig.~\ref{sta}(a)) \cite{Pizzocchero2016} is used to stack the 2D flakes in this work.
Detailed processes could be found in previuos works \cite{Huang2020,Pizzocchero2016} thus are not discussed here.
SFigure~\ref{sta}(b)-(d) are images during stacking, corresponding to the 2D flakes in Sec.~\ref{sec3a} SFig.~\ref{fla}.
SFigure~\ref{sta}(e) is the image after 2D flakes are released on the top of nanotrenches.
Some contaminates are observed during the stacking processes, which might originate from residual PVA or water drops.
The next wet etching step could also introduce contaminates.
These contaminates between 2D flakes will result in environment disorders for TMD excitons.
Therefore a vacuum annealing is used to remove the contaminates after the fabrication as shown next in Sec.~\ref{sec3d}.

\subsection{\label{sec3d}Wet Etching and Annealing}

\begin{figure}
	\includegraphics[width=\linewidth]{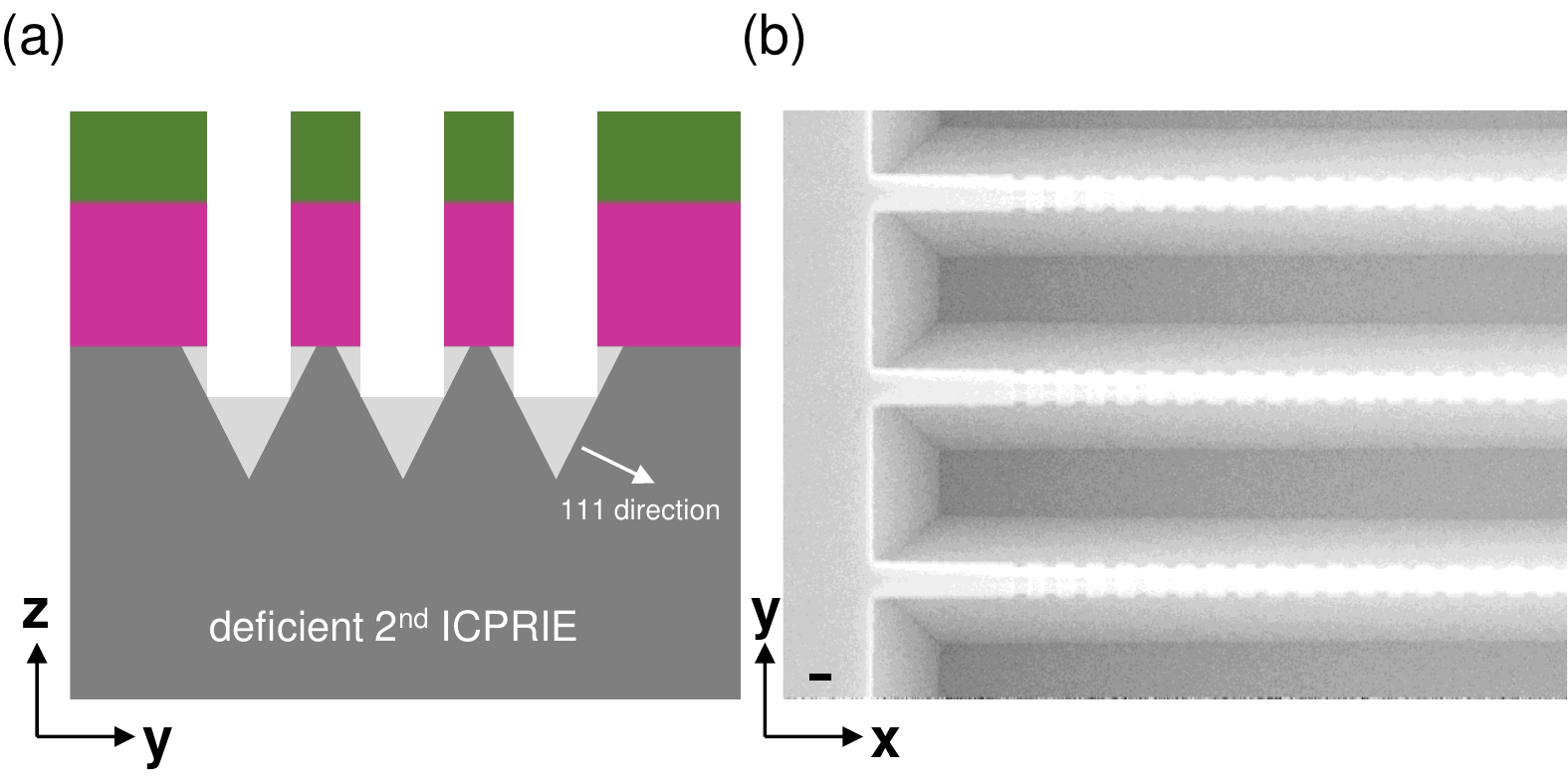}
	\caption{\label{und}
		Deficient etching in the 2nd nanofabrication.
		(a) Schematic. Dark grey is the Si substrate, purple is Si$_3$N$_4$ and green is 2D flakes. White part in materials is the etched part in the 2nd ICPRIE, and light grey is the part etched by the KOH solution. The wet etching stops due to KOH can hardly etch Si in 111 direction. In this situation TMAH solution is used to complete the under etching.
		(b) SEM image of cavities where KOH solution could not under etch completely. The black scale bar on bottom left corner is 200 nm.
	}
\end{figure}

The wet under etching of silicon is mainly implemented with Potassium Hydroxide (KOH) or Tetramethylammonium Hydroxide (TMAH) solution.
Si fabrication is a very mature technology with many literature where the details could be found, thus is not specifically discussed here.
Generally, the KOH solution is safe and fast, but will introduce K$^+$ ion contamination.
In contrast, the TMAH solution is toxic but has better etching result without ion contamination.
Furthermore, the bad etching anisotropy of TMAH is an advantage for under etching, because the anisotropy KOH etching cannot remove all Si below cavities if the ICPRIE etching in the second nanofabrication is not enough, as shown in SFig.~\ref{und}.
In our work, we usually first use 50$\%$ KOH solution for a safe wet etching.
If the etching is not enough as shown in SFig.~\ref{und}, we will use 25$\%$ TMAH solution to complete the wet under etching.

\begin{figure}
	\includegraphics[width=\linewidth]{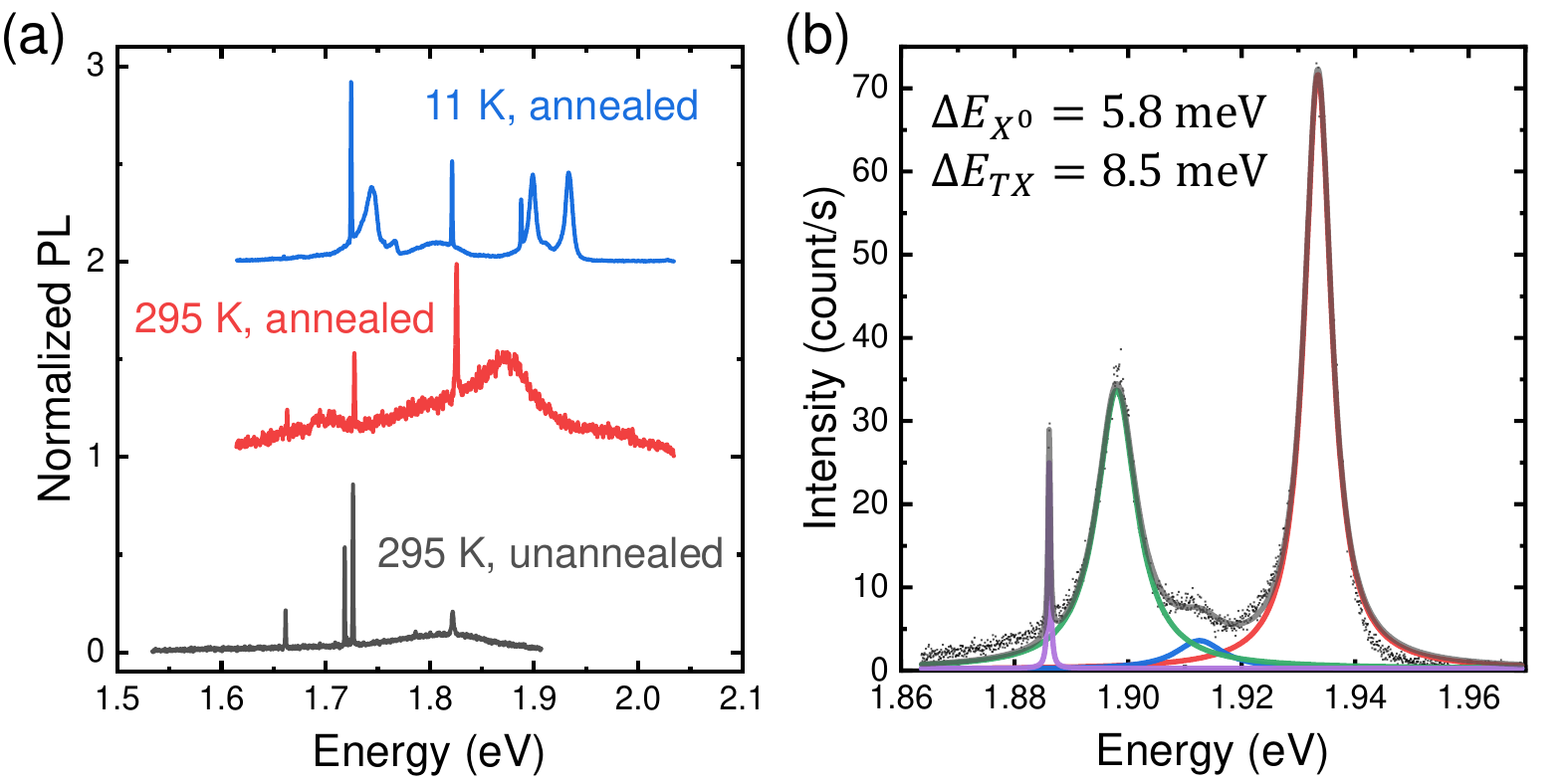}
	\caption{\label{ann}
		Improvement after annealing.
		(a) Normalized PL spectra (300 grating) of cavity B2, measured (grey) at 295 K before annealing with the excitation power of $165\ \mathrm{\mu W}$, (red) at 295 K after annealing with the excitation power of $41\ \mathrm{\mu W}$, and (blue) at 11 K after annealing with the excitation power of $24\ \mathrm{\mu W}$.
		(b) PL spectrum measured at 11 K after annealing with the excitation power of $14\ \mathrm{\mu W}$. X$^0$ (red), TX (green) and cavity mode M1 (purple) can be observed. The small peak (blue) should be inter-valley trions. The linewidth of X$^0$ is 5.8 meV.
	}
\end{figure}

Contaminates are generated during the fabrication processes, such as residual ebeam resist, residual PVA, residual KOH and residual organic solvent.
Therefore, a vacuum annealing \cite{Jain_2018} is used to clean the cavities, and significant improvement is observed by comparison the PL spectra of cavity B2 before and after annealing as shown in SFig.~\ref{ann}(a).
Before annealing (grey line), the exciton peak energy at room temperature (295 K) is at 1.81 eV with a huge linewidth over 100 meV, which is quite different compare to clean samples \cite{PhysRevX.7.021026}.
Furthermore, the cavity mode (1.822 eV) is observed with the energy higher than that of exciton, indicating the coupling between cavity and TMD monolayer is not good.
In contrast after annealing, cavity modes can only be observed with the energy smaller than TX.
This absorption reflects the complex exciton dynamics and the impact of reabsorption in the TMD monolayer as discussed in the main paper.
Here we believe that the hBN-encapsulation, which improved by the annealing, enhances this absorption by improving the cavity-TMD coupling.

Significant improvements of excitonic properties are observed after annealing, where the free exciton at 295 K is at 1.86 eV with the linewidth of 37 meV (red line in SFig.~\ref{ann}(a)), exactly the standard values for clean samples \cite{PhysRevX.7.021026,Shree_2019}.
The asymmetric free exciton peak indicates it contains both X$^0$ and TX.
At low temperature 11 K, we could clearly see the peaks from the variety of excitons in the hBN-encapsulated MoS$_2$ monolayer (blue line in SFig.~\ref{ann}(a)).
SFigure~\ref{ann}(b) shows the spectrum with 1200 grating where X$^0$ at 1.933 eV, TX at 1.898 eV and cavity mode M1 at 1.886 eV are observed.
The linewidth is $\mathrm{5.8\ meV}$ of X$^0$ and $\mathrm{8.5\ meV}$ of TX.
The relative strong TX peak might originate from the ion contamination during the etching and the extrinsic doping from thick hBN \cite{Ju2014}.
Another small peak at 1.912 eV (blue line in SFig.~\ref{ann}(b)) should be the intervalley trions \cite{Lyons2019}.
These parameters of free excitons are similar to previous literature \cite{PhysRevX.7.021026,Shree_2019}, indicating the good quality of 2D flakes and the good attachment between the hBN/MoS$_2$/hBN layers after annealing.

\section{\label{sec4}Measurement and Additional Results}

The PL spectra in this work are measured by a confocal micro-PL system.
The objective has a magnification of 100 and a NA of 0.75.
The size of laser spot is $\sim1\ \mathrm{\mu m^2}$.
The temperature of sample is jointly controlled by the liquid helium flow and the heater.
The sample position is aligned with a three-dimensional xyz nanopositioner.
A narrow linewidth solid laser with a wavelength of 532 nm is used to excite the TMD excitons.
The laser power is adjusted by a filter with motor controlled by the computer.
The PL spectra is collected by a matrix array Si CCD detector in a spectrometer with a focal length of 0.55 m.
Unless specified the grating for spectroscopy has 1200 grooves per mm.

In this section we present additional results on the properties of cavities, excitons and the interaction between them.
The cavity Q is extracted by the deconvolution of measured PL spectra. The measured linewidths are very close to the instrumental broadening, demonstrated by the high-resolution reflectivity measurement.
The temperature dependence of bare cavity modes indicate that the energy slightly shift due to condensation of atmospheric gases while the linewidth (Q) keeps almost same.
The temperature and power dependence of excitonic properties are consistent to previous literature.
These results further supports our conclusions on the novel exciton-photon interactions discussed in the main paper.

\subsection{\label{sec4a}Bare Cavity Properties}

\begin{figure}
	\includegraphics[width=\linewidth]{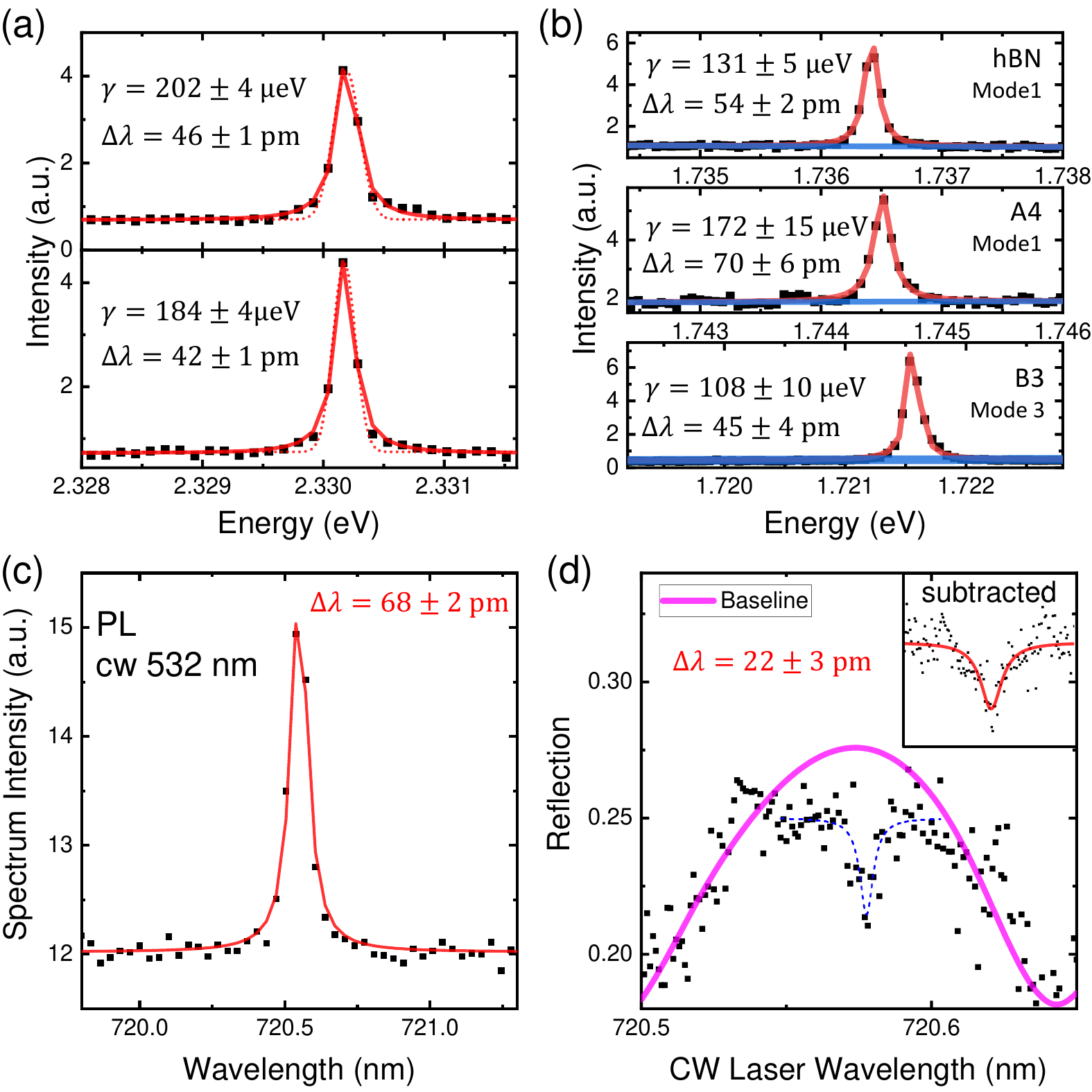}
	\caption{\label{las}
		Cavity Q and instrumental broadening.
		(a) Typical spectra of the excitation laser at 532 nm ($\sim$ kHz actual width). Solid red lines are Lorentz fittings, and dashed red lines are Gaussian fittings.
		(b) Three typical PL spectra of cavities, from the M1 of a hBN/Si$_3$N$_4$ cavity without TMD (SFig.~\ref{fab}(d)), the M1 of cavity A4 and the M3 of cavity B3. Blue line is the background and red line is the background plus Lorentz fitting.
		(c) PL spectrum of one cavity mode.
		(d) The reflectivity spectra of the same cavity mode in (c), measured consecutively on the same day.
	}
\end{figure}

The resolution of PL spectra is limited by the instrumental broadening in the measurement setup.
The resolution of the 0.55 m focal length spectrometer with 1200 grating is usually 0.05 nm, corresponding to $\mathrm{0.13\ meV}$ for signal around 1.8 eV the value we used for deconvolution in the main paper.
SFigure~\ref{las}(a) shows the instrumental broadening function by measuring the spectra of the excitation laser, of which the actual linewidth $\sim$ kHz is negligible.
By comparing the Lorentz (solid) and Gaussian (dash) fittings, we find that the broadening function of our setup is Lorentz shape.
For the deconvolution with the Lorentz broadening, we just need to decrease the measured PL linewidth by the instrumental broadening width as used in the main paper.
SFigure~\ref{las}(b) shows three typical cavity peaks and the fitting linewidths are $131\ \mathrm{\mu eV}$ (54 pm), $172\ \mathrm{\mu eV}$ (70 pm) and $108\ \mathrm{\mu eV}$ (45 pm).
The peaks are such narrow that the Q is already over 10$^4$ before deconvolution.
Especially for M3 in cavity B3, the measured linewidth 45 pm is almost same to the instrumental broadening, indicating a much higher actual Q value.

To further demonstrate the instrumental broadening, we measure the differential reflectivity on one cavity to achieve the high-resolution Q value and compare the results to the PL spectrum.
The PL in SFig.~\ref{las}(c) and differential reflectivity spectra in SFig.~\ref{las}(d) are from the same cavity mode measured consecutively on the same day.
In PL data, the linewidth of the cavity mode is 68 pm whereas the reflectivity spectrum exhibits the actual linewidth of 22 pm (Q=32750) for this particular cavity, superimposed on thin film interference.
Thus, the cavity Q values that we estimated by deconvoluting measured PL spectra with the spectra response of the system ($\gamma_{real}=\gamma_{PL\ measured}-\gamma_{broadening}$, $\gamma_{broadening}\approx50\mathrm{\ pm}$) are very close to those measured using high-resolution spectroscopy.
Furthermore, the conclusions in the main paper are not affected by the exact Q values.
Therefore, we did not measure the reflectivity of all cavity modes one by one.

\begin{figure}
	\includegraphics[width=\linewidth]{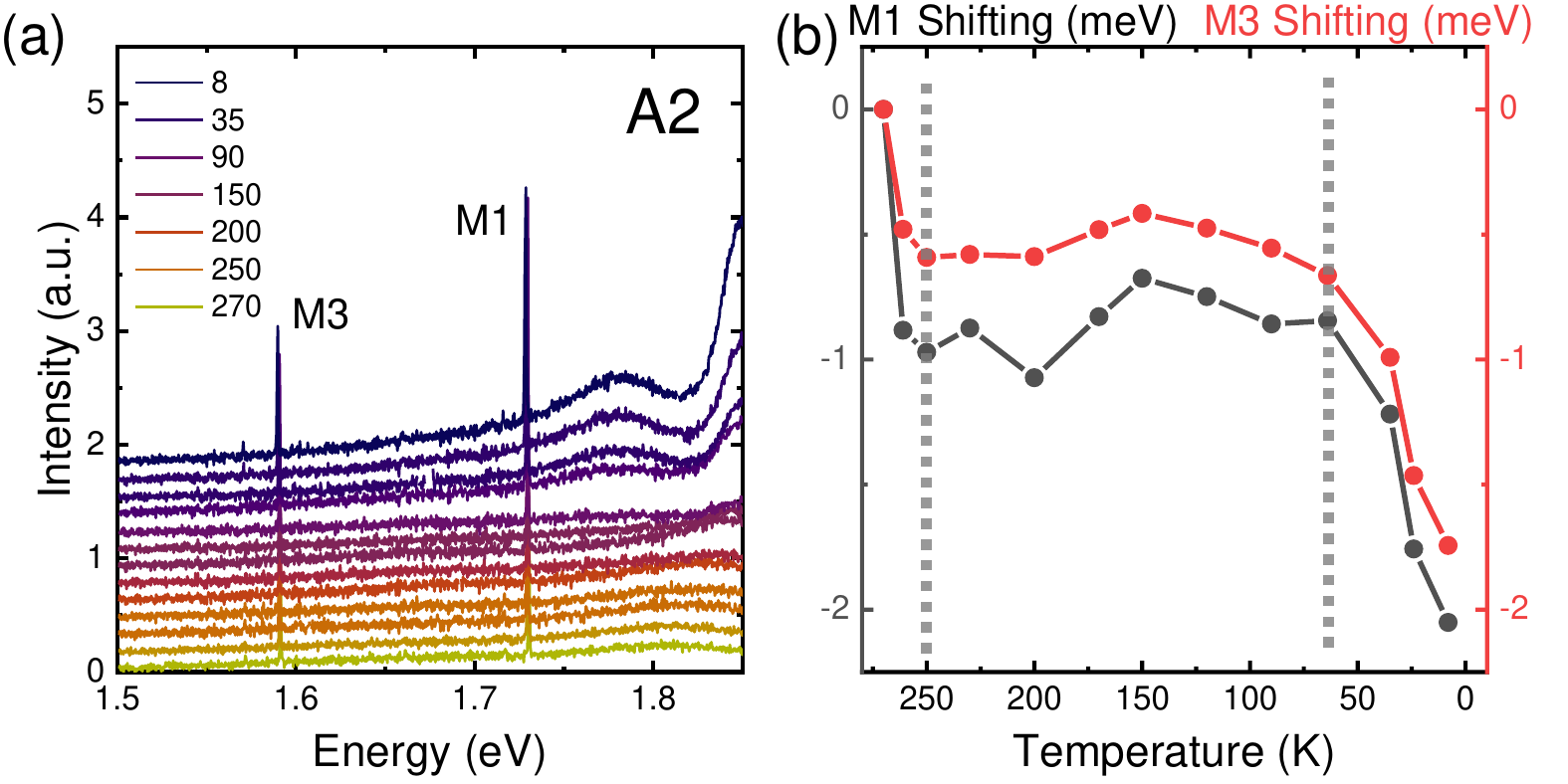}
	\caption{\label{cat}
		Temperature dependence of the bare cavity mode in cavity A2 (unannealed).
		(a) PL spectra measured with the laser power of 25 $\mathrm{\mu W}$ and collected with the 300 grating.
		M2 cannot be observed due to the cavity mode function $\bm{\alpha}$ is zero at center.
		(b) Cavity energy shift.
	}
\end{figure}

As mentioned in Sec.~\ref{sec3d}, the TMD excitons without annealing have bad optical quality thus the cavity peak measured in sample A is very close to the bare cavity mode.
SFigure~\ref{cat}(a) shows the temperature-dependent (descending) PL spectra of cavity A2.
The temperature dependence of cavity mode energies is presented in SFig.~\ref{cat}(b).
The energy shift of bare cavity mode depends on two factors: the condensation of atmospheric gases and the change of refractive index.
The gas condensation at low temperature $T$ will red shift the cavity mode \cite{doi:10.1063/1.2076435}.
The refractive index of Si$_3$N$_4$ $n_{Si_3N_4}$ decreases as $T$ decreases \cite{7463458}.
The refractive index of hBN $n_{hBN}$ is not well known yet as mentioned in the main paper, but the change of $n_{hBN}$ could be expected small due to the large band gap of hBN.
Thus generally the decrease of $n_{Si_3N_4}$ will blue shift the cavity mode as $T$ decreases.
The experimental results in SFig.~\ref{cat}(b) shows a rapid red shift when $250<T<295\ \mathrm{K}$ and $T<70\ \mathrm{K}$ corresponding to the gas condensation.
In contrast when $T$ is between 250 and 70 K, the cavity energy is relatively stable, indicating both the two effects are not significant within this range.

\begin{figure}
	\includegraphics[width=\linewidth]{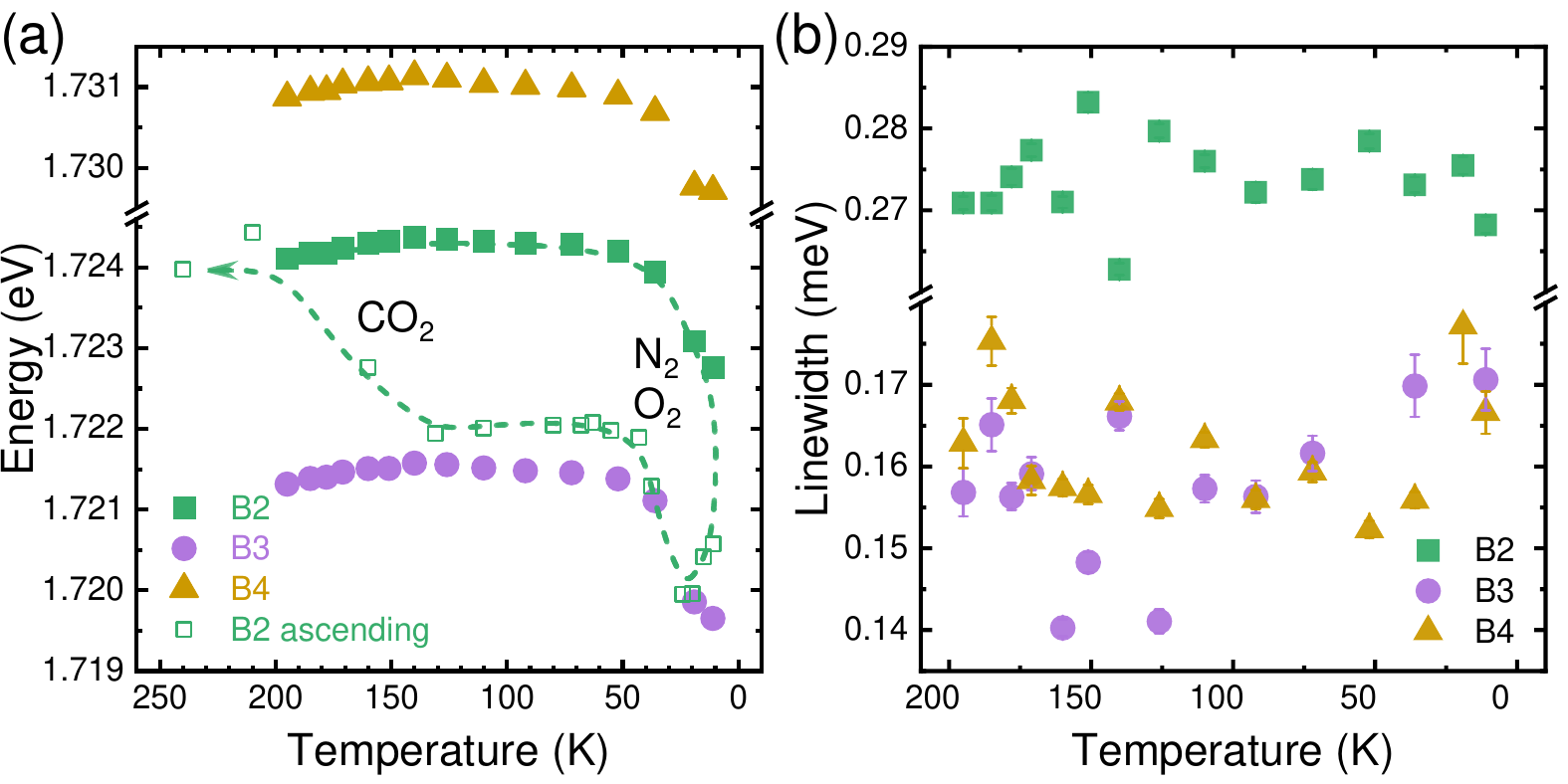}
	\caption{\label{ct2}
		Temperature dependence of the bare cavity mode M3 in cavity B2-4.
		(a) The M3 peak energies. Solid dots are measured with descending $T$ while hollow dots are measured with ascending $T$. The comparison between descending and ascending results indicates condensations of different gases at different $T$.
		(b) The M3 peak linewidths. The temperature-dependent change is very little ($\sim 20\ \mathrm{\mu eV}$), indicating a constant $\gamma_C$ as the temperature changes.
	}
\end{figure}

The bare cavity mode M3 is also observed in the cavity-LX interaction in sample B, and the properties of bare cavity modes are extracted and presented in SFig.~\ref{ct2}.
The energy shifts in SFig.~\ref{ct2}(a) are in good agreement with the results in SFig.~\ref{cat}(b).
The linewidths shown in SFig.~\ref{ct2}(b) are very stable with temperature (vibration $\sim 20\ \mathrm{\mu eV}$), indicating neither the gas condensation nor the refractive index change will significantly affect the cavity Q.
Additionally, the energy shift is different with the ascending and descending temperature (green dashed line in SFig.~\ref{ct2}(a)).
This is due to two points.
Firstly, the condensation or vaporization of gas is determined by the value of $T$ (below or above the sublimation point) rather than the ascending or descending.
Thus when $T<25\ \mathrm{K}$, whether $T$ is ascending or descending the gas condensation continuously red shift the cavity mode.
This red shift rate is $\sim$ 1 meV per hour when $T$ just reaches 11K, and $\sim$ 0.1 meV per hour after the sample is cooled down a long time (gas condensation also has its limit \cite{doi:10.1063/1.2076435}).
Therefore, the PL spectra measured at low temperature are usually taken after the sample is cooled down a long time, and the integration time is usually around 20 seconds to suppress effects from the peak shifting.
Secondly, the condensation of gas is slow (depending on the vacuity) while the vaporization is fast (quickly pumped away).
The condensation rate of different gases depends on their concentration in atmosphere.
By comparison between the descending and ascending data in SFig.~\ref{ct2}(a), we find that the relatively rapid red shift (descending) and rapid blue shift (ascending) when $T<60\ \mathrm{K}$ is from gases which have the sublimation point within this $T$ range and are rich in the air (thus relatively rapid red shift with descending $T$), e.g., N$_2$ and O$_2$ \cite{STakayoshi2009TheBS}.
In contrast, the little shift (descending) and rapid blue shift (ascending) when $150<T<200\ \mathrm{K}$ is from the gas which has the sublimation point within this $T$ range but is rare in the air (thus very slow red shift with descending $T$), e.g., CO$_2$ \cite{Witkowski2014}.
The good news is that the bare cavity mode is always stable between 70 and 150 K, where we focus in the temperature-dependent measurement in this work.

\subsection{\label{sec4b}Excitonic Properties in hBN-encapsulated MoS\texorpdfstring{$_2$}{2} Monolayer}

\begin{figure}
	\includegraphics[width=\linewidth]{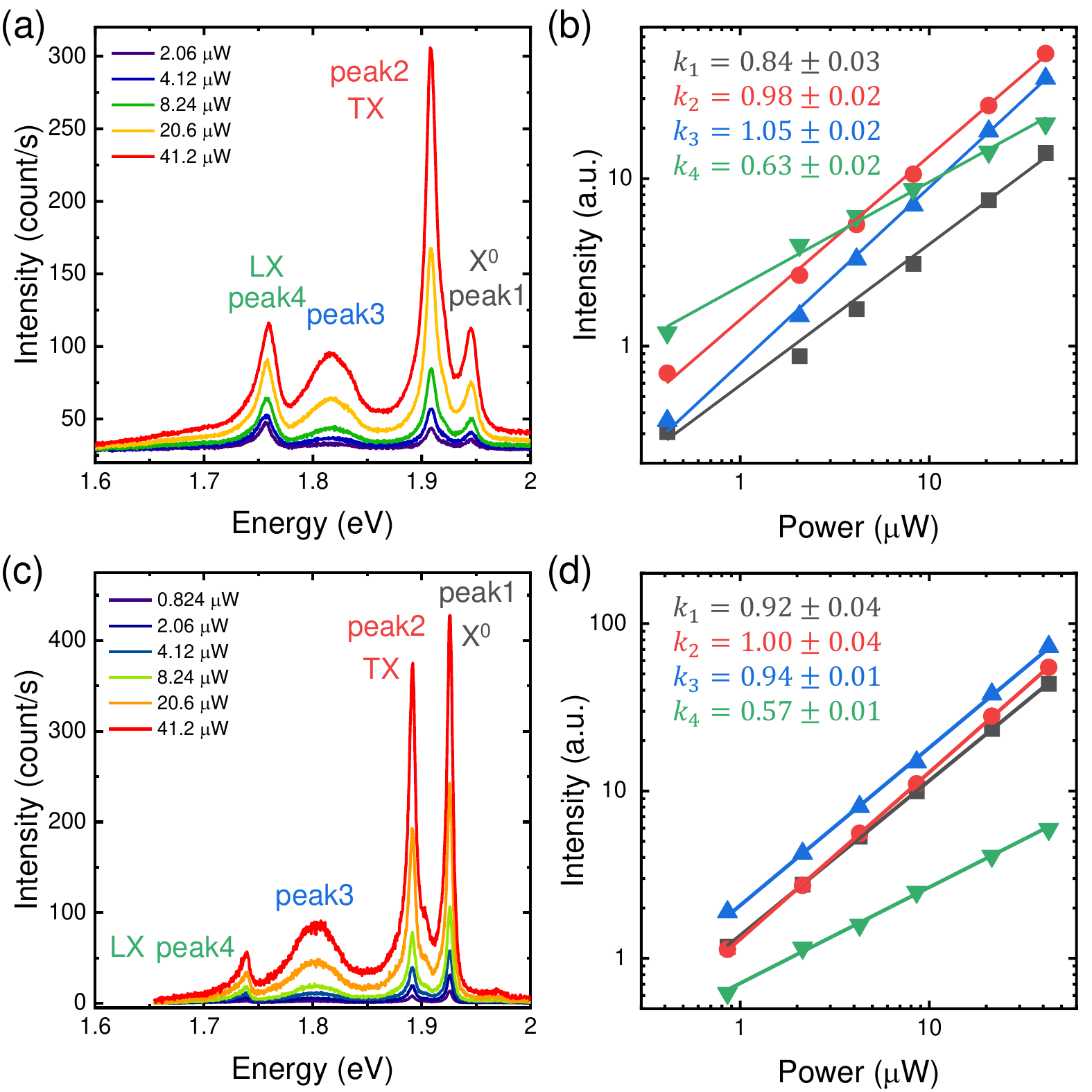}
	\caption{\label{une}
		Excitons in unetched hBN/MoS$_2$/hBN heterostructures, measured at 11 K.
		(a) Power-dependent PL spectra of an unetched hBN/MoS$_2$/hBN heterostructure annealed at 400 $^\circ$C. Peak 1 is X$^0$. Peak 2 is TX. Peak 3 should be dark trions. Peak 4 is LX from atomistic defects. The linewidth of X$^0$ is 11 meV.
		(b) Power dependence of four peaks in (a).
		(c) Power-dependent PL spectra of another unetched hBN/MoS$_2$/hBN heterostructure annealed at 240 $^\circ$C. The linewidth of X$^0$ is 5.5 meV.
		(d) Power dependence of four peaks in (c).
	}
\end{figure}

SFigure~\ref{une} shows the power dependence of various TMD excitons at 11 K in hBN-encapsulated MoS$_2$ monolayers before the 2nd nanofabrication (2D flakes are not etched yet).
In SFig.~\ref{une}(a) the linewidth of X$^0$ is $\gamma_{X^0}=11\ \mathrm{meV}$, due to the high annealing temperature 400 $^\circ$C might damage the TMD a bit which also results in the power dependence $k_1=0.84$ smaller than the theoretical value of 1 \cite{doi:10.1021/acs.nanolett.9b05323,Shree_2019}.
In contrast in another sample annealed at 240 $^\circ$C (SFig.~\ref{une}(c)), $\gamma_{X^0}=5.5\ \mathrm{meV}$ and $k_1=0.92$ agrees well with the values in previous literature \cite{PhysRevX.7.021026,Shree_2019,doi:10.1021/acs.nanolett.9b05323}.
On both samples the power dependence of TX ($k_2$) is around 1 which is the typical value for free excitons.

Peak 4 is the LX peak from atomistic defects, with the energy $\sim$ 1.75 eV and the power dependence $k_4\sim 0.6$ which is similar to previous literature \cite{doi:10.1021/acs.nanolett.9b05323,Shree_2019}.
We would expect the Peak 3 between TX and LX to be the dark trions \cite{He2020}, because the power dependence $k_3 \sim 1$ exclude the disorder bound excitons ($k<1$) \cite{doi:10.1021/acs.nanolett.9b05323} or charged biexcitons ($k\sim1.55$) \cite{Barbone2018} which are in the similar energy range.
The observation of dark excitons should originate from the strong tensile stress in Si$_3$N$_4$ below the 2D layers which breaks the valley symmetry  \cite{Chai2017,Lee2017,GE2021109338}.
Meanwhile, the out-of-plane cavity electric field $E_y$ could also enhances the dark excitons which have the out-of-plane dipole \cite{Zhou2017}.
These effects make it reasonable to observe the dark trions.

\subsection{\label{sec4e}Cavity-TX Interaction}

\begin{figure}
	\includegraphics[width=\linewidth]{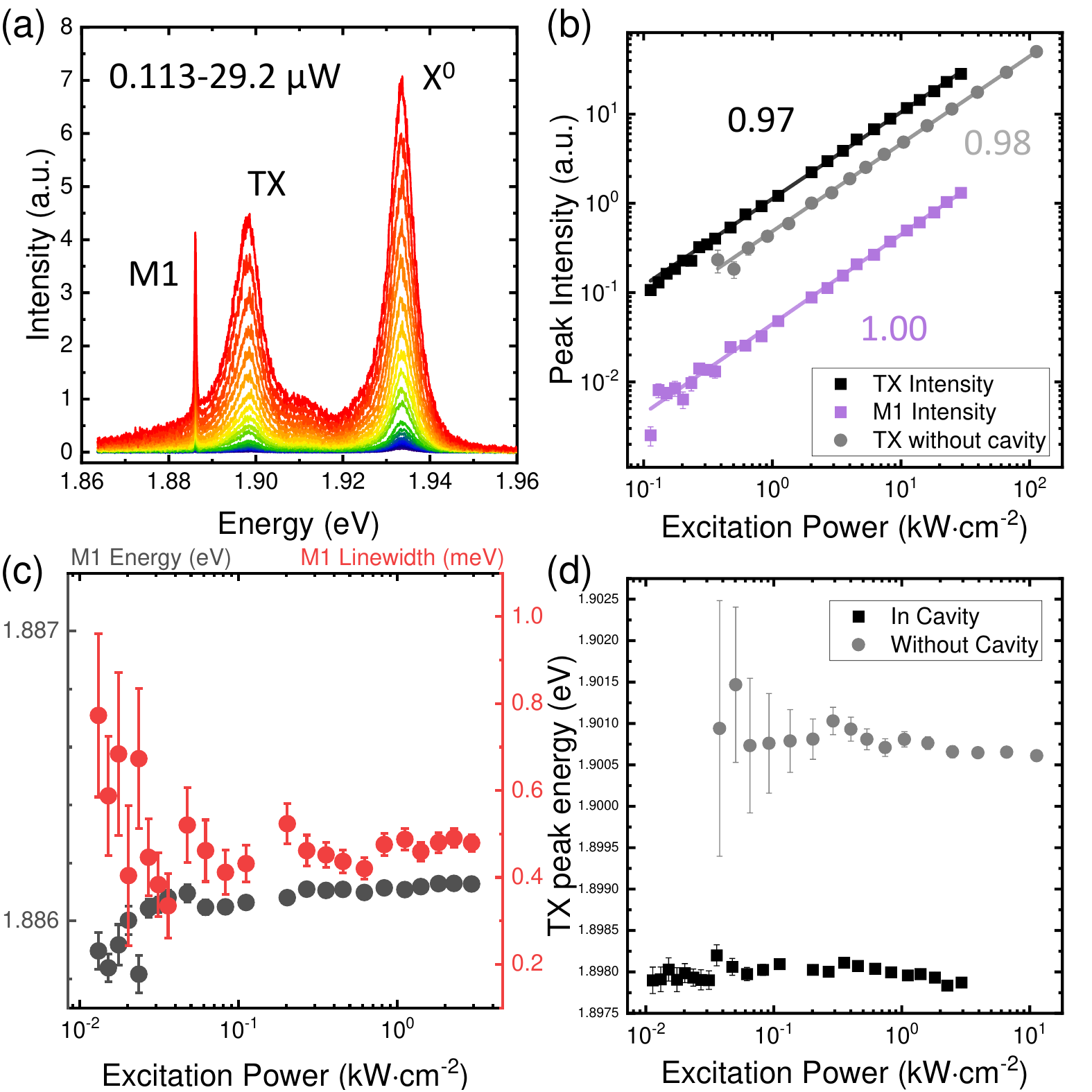}
	\caption{\label{ctx}
		Power dependence of cavity-TX interaction.
		(a) Power-dependent PL spectra at low temperature. The laser spot is also $1\ \mathrm{\mu m^2}$ here.
		(b) The intensity-power dependence of TX and M1 both have the slope of 1.
		Gray is the TX measured away from the cavity as a comparison, and the same linearity is observed.
		(c) The energy and linewidth of M1 peak is also stable, except some low-power points with large error bar.
		(d) No Coulomb repulsion introduced blue shifting is observed as pump power increases.
		Due to low SNR, some low-power data points in (b)(c)(d) have large fitting errors.
	}
\end{figure}

As discussed in the main paper, the cavity-TX interaction is measured at the linear low excitation level as presented in the spectra in SFig.~\ref{ctx}(a).
Within the pumping power range of $0.01-2.92\ \mathrm{kW\cdot cm^{-2}}$ ($2.1\ \mathrm{kW\cdot cm^{-2}}$ is used for Fig.~\ref{f2}(b) in the main paper), the intensity-power slope is quite linear as shown in SFig.~\ref{ctx}(b).
The energy and wavelength of the cavity-branch peak is also stable as shown in SFig.~\ref{ctx}(c).
These linear observations demonstrate the interaction at the low excitation level.
Furthermore, at high excitation level the high carrier density will introduce Coulomb repulsion which results in a blue shift as pump power increases \cite{doi.org:10.1002/pssr.202000222}.
However, no blue shift of TX is observed as shown in SFig.~\ref{ctx}(d).
Some vibrations in the data with large error bar at low power is reasonable due to the low fitting accuracy with the small signal to noise ratio (SNR).

The experimental interaction strength $g$ of the cavity-TX interaction (Fig.~\ref{f3} in the main paper) is extracted from the experimental data with the Jaynes-Cummings model.
The emitter has frequency $\Omega_X=\omega_X-\frac{i}{2}\gamma_X$ and the cavity has frequency $\Omega_C=\omega_C-\frac{i}{2}\gamma_C$, where $\omega_X$ ($\Omega_C$) and $\gamma_X$ ($\gamma_C$) are the energy and linewidth of the exciton (cavity mode).
The eigenvalues of exciton-photon interaction system in the Jaynes-Cummings model is well known as as
\begin{eqnarray}
	\Omega_\pm=\frac{\Omega_X+\Omega_C}{2}\pm\sqrt{g^2+\left(\frac{\Omega_X-\Omega_C}{2}\right)^2}\nonumber.
\end{eqnarray}
$\Omega_+=\omega_+-\frac{i}{2}\gamma_+$ and $\Omega_-=\omega_--\frac{i}{2}\gamma_-$ are the frequency of exciton-like and cavity-like branch polaritons, which are experimentally extracted from the $T$-dependent PL in Fig.~\ref{f2}(b) in the main paper.
For brevity $\hbar$ is neglected.
After conversion we can get
\begin{eqnarray}
	\Omega_X=\frac{\Omega_+ +\Omega_-}{2}+\sqrt{\left(\frac{\Omega_+-\Omega_-}{2}\right)^2-g^2}\nonumber \\
	\Omega_C=\frac{\Omega_+ +\Omega_-}{2}-\sqrt{\left(\frac{\Omega_+-\Omega_-}{2}\right)^2-g^2}\nonumber ,
\end{eqnarray}
In weak interaction regime, $g$ is small compared to $\Omega_+-\Omega_-$, thus we could treat $g$ as a small quantity and use the approximation for the square root item to get
\begin{eqnarray}
	\Omega_C&=&\frac{\Omega_++\Omega_-}{2}-\left(\frac{\Omega_+-\Omega_-}{2}\right)\left(1-\frac{g^2}{2\left(\frac{\Omega_+-\Omega_-}{2}\right)^2}\right) \nonumber\\
	&=&\Omega_-+\frac{g^2}{\Omega_+-\Omega_-}\nonumber\\
	&=&\Omega_-+\frac{g^2\left(\Delta\omega'+\frac{i}{2}\Delta\gamma'\right)}{\Delta\omega'^2+\frac{1}{4}\Delta\gamma'^2}\nonumber
\end{eqnarray}
where $\Delta\omega'=\omega_+-\omega_-$ and $\Delta\gamma'=\gamma_+-\gamma_-$.
For the cavity-TX interaction system, $g\ll\Delta\omega'$ at low $T$ and $g\ll\Delta\gamma'$ at high $T$, thus the approximation discussed above always works.
The linewidth of cavity-like branch polariton is
\begin{eqnarray}
	\gamma_C=\gamma_--\frac{g^2\Delta\gamma'}{\Delta\omega'^2+\frac{1}{4}\Delta\gamma'^2}\nonumber
\end{eqnarray}
thus we could get interact strength
\begin{eqnarray}
	g=\sqrt{\frac{\gamma_--\gamma_C}{\Delta\gamma'}\left(\Delta\omega'^2+\frac{1}{4}\Delta\gamma'^2\right)} \nonumber.
\end{eqnarray}
We extract the frequency of polaritons $\Omega_\pm$ by multi-Lorentz fitting of the experimental PL spectra and the results are presented in Fig.~\ref{f3}(b)(c) in the main paper.
The bare cavity linewidth before interaction $\gamma_C$ cannot be measured directly.
Nonetheless, due to the nonlocal interaction theory, as $T$ increases the interaction strength $g$ will first decrease to zero and then rapidly increase.
Meanwhile, $\gamma_C$ changes little with $T$ as discussed in SFig.~\ref{ct2}.
Therefore, to fit the zero interaction strength we use the smallest measured linewidth of the cavity peak $375\ \mathrm{\mu eV}$ as $\gamma_C$, and the corresponding results of $g$ is presented in Fig.~\ref{f3}(d) in the main paper.
We note that although $\gamma_C$ cannot be directly measured, $g$ is proportional to $\sqrt{\gamma_--\gamma_C}$.
Thus, even if the actual value of $\gamma_C$ is a bit larger or smaller than the value we chosen, the nonmonotonic temperature-dependence of $g$ will keep the same.

Theoretically we could also get the energy of the cavity-like branch polariton
\begin{eqnarray}
	\omega_C=\omega_-+\frac{g^2\Delta\omega'}{\Delta\omega'^2+\frac{1}{4}\Delta\gamma'^2}\nonumber.
\end{eqnarray}
or the energy or linewidth of the exciton-like branch polariton
\begin{eqnarray}
	\Omega_X=\Omega_+-\frac{g^2\left(\Delta\omega'+\frac{i}{2}\Delta\gamma'\right)}{\Delta\omega'^2+\frac{1}{4}\Delta\gamma'^2}\nonumber
\end{eqnarray}
but $\omega_C$, $\omega_X$ and $\gamma_X$ is hard to directly measure.
These three values also change with temperature, e.g., the bare cavity mode energy $\omega_C$ shift with $T$ as discussed in SFig.~\ref{cat}.
Therefore, the linewidth $\gamma_-$ is used to extract $g$ from the experimental data as discussed before.
Additionally, we note that not all excitons in the TX peak interact with the cavity (e.g., due to polarization), thus the TX peak in spectra contains both bare TXs and the exciton-like branch polariton.
This is not a problem.
We can set the frequency of "real" exciton-like polariton as $\Omega_+^r=\omega_+^r-\frac{i}{2}\gamma_+^r$ in contrast to the measured $\Omega_+$ and their difference as $\Delta\omega^r=\omega_+^r-\omega_+$ and $\Delta\gamma^r=\gamma_+^r-\gamma_+$.
As the measured value $\Omega_+$ is between the bare exciton $\Omega_X$ and the "real" polcaiton $\Omega_+^r$, we could get $\Omega_+^r-\Omega_+ < \Omega_+^r-\Omega_X$.
As discussed above, $\Omega_+^r-\Omega_X \propto g^2\nonumber$ is a second order small quantity, thus $\Delta\omega^r$ and $\Delta\gamma^r$ are both at least second order small quantity.
If correct the previous equations by replacing $\Omega_+$ with $\Omega_+^r$, we will have
\begin{eqnarray}
	g^2=\frac{\gamma_--\gamma_C}{\Delta\gamma'+\Delta\gamma^r}\left(\left(\Delta\omega'+\Delta\omega^r\right)^2+\frac{1}{4}\left(\Delta\gamma'+\Delta\gamma^r\right)^2\right) \nonumber
\end{eqnarray}
where $g^2$, $\gamma_--\gamma_C$, $\Delta\omega^r$ and $\Delta\gamma^r$ are second order small quantities, while $\Delta\omega'$ and $\Delta\gamma'$ are not small quantity.
By approximation, we retain only second order small quantites on both sides and get
\begin{eqnarray}
	g^2=\frac{\gamma_--\gamma_C}{\Delta\gamma'}\left(\Delta\omega'^2+\frac{1}{4}\Delta\gamma'^2\right) \nonumber.
\end{eqnarray}
which is same as before.

The theoretical nonlocal interaction strength in the main paper is calculated with excitons in the weak confinement regime \cite{PhysRevB.86.085304}.
The weak confinement means the electron-hole are strongly bounded as a exciton and the exciton could move "freely" due to the weak confinement.
The exciton envelope function could be written as
\begin{eqnarray}
	\chi\left(\mathbf{R},\mathbf{r}\right)=\chi_{CM}\left(\mathbf{R}\right)\chi_{rel}\left(\mathbf{r'}\right) \nonumber
\end{eqnarray}
where
\begin{eqnarray}
	\mathbf{R}&=&\dfrac{m_e\mathbf{r_e}+m_h\mathbf{r_h}}{m_e+m_h} \nonumber\\
	\mathbf{r'}&=&\mathbf{r_e}-\mathbf{r_h} \nonumber
\end{eqnarray}
corresponding to the center-of-mass position and the electron-hole relative position.
$m_e,\mathbf{r_e}$ ($m_h,\mathbf{r_h}$) mean the effective mass and position of electron (hole).
Due to the large binding energy, the relative motion wavefunction $\chi_{rel}\left(\mathbf{r'}\right)$ is spatially small and will not change much with the temperature.
While the center-of-mass wavefunction $\chi_{CM}\left(\mathbf{R}\right)$ is quite large in spatial and will change with the temperature, thus plays the main role in the nonlocal effect.
For a brief simulation, we use a Gaussian function \cite{PhysRevB.86.085304}
\begin{eqnarray}
	\chi_{CM}\left(x\right)=\left(\frac{1}{\pi}\right)^\frac{1}{4}\left(\frac{1}{L}\right)^\frac{1}{2}e^{-\frac{x^2}{2L^2}} \nonumber
\end{eqnarray}
as $\chi_{CM}\left(x\right)$ with the spatial extent $L$.

\begin{figure}
	\includegraphics[width=\linewidth]{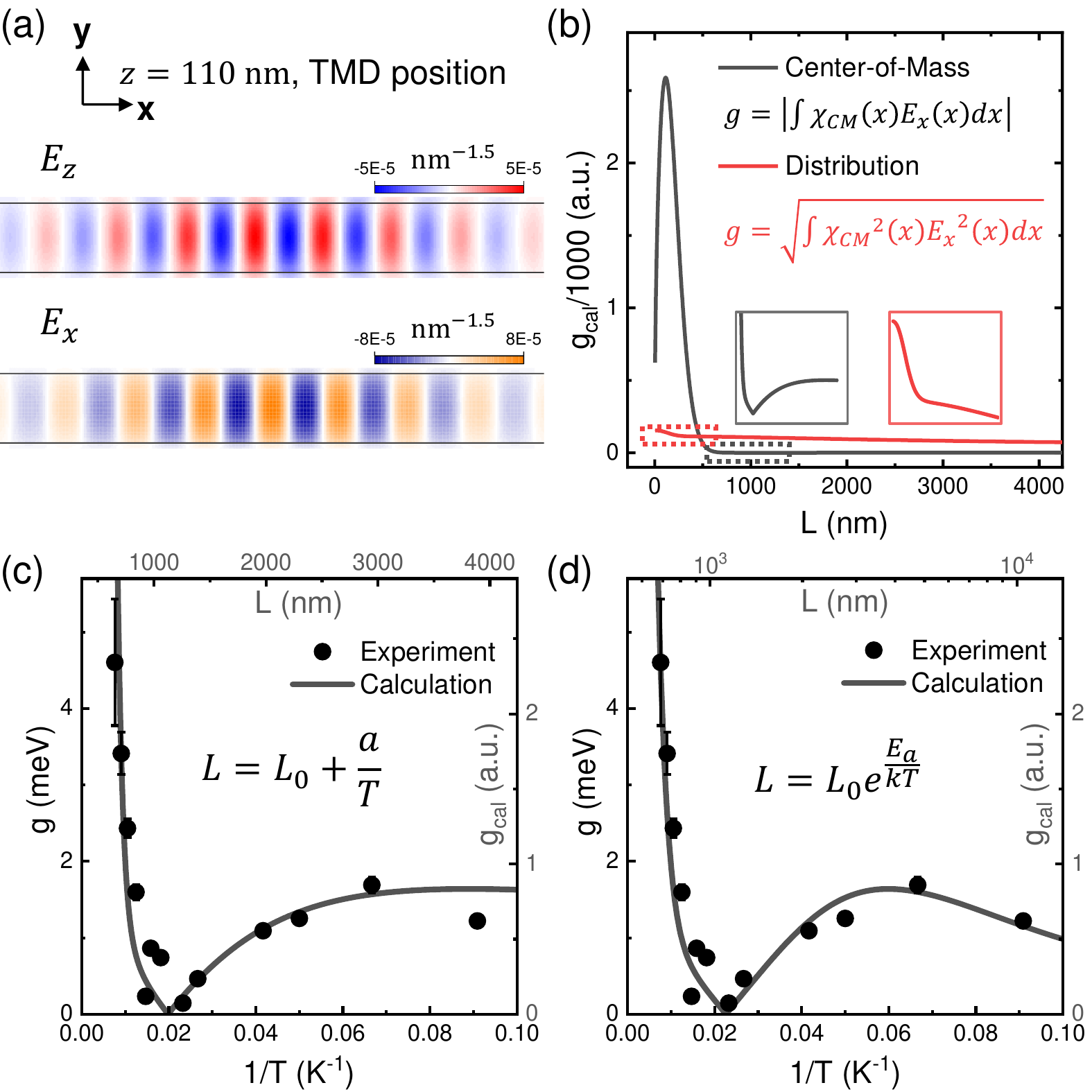}
	\caption{\label{dif}
		Nonlocal interaction strength calculation.
		(a) The cavity electric field distribution at the TMD position.
		(b) Calculated interaction strength $g$ with different $L$. Black line is for excitons with a center-of-mass wavefunction, and red line is for excitons with a Gaussian spatial distribution as a comparison.
		(c) Fitting by $L$ linear to $1/T$.
		(d) Fitting by $\mathrm{ln}\left(L\right)$ linear to $1/T$.
	}
\end{figure}

As our cavity is 1D photonic crystal nanobeam, the non-trivial electric field distribution is along x direction while the electric field distribution along y direction is trivial, as shown in SFig.~\ref{dif}(a).
Therefore, in the interaction strength we only calculate the integral in x direction by
\begin{eqnarray}
	g_{cal}=\vert\int \chi_{CM}(x)E_x(x)dx\vert \nonumber
\end{eqnarray}
where the electric field of cavity mode $E_x$ is simulated by the FDTD method as shown in Sec.~\ref{sec2a}.
Due to the cavity electric field $E_x$ has both positive and negative value at different positions, the integral can also have positive or negative value with different $\chi_{CM}$ function, resulting in a discontinuous $g$ when the integral value passes through the zero point.
As shown by the calculation results (black line in SFig.~\ref{dif}(b)), for the Gaussian $\chi_{CM}$ with varying $L$, there is a zero point of $g$ around $L=800\ \mathrm{nm}$.
In contrast, if we treat the the multiple excitons with a Gaussian spatial distribution as mentioned in the main paper, the collective interaction strength of multiple excitons is \cite{PhysRevA.96.011801}
\begin{eqnarray}
	g_{cal}&=&\left(\sum g_i^2\right)^{1/2} \nonumber \\
	g_i&=&d\cdot E_x(x_i) \nonumber
\end{eqnarray}
where $g_i$ is the interaction strength between one exciton and the cavity mode in the dipole approximation given by the dipole moment $d$ (same for all excitons).
By changing the discrete sum to distributed integral, we can get
\begin{eqnarray}
	g_{cal}=d\cdot\left(\int \chi^2_{CM}(x)E_x^2(x)dx\right)^{1/2} \nonumber
\end{eqnarray}
and the result is shown by the red line in SFig.~\ref{dif}(b).
As $E_x^2$ is always positive, the interaction strength $g$ is always continuous and only a monotonic decreasing is observed as $L$ increases.

The temperature dependence of the spatial extent $L$ is then the key point to contact the experimental and calculation results discussed above.
In our cavity the MoS$_2$ monolayer is encapsulated with hBN, thus the spatial extent $L$ is mainly limited by the exciton-phonon coupling.
For brief in the main paper, we use the simple model with $L$ linear to $1/T$ as
\begin{eqnarray}
	L=L_0+\frac{a}{T} \nonumber
\end{eqnarray}
where $L_0$ and $a$ are corresponding coefficients.
The corresponding fitting is shown in SFig.~\ref{dif}(c) with the coefficients of $L_0=354\pm24\ \mathrm{nm}$ and $a=38.9\pm2.4\ \mathrm{\mu m\cdot K}$.
$L$ at room temperature $T=300\ \mathrm{K}$ is $483\pm32\ \mathrm{nm}$.
Additionally, here we also present the result with a superlinear model
\begin{eqnarray}
	L=L_0e^{\frac{E_a}{kT}} \nonumber
\end{eqnarray}
which has also been reported in some literature \cite{KANGAWA2002517}.
$L_0$ and $E_a$ are corresponding coefficients and the fitting result is shown in SFig.~\ref{dif}(d) with $L_0=537\ \mathrm{nm}$ and $E_a=2.82\ \mathrm{meV}$ (approximated to $17.5\ \mathrm{\mu m\cdot K}$ for large $T$).
$L$ at room temperature $T=300\ \mathrm{K}$ is $600\ \mathrm{nm}$.
By comparison, the fitting result with the superlinear model seems a bit better, but the difference is not so much.
The actual motion of free excitons in TMD monolayer should be complex, depending on multiple factors.
Besides the exciton-phonon scattering discussed above, excitation laser power, ionized impurity scattering, impurity and doping in semiconductor will also codetermine \cite{doi:10.1021/acsnano.6b05580,Wang2016,doi:10.1021/acsnano.0c05305}.
Nonetheless, the resulting $L$-dependence of calculated interaction strength presented by the grey lines in SFig.~\ref{dif}(c)(d) reproduces our experimental findings (dots) remarkably well, due to the general principles in this work including the nonlocal effects and the shrinking center-of-mass wavefunction is theoretically correct.

In addition, as presented in SFig.~\ref{dif}(b), theoretically there is a maximum interaction strength when $L$ is around 200 nm.
This is also an interesting point and what we plan to achieve next.
Similar phenomena has been reported in the quantum dot system \cite{PhysRevLett.122.087401}.
However, the difficult point here is that, $\chi_{CM}\left(\mathbf{R}\right)$ in 2D materials is far from fully explored \cite{C9NR07056G}.
The stacking of a 2D-flake below 1 $\mu$m size is obviously difficult.
The external electric or magnetic field would be a potential way to control $\chi_{CM}\left(\mathbf{R}\right)$, but further investigations are need.

\section{\label{sec5}Summary}

In summary, we here report on the theoretical design, fabrication methods and additional experiment results of the 2D-material nanophotonic system.
With the considering of surface roughness, etching angle and misalignment during the fabrication, our cavity structure is demonstrated with high quality and high robustness.
Meanwhile, the high feasibility is ensured by that the fabrication is all implemented with common equipment and chemicals.
Furthermore, the additional experimental data not only strengthens the conclusion of novel exciton-photon interactions discussed in the main paper, but also expands the discussions such as the control of center-of-mass wavefunction as interesting topics in future work.


\begin{thebibliography}{52}%
\makeatletter
\providecommand \@ifxundefined [1]{%
 \@ifx{#1\undefined}
}%
\providecommand \@ifnum [1]{%
 \ifnum #1\expandafter \@firstoftwo
 \else \expandafter \@secondoftwo
 \fi
}%
\providecommand \@ifx [1]{%
 \ifx #1\expandafter \@firstoftwo
 \else \expandafter \@secondoftwo
 \fi
}%
\providecommand \natexlab [1]{#1}%
\providecommand \enquote  [1]{``#1''}%
\providecommand \bibnamefont  [1]{#1}%
\providecommand \bibfnamefont [1]{#1}%
\providecommand \citenamefont [1]{#1}%
\providecommand \href@noop [0]{\@secondoftwo}%
\providecommand \href [0]{\begingroup \@sanitize@url \@href}%
\providecommand \@href[1]{\@@startlink{#1}\@@href}%
\providecommand \@@href[1]{\endgroup#1\@@endlink}%
\providecommand \@sanitize@url [0]{\catcode `\\12\catcode `\$12\catcode
  `\&12\catcode `\#12\catcode `\^12\catcode `\_12\catcode `\%12\relax}%
\providecommand \@@startlink[1]{}%
\providecommand \@@endlink[0]{}%
\providecommand \url  [0]{\begingroup\@sanitize@url \@url }%
\providecommand \@url [1]{\endgroup\@href {#1}{\urlprefix }}%
\providecommand \urlprefix  [0]{URL }%
\providecommand \Eprint [0]{\href }%
\providecommand \doibase [0]{https://doi.org/}%
\providecommand \selectlanguage [0]{\@gobble}%
\providecommand \bibinfo  [0]{\@secondoftwo}%
\providecommand \bibfield  [0]{\@secondoftwo}%
\providecommand \translation [1]{[#1]}%
\providecommand \BibitemOpen [0]{}%
\providecommand \bibitemStop [0]{}%
\providecommand \bibitemNoStop [0]{.\EOS\space}%
\providecommand \EOS [0]{\spacefactor3000\relax}%
\providecommand \BibitemShut  [1]{\csname bibitem#1\endcsname}%
\let\auto@bib@innerbib\@empty
\bibitem [{\citenamefont {Wang}\ \emph {et~al.}(2018)\citenamefont {Wang},
  \citenamefont {Chernikov}, \citenamefont {Glazov}, \citenamefont {Heinz},
  \citenamefont {Marie}, \citenamefont {Amand},\ and\ \citenamefont
  {Urbaszek}}]{RevModPhys.90.021001}%
  \BibitemOpen
  \bibfield  {author} {\bibinfo {author} {\bibfnamefont {G.}~\bibnamefont
  {Wang}}, \bibinfo {author} {\bibfnamefont {A.}~\bibnamefont {Chernikov}},
  \bibinfo {author} {\bibfnamefont {M.~M.}\ \bibnamefont {Glazov}}, \bibinfo
  {author} {\bibfnamefont {T.~F.}\ \bibnamefont {Heinz}}, \bibinfo {author}
  {\bibfnamefont {X.}~\bibnamefont {Marie}}, \bibinfo {author} {\bibfnamefont
  {T.}~\bibnamefont {Amand}},\ and\ \bibinfo {author} {\bibfnamefont
  {B.}~\bibnamefont {Urbaszek}},\ }\bibfield  {title} {\bibinfo {title}
  {{Colloquium: Excitons in atomically thin transition metal
  dichalcogenides}},\ }\href {https://doi.org/10.1103/RevModPhys.90.021001}
  {\bibfield  {journal} {\bibinfo  {journal} {Rev. Mod. Phys.}\ }\textbf
  {\bibinfo {volume} {90}},\ \bibinfo {pages} {021001} (\bibinfo {year}
  {2018})}\BibitemShut {NoStop}%
\bibitem [{\citenamefont {Ardizzone}\ \emph {et~al.}(2019)\citenamefont
  {Ardizzone}, \citenamefont {Marco}, \citenamefont {Giorgi}, \citenamefont
  {Dominici}, \citenamefont {Ballarini},\ and\ \citenamefont
  {Sanvitto}}]{Ardizzone_2019}%
  \BibitemOpen
  \bibfield  {author} {\bibinfo {author} {\bibfnamefont {V.}~\bibnamefont
  {Ardizzone}}, \bibinfo {author} {\bibfnamefont {L.~D.}\ \bibnamefont
  {Marco}}, \bibinfo {author} {\bibfnamefont {M.~D.}\ \bibnamefont {Giorgi}},
  \bibinfo {author} {\bibfnamefont {L.}~\bibnamefont {Dominici}}, \bibinfo
  {author} {\bibfnamefont {D.}~\bibnamefont {Ballarini}},\ and\ \bibinfo
  {author} {\bibfnamefont {D.}~\bibnamefont {Sanvitto}},\ }\bibfield  {title}
  {\bibinfo {title} {{Emerging 2D materials for room-temperature
  polaritonics}},\ }\href {https://doi.org/10.1515/nanoph-2019-0114} {\bibfield
   {journal} {\bibinfo  {journal} {Nanophotonics}\ }\textbf {\bibinfo {volume}
  {8}},\ \bibinfo {pages} {1547} (\bibinfo {year} {2019})}\BibitemShut
  {NoStop}%
\bibitem [{\citenamefont {Stier}\ \emph {et~al.}(2018)\citenamefont {Stier},
  \citenamefont {Wilson}, \citenamefont {Velizhanin}, \citenamefont {Kono},
  \citenamefont {Xu},\ and\ \citenamefont {Crooker}}]{PhysRevLett.120.057405}%
  \BibitemOpen
  \bibfield  {author} {\bibinfo {author} {\bibfnamefont {A.~V.}\ \bibnamefont
  {Stier}}, \bibinfo {author} {\bibfnamefont {N.~P.}\ \bibnamefont {Wilson}},
  \bibinfo {author} {\bibfnamefont {K.~A.}\ \bibnamefont {Velizhanin}},
  \bibinfo {author} {\bibfnamefont {J.}~\bibnamefont {Kono}}, \bibinfo {author}
  {\bibfnamefont {X.}~\bibnamefont {Xu}},\ and\ \bibinfo {author}
  {\bibfnamefont {S.~A.}\ \bibnamefont {Crooker}},\ }\bibfield  {title}
  {\bibinfo {title} {Magnetooptics of exciton rydberg states in a monolayer
  semiconductor},\ }\href {https://doi.org/10.1103/PhysRevLett.120.057405}
  {\bibfield  {journal} {\bibinfo  {journal} {Phys. Rev. Lett.}\ }\textbf
  {\bibinfo {volume} {120}},\ \bibinfo {pages} {057405} (\bibinfo {year}
  {2018})}\BibitemShut {NoStop}%
\bibitem [{\citenamefont {Goryca}\ \emph {et~al.}(2019)\citenamefont {Goryca},
  \citenamefont {Li}, \citenamefont {Stier}, \citenamefont {Taniguchi},
  \citenamefont {Watanabe}, \citenamefont {Courtade}, \citenamefont {Shree},
  \citenamefont {Robert}, \citenamefont {Urbaszek}, \citenamefont {Marie},\
  and\ \citenamefont {Crooker}}]{Goryca2019}%
  \BibitemOpen
  \bibfield  {author} {\bibinfo {author} {\bibfnamefont {M.}~\bibnamefont
  {Goryca}}, \bibinfo {author} {\bibfnamefont {J.}~\bibnamefont {Li}}, \bibinfo
  {author} {\bibfnamefont {A.~V.}\ \bibnamefont {Stier}}, \bibinfo {author}
  {\bibfnamefont {T.}~\bibnamefont {Taniguchi}}, \bibinfo {author}
  {\bibfnamefont {K.}~\bibnamefont {Watanabe}}, \bibinfo {author}
  {\bibfnamefont {E.}~\bibnamefont {Courtade}}, \bibinfo {author}
  {\bibfnamefont {S.}~\bibnamefont {Shree}}, \bibinfo {author} {\bibfnamefont
  {C.}~\bibnamefont {Robert}}, \bibinfo {author} {\bibfnamefont
  {B.}~\bibnamefont {Urbaszek}}, \bibinfo {author} {\bibfnamefont
  {X.}~\bibnamefont {Marie}},\ and\ \bibinfo {author} {\bibfnamefont {S.~A.}\
  \bibnamefont {Crooker}},\ }\bibfield  {title} {\bibinfo {title} {Revealing
  exciton masses and dielectric properties of monolayer semiconductors with
  high magnetic fields},\ }\href {https://doi.org/10.1038/s41467-019-12180-y}
  {\bibfield  {journal} {\bibinfo  {journal} {Nat. Commun.}\ }\textbf {\bibinfo
  {volume} {10}},\ \bibinfo {pages} {4172} (\bibinfo {year}
  {2019})}\BibitemShut {NoStop}%
\bibitem [{\citenamefont {Wierzbowski}\ \emph {et~al.}(2017)\citenamefont
  {Wierzbowski}, \citenamefont {Klein}, \citenamefont {Sigger}, \citenamefont
  {Straubinger}, \citenamefont {Kremser}, \citenamefont {Taniguchi},
  \citenamefont {Watanabe}, \citenamefont {Wurstbauer}, \citenamefont
  {Holleitner}, \citenamefont {Kaniber}, \citenamefont {Müller},\ and\
  \citenamefont {Finley}}]{Wierzbowski_2017}%
  \BibitemOpen
  \bibfield  {author} {\bibinfo {author} {\bibfnamefont {J.}~\bibnamefont
  {Wierzbowski}}, \bibinfo {author} {\bibfnamefont {J.}~\bibnamefont {Klein}},
  \bibinfo {author} {\bibfnamefont {F.}~\bibnamefont {Sigger}}, \bibinfo
  {author} {\bibfnamefont {C.}~\bibnamefont {Straubinger}}, \bibinfo {author}
  {\bibfnamefont {M.}~\bibnamefont {Kremser}}, \bibinfo {author} {\bibfnamefont
  {T.}~\bibnamefont {Taniguchi}}, \bibinfo {author} {\bibfnamefont
  {K.}~\bibnamefont {Watanabe}}, \bibinfo {author} {\bibfnamefont
  {U.}~\bibnamefont {Wurstbauer}}, \bibinfo {author} {\bibfnamefont {A.~W.}\
  \bibnamefont {Holleitner}}, \bibinfo {author} {\bibfnamefont
  {M.}~\bibnamefont {Kaniber}}, \bibinfo {author} {\bibfnamefont
  {K.}~\bibnamefont {Müller}},\ and\ \bibinfo {author} {\bibfnamefont {J.~J.}\
  \bibnamefont {Finley}},\ }\bibfield  {title} {\bibinfo {title} {{Direct
  exciton emission from atomically thin transition metal dichalcogenide
  heterostructures near the lifetime limit}},\ }\href
  {https://doi.org/10.1038/s41598-017-09739-4} {\bibfield  {journal} {\bibinfo
  {journal} {Sci. Rep.}\ }\textbf {\bibinfo {volume} {7}},\ \bibinfo {pages}
  {12383} (\bibinfo {year} {2017})}\BibitemShut {NoStop}%
\bibitem [{\citenamefont {Cadiz}\ \emph {et~al.}(2017)\citenamefont {Cadiz},
  \citenamefont {Courtade}, \citenamefont {Robert}, \citenamefont {Wang},
  \citenamefont {Shen}, \citenamefont {Cai}, \citenamefont {Taniguchi},
  \citenamefont {Watanabe}, \citenamefont {Carrere}, \citenamefont {Lagarde},
  \citenamefont {Manca}, \citenamefont {Amand}, \citenamefont {Renucci},
  \citenamefont {Tongay}, \citenamefont {Marie},\ and\ \citenamefont
  {Urbaszek}}]{PhysRevX.7.021026}%
  \BibitemOpen
  \bibfield  {author} {\bibinfo {author} {\bibfnamefont {F.}~\bibnamefont
  {Cadiz}}, \bibinfo {author} {\bibfnamefont {E.}~\bibnamefont {Courtade}},
  \bibinfo {author} {\bibfnamefont {C.}~\bibnamefont {Robert}}, \bibinfo
  {author} {\bibfnamefont {G.}~\bibnamefont {Wang}}, \bibinfo {author}
  {\bibfnamefont {Y.}~\bibnamefont {Shen}}, \bibinfo {author} {\bibfnamefont
  {H.}~\bibnamefont {Cai}}, \bibinfo {author} {\bibfnamefont {T.}~\bibnamefont
  {Taniguchi}}, \bibinfo {author} {\bibfnamefont {K.}~\bibnamefont {Watanabe}},
  \bibinfo {author} {\bibfnamefont {H.}~\bibnamefont {Carrere}}, \bibinfo
  {author} {\bibfnamefont {D.}~\bibnamefont {Lagarde}}, \bibinfo {author}
  {\bibfnamefont {M.}~\bibnamefont {Manca}}, \bibinfo {author} {\bibfnamefont
  {T.}~\bibnamefont {Amand}}, \bibinfo {author} {\bibfnamefont
  {P.}~\bibnamefont {Renucci}}, \bibinfo {author} {\bibfnamefont
  {S.}~\bibnamefont {Tongay}}, \bibinfo {author} {\bibfnamefont
  {X.}~\bibnamefont {Marie}},\ and\ \bibinfo {author} {\bibfnamefont
  {B.}~\bibnamefont {Urbaszek}},\ }\bibfield  {title} {\bibinfo {title}
  {{Excitonic Linewidth Approaching the Homogeneous Limit in
  ${\mathrm{MoS}}_{2}$-Based van der Waals Heterostructures}},\ }\href
  {https://doi.org/10.1103/PhysRevX.7.021026} {\bibfield  {journal} {\bibinfo
  {journal} {Phys. Rev. X}\ }\textbf {\bibinfo {volume} {7}},\ \bibinfo {pages}
  {021026} (\bibinfo {year} {2017})}\BibitemShut {NoStop}%
\bibitem [{\citenamefont {Raja}\ \emph {et~al.}(2019)\citenamefont {Raja},
  \citenamefont {Waldecker}, \citenamefont {Zipfel}, \citenamefont {Cho},
  \citenamefont {Brem}, \citenamefont {Ziegler}, \citenamefont {Kulig},
  \citenamefont {Taniguchi}, \citenamefont {Watanabe}, \citenamefont {Malic},
  \citenamefont {Heinz}, \citenamefont {Berkelbach},\ and\ \citenamefont
  {Chernikov}}]{Raja2019}%
  \BibitemOpen
  \bibfield  {author} {\bibinfo {author} {\bibfnamefont {A.}~\bibnamefont
  {Raja}}, \bibinfo {author} {\bibfnamefont {L.}~\bibnamefont {Waldecker}},
  \bibinfo {author} {\bibfnamefont {J.}~\bibnamefont {Zipfel}}, \bibinfo
  {author} {\bibfnamefont {Y.}~\bibnamefont {Cho}}, \bibinfo {author}
  {\bibfnamefont {S.}~\bibnamefont {Brem}}, \bibinfo {author} {\bibfnamefont
  {J.~D.}\ \bibnamefont {Ziegler}}, \bibinfo {author} {\bibfnamefont
  {M.}~\bibnamefont {Kulig}}, \bibinfo {author} {\bibfnamefont
  {T.}~\bibnamefont {Taniguchi}}, \bibinfo {author} {\bibfnamefont
  {K.}~\bibnamefont {Watanabe}}, \bibinfo {author} {\bibfnamefont
  {E.}~\bibnamefont {Malic}}, \bibinfo {author} {\bibfnamefont {T.~F.}\
  \bibnamefont {Heinz}}, \bibinfo {author} {\bibfnamefont {T.~C.}\ \bibnamefont
  {Berkelbach}},\ and\ \bibinfo {author} {\bibfnamefont {A.}~\bibnamefont
  {Chernikov}},\ }\bibfield  {title} {\bibinfo {title} {Dielectric disorder in
  two-dimensional materials},\ }\href
  {https://doi.org/10.1038/s41565-019-0520-0} {\bibfield  {journal} {\bibinfo
  {journal} {Nat. Nanotechnol.}\ }\textbf {\bibinfo {volume} {14}},\ \bibinfo
  {pages} {832} (\bibinfo {year} {2019})}\BibitemShut {NoStop}%
\bibitem [{\citenamefont {Li}\ \emph {et~al.}(2014)\citenamefont {Li},
  \citenamefont {Chernikov}, \citenamefont {Zhang}, \citenamefont {Rigosi},
  \citenamefont {Hill}, \citenamefont {Zande}, \citenamefont {Chenet},
  \citenamefont {Shih}, \citenamefont {Hone},\ and\ \citenamefont
  {Heinz}}]{Li_PRB2014}%
  \BibitemOpen
  \bibfield  {author} {\bibinfo {author} {\bibfnamefont {Y.}~\bibnamefont
  {Li}}, \bibinfo {author} {\bibfnamefont {A.}~\bibnamefont {Chernikov}},
  \bibinfo {author} {\bibfnamefont {X.}~\bibnamefont {Zhang}}, \bibinfo
  {author} {\bibfnamefont {A.}~\bibnamefont {Rigosi}}, \bibinfo {author}
  {\bibfnamefont {H.~M.}\ \bibnamefont {Hill}}, \bibinfo {author}
  {\bibfnamefont {A.~M. v.~d.}\ \bibnamefont {Zande}}, \bibinfo {author}
  {\bibfnamefont {D.~A.}\ \bibnamefont {Chenet}}, \bibinfo {author}
  {\bibfnamefont {E.-M.}\ \bibnamefont {Shih}}, \bibinfo {author}
  {\bibfnamefont {J.}~\bibnamefont {Hone}},\ and\ \bibinfo {author}
  {\bibfnamefont {T.~F.}\ \bibnamefont {Heinz}},\ }\bibfield  {title} {\bibinfo
  {title} {{Measurement of the optical dielectric function of monolayer
  transition-metal dichalcogenides: MoS2, MoSe2, WS2, and WSe2}},\ }\href
  {https://doi.org/10.1103/physrevb.90.205422} {\bibfield  {journal} {\bibinfo
  {journal} {Phys. Rev. B}\ }\textbf {\bibinfo {volume} {90}},\ \bibinfo
  {pages} {205422} (\bibinfo {year} {2014})}\BibitemShut {NoStop}%
\bibitem [{\citenamefont {Castellanos-Gomez}\ \emph {et~al.}(2014)\citenamefont
  {Castellanos-Gomez}, \citenamefont {Buscema}, \citenamefont {Molenaar},
  \citenamefont {Singh}, \citenamefont {Janssen}, \citenamefont {Zant},\ and\
  \citenamefont {Steele}}]{Castellanos-Gomez_2014}%
  \BibitemOpen
  \bibfield  {author} {\bibinfo {author} {\bibfnamefont {A.}~\bibnamefont
  {Castellanos-Gomez}}, \bibinfo {author} {\bibfnamefont {M.}~\bibnamefont
  {Buscema}}, \bibinfo {author} {\bibfnamefont {R.}~\bibnamefont {Molenaar}},
  \bibinfo {author} {\bibfnamefont {V.}~\bibnamefont {Singh}}, \bibinfo
  {author} {\bibfnamefont {L.}~\bibnamefont {Janssen}}, \bibinfo {author}
  {\bibfnamefont {H.~S. J. v.~d.}\ \bibnamefont {Zant}},\ and\ \bibinfo
  {author} {\bibfnamefont {G.~A.}\ \bibnamefont {Steele}},\ }\bibfield  {title}
  {\bibinfo {title} {{Deterministic transfer of two-dimensional materials by
  all-dry viscoelastic stamping}},\ }\href
  {https://doi.org/10.1088/2053-1583/1/1/011002} {\bibfield  {journal}
  {\bibinfo  {journal} {2D Mater.}\ }\textbf {\bibinfo {volume} {1}},\ \bibinfo
  {pages} {011002} (\bibinfo {year} {2014})}\BibitemShut {NoStop}%
\bibitem [{\citenamefont {Schneider}\ \emph {et~al.}(2018)\citenamefont
  {Schneider}, \citenamefont {Glazov}, \citenamefont {Korn}, \citenamefont
  {Höfling},\ and\ \citenamefont {Urbaszek}}]{Schneider_2018}%
  \BibitemOpen
  \bibfield  {author} {\bibinfo {author} {\bibfnamefont {C.}~\bibnamefont
  {Schneider}}, \bibinfo {author} {\bibfnamefont {M.~M.}\ \bibnamefont
  {Glazov}}, \bibinfo {author} {\bibfnamefont {T.}~\bibnamefont {Korn}},
  \bibinfo {author} {\bibfnamefont {S.}~\bibnamefont {Höfling}},\ and\
  \bibinfo {author} {\bibfnamefont {B.}~\bibnamefont {Urbaszek}},\ }\bibfield
  {title} {\bibinfo {title} {{Two-dimensional semiconductors in the regime of
  strong light-matter coupling}},\ }\href
  {https://doi.org/10.1038/s41467-018-04866-6} {\bibfield  {journal} {\bibinfo
  {journal} {Nat. Commun.}\ }\textbf {\bibinfo {volume} {9}},\ \bibinfo {pages}
  {2695} (\bibinfo {year} {2018})}\BibitemShut {NoStop}%
\bibitem [{\citenamefont {Dufferwiel}\ \emph {et~al.}(2015)\citenamefont
  {Dufferwiel}, \citenamefont {Schwarz}, \citenamefont {Withers}, \citenamefont
  {Trichet}, \citenamefont {Li}, \citenamefont {Sich}, \citenamefont
  {Pozo-Zamudio}, \citenamefont {Clark}, \citenamefont {Nalitov}, \citenamefont
  {Solnyshkov}, \citenamefont {Malpuech}, \citenamefont {Novoselov},
  \citenamefont {Smith}, \citenamefont {Skolnick}, \citenamefont
  {Krizhanovskii},\ and\ \citenamefont {Tartakovskii}}]{Dufferwiel_2015}%
  \BibitemOpen
  \bibfield  {author} {\bibinfo {author} {\bibfnamefont {S.}~\bibnamefont
  {Dufferwiel}}, \bibinfo {author} {\bibfnamefont {S.}~\bibnamefont {Schwarz}},
  \bibinfo {author} {\bibfnamefont {F.}~\bibnamefont {Withers}}, \bibinfo
  {author} {\bibfnamefont {A.~A.~P.}\ \bibnamefont {Trichet}}, \bibinfo
  {author} {\bibfnamefont {F.}~\bibnamefont {Li}}, \bibinfo {author}
  {\bibfnamefont {M.}~\bibnamefont {Sich}}, \bibinfo {author} {\bibfnamefont
  {O.~D.}\ \bibnamefont {Pozo-Zamudio}}, \bibinfo {author} {\bibfnamefont
  {C.}~\bibnamefont {Clark}}, \bibinfo {author} {\bibfnamefont
  {A.}~\bibnamefont {Nalitov}}, \bibinfo {author} {\bibfnamefont {D.~D.}\
  \bibnamefont {Solnyshkov}}, \bibinfo {author} {\bibfnamefont
  {G.}~\bibnamefont {Malpuech}}, \bibinfo {author} {\bibfnamefont {K.~S.}\
  \bibnamefont {Novoselov}}, \bibinfo {author} {\bibfnamefont {J.~M.}\
  \bibnamefont {Smith}}, \bibinfo {author} {\bibfnamefont {M.~S.}\ \bibnamefont
  {Skolnick}}, \bibinfo {author} {\bibfnamefont {D.~N.}\ \bibnamefont
  {Krizhanovskii}},\ and\ \bibinfo {author} {\bibfnamefont {A.~I.}\
  \bibnamefont {Tartakovskii}},\ }\bibfield  {title} {\bibinfo {title}
  {{Exciton–polaritons in van der Waals heterostructures embedded in tunable
  microcavities}},\ }\href {https://doi.org/10.1038/ncomms9579} {\bibfield
  {journal} {\bibinfo  {journal} {Nat. Commun.}\ }\textbf {\bibinfo {volume}
  {6}},\ \bibinfo {pages} {8579} (\bibinfo {year} {2015})}\BibitemShut
  {NoStop}%
\bibitem [{\citenamefont {Zhang}\ \emph {et~al.}(2018)\citenamefont {Zhang},
  \citenamefont {Gogna}, \citenamefont {Burg}, \citenamefont {Tutuc},\ and\
  \citenamefont {Deng}}]{Zhang2018}%
  \BibitemOpen
  \bibfield  {author} {\bibinfo {author} {\bibfnamefont {L.}~\bibnamefont
  {Zhang}}, \bibinfo {author} {\bibfnamefont {R.}~\bibnamefont {Gogna}},
  \bibinfo {author} {\bibfnamefont {W.}~\bibnamefont {Burg}}, \bibinfo {author}
  {\bibfnamefont {E.}~\bibnamefont {Tutuc}},\ and\ \bibinfo {author}
  {\bibfnamefont {H.}~\bibnamefont {Deng}},\ }\bibfield  {title} {\bibinfo
  {title} {{Photonic-crystal exciton-polaritons in monolayer semiconductors}},\
  }\href {https://doi.org/10.1038/s41467-018-03188-x} {\bibfield  {journal}
  {\bibinfo  {journal} {Nat. Commun.}\ }\textbf {\bibinfo {volume} {9}},\
  \bibinfo {pages} {713} (\bibinfo {year} {2018})}\BibitemShut {NoStop}%
\bibitem [{\citenamefont {Ma}\ \emph {et~al.}(2020)\citenamefont {Ma},
  \citenamefont {Youngblood}, \citenamefont {Liu}, \citenamefont {Cheng},
  \citenamefont {Cunha}, \citenamefont {Kudtarkar}, \citenamefont {Wang},\ and\
  \citenamefont {Lan}}]{Ma_2020}%
  \BibitemOpen
  \bibfield  {author} {\bibinfo {author} {\bibfnamefont {X.}~\bibnamefont
  {Ma}}, \bibinfo {author} {\bibfnamefont {N.}~\bibnamefont {Youngblood}},
  \bibinfo {author} {\bibfnamefont {X.}~\bibnamefont {Liu}}, \bibinfo {author}
  {\bibfnamefont {Y.}~\bibnamefont {Cheng}}, \bibinfo {author} {\bibfnamefont
  {P.}~\bibnamefont {Cunha}}, \bibinfo {author} {\bibfnamefont
  {K.}~\bibnamefont {Kudtarkar}}, \bibinfo {author} {\bibfnamefont
  {X.}~\bibnamefont {Wang}},\ and\ \bibinfo {author} {\bibfnamefont
  {S.}~\bibnamefont {Lan}},\ }\bibfield  {title} {\bibinfo {title}
  {{Engineering photonic environments for two-dimensional materials}},\ }\href
  {https://doi.org/10.1515/nanoph-2020-0524} {\bibfield  {journal} {\bibinfo
  {journal} {Nanophotonics}\ }\textbf {\bibinfo {volume} {10}},\ \bibinfo
  {pages} {1031} (\bibinfo {year} {2020})}\BibitemShut {NoStop}%
\bibitem [{\citenamefont {Rosser}\ \emph {et~al.}(2021)\citenamefont {Rosser},
  \citenamefont {Gerace}, \citenamefont {Andreani},\ and\ \citenamefont
  {Majumdar}}]{2107.00078}%
  \BibitemOpen
  \bibfield  {author} {\bibinfo {author} {\bibfnamefont {D.}~\bibnamefont
  {Rosser}}, \bibinfo {author} {\bibfnamefont {D.}~\bibnamefont {Gerace}},
  \bibinfo {author} {\bibfnamefont {L.~C.}\ \bibnamefont {Andreani}},\ and\
  \bibinfo {author} {\bibfnamefont {A.}~\bibnamefont {Majumdar}},\ }\bibfield
  {title} {\bibinfo {title} {Optimal condition to probe strong coupling of
  two-dimensional excitons and zero-dimensional cavity modes},\ }\href
  {https://doi.org/10.1103/PhysRevB.104.235436} {\bibfield  {journal} {\bibinfo
   {journal} {Phys. Rev. B}\ }\textbf {\bibinfo {volume} {104}},\ \bibinfo
  {pages} {235436} (\bibinfo {year} {2021})}\BibitemShut {NoStop}%
\bibitem [{\citenamefont {Lundt}\ \emph {et~al.}(2016)\citenamefont {Lundt},
  \citenamefont {Mary{\'{n}}ski}, \citenamefont {Cherotchenko}, \citenamefont
  {Pant}, \citenamefont {Fan}, \citenamefont {Tongay}, \citenamefont {Sek},
  \citenamefont {Kavokin}, \citenamefont {Höfling},\ and\ \citenamefont
  {Schneider}}]{Lundt_2016}%
  \BibitemOpen
  \bibfield  {author} {\bibinfo {author} {\bibfnamefont {N.}~\bibnamefont
  {Lundt}}, \bibinfo {author} {\bibfnamefont {A.}~\bibnamefont
  {Mary{\'{n}}ski}}, \bibinfo {author} {\bibfnamefont {E.}~\bibnamefont
  {Cherotchenko}}, \bibinfo {author} {\bibfnamefont {A.}~\bibnamefont {Pant}},
  \bibinfo {author} {\bibfnamefont {X.}~\bibnamefont {Fan}}, \bibinfo {author}
  {\bibfnamefont {S.}~\bibnamefont {Tongay}}, \bibinfo {author} {\bibfnamefont
  {G.}~\bibnamefont {Sek}}, \bibinfo {author} {\bibfnamefont {A.~V.}\
  \bibnamefont {Kavokin}}, \bibinfo {author} {\bibfnamefont {S.}~\bibnamefont
  {Höfling}},\ and\ \bibinfo {author} {\bibfnamefont {C.}~\bibnamefont
  {Schneider}},\ }\bibfield  {title} {\bibinfo {title} {Monolayered {MoSe}2: a
  candidate for room temperature polaritonics},\ }\href
  {https://doi.org/10.1088/2053-1583/4/1/015006} {\bibfield  {journal}
  {\bibinfo  {journal} {2D Mater.}\ }\textbf {\bibinfo {volume} {4}},\
  \bibinfo {pages} {015006} (\bibinfo {year} {2016})}\BibitemShut {NoStop}%
\bibitem [{\citenamefont {Hu}\ \emph {et~al.}(2017)\citenamefont {Hu},
  \citenamefont {Wang}, \citenamefont {Wu}, \citenamefont {Zhang},
  \citenamefont {Shan}, \citenamefont {Lu}, \citenamefont {Wang}, \citenamefont
  {Luo}, \citenamefont {Zhang}, \citenamefont {Liao}, \citenamefont {Wu},
  \citenamefont {Shen},\ and\ \citenamefont {Chen}}]{Hu_2016}%
  \BibitemOpen
  \bibfield  {author} {\bibinfo {author} {\bibfnamefont {T.}~\bibnamefont
  {Hu}}, \bibinfo {author} {\bibfnamefont {Y.}~\bibnamefont {Wang}}, \bibinfo
  {author} {\bibfnamefont {L.}~\bibnamefont {Wu}}, \bibinfo {author}
  {\bibfnamefont {L.}~\bibnamefont {Zhang}}, \bibinfo {author} {\bibfnamefont
  {Y.}~\bibnamefont {Shan}}, \bibinfo {author} {\bibfnamefont {J.}~\bibnamefont
  {Lu}}, \bibinfo {author} {\bibfnamefont {J.}~\bibnamefont {Wang}}, \bibinfo
  {author} {\bibfnamefont {S.}~\bibnamefont {Luo}}, \bibinfo {author}
  {\bibfnamefont {Z.}~\bibnamefont {Zhang}}, \bibinfo {author} {\bibfnamefont
  {L.}~\bibnamefont {Liao}}, \bibinfo {author} {\bibfnamefont {S.}~\bibnamefont
  {Wu}}, \bibinfo {author} {\bibfnamefont {X.}~\bibnamefont {Shen}},\ and\
  \bibinfo {author} {\bibfnamefont {Z.}~\bibnamefont {Chen}},\ }\bibfield
  {title} {\bibinfo {title} {{Strong coupling between Tamm plasmon polariton
  and two dimensional semiconductor excitons}},\ }\href
  {https://doi.org/10.1063/1.4974901} {\bibfield  {journal} {\bibinfo
  {journal} {Appl. Phys. Lett.}\ }\textbf {\bibinfo {volume} {110}},\ \bibinfo
  {pages} {051101} (\bibinfo {year} {2017})}\BibitemShut {NoStop}%
\bibitem [{\citenamefont {Flatten}\ \emph {et~al.}(2016)\citenamefont
  {Flatten}, \citenamefont {He}, \citenamefont {Coles}, \citenamefont
  {Trichet}, \citenamefont {Powell}, \citenamefont {Taylor}, \citenamefont
  {Warner},\ and\ \citenamefont {Smith}}]{Flatten_2016}%
  \BibitemOpen
  \bibfield  {author} {\bibinfo {author} {\bibfnamefont {L.~C.}\ \bibnamefont
  {Flatten}}, \bibinfo {author} {\bibfnamefont {Z.}~\bibnamefont {He}},
  \bibinfo {author} {\bibfnamefont {D.~M.}\ \bibnamefont {Coles}}, \bibinfo
  {author} {\bibfnamefont {A.~A.~P.}\ \bibnamefont {Trichet}}, \bibinfo
  {author} {\bibfnamefont {A.~W.}\ \bibnamefont {Powell}}, \bibinfo {author}
  {\bibfnamefont {R.~A.}\ \bibnamefont {Taylor}}, \bibinfo {author}
  {\bibfnamefont {J.~H.}\ \bibnamefont {Warner}},\ and\ \bibinfo {author}
  {\bibfnamefont {J.~M.}\ \bibnamefont {Smith}},\ }\bibfield  {title} {\bibinfo
  {title} {{Room-temperature exciton-polaritons with two-dimensional WS2}},\
  }\href {https://doi.org/10.1038/srep33134} {\bibfield  {journal} {\bibinfo
  {journal} {Sci. Rep.}\ }\textbf {\bibinfo {volume} {6}},\ \bibinfo {pages}
  {33134} (\bibinfo {year} {2016})}\BibitemShut {NoStop}%
\bibitem [{\citenamefont {Wu}\ \emph {et~al.}(2015)\citenamefont {Wu},
  \citenamefont {Buckley}, \citenamefont {Schaibley}, \citenamefont {Feng},
  \citenamefont {Yan}, \citenamefont {Mandrus}, \citenamefont {Hatami},
  \citenamefont {Yao}, \citenamefont {Vu{\v{c}}kovi{\'{c}}}, \citenamefont
  {Majumdar},\ and\ \citenamefont {Xu}}]{Wu2015}%
  \BibitemOpen
  \bibfield  {author} {\bibinfo {author} {\bibfnamefont {S.}~\bibnamefont
  {Wu}}, \bibinfo {author} {\bibfnamefont {S.}~\bibnamefont {Buckley}},
  \bibinfo {author} {\bibfnamefont {J.~R.}\ \bibnamefont {Schaibley}}, \bibinfo
  {author} {\bibfnamefont {L.}~\bibnamefont {Feng}}, \bibinfo {author}
  {\bibfnamefont {J.}~\bibnamefont {Yan}}, \bibinfo {author} {\bibfnamefont
  {D.~G.}\ \bibnamefont {Mandrus}}, \bibinfo {author} {\bibfnamefont
  {F.}~\bibnamefont {Hatami}}, \bibinfo {author} {\bibfnamefont
  {W.}~\bibnamefont {Yao}}, \bibinfo {author} {\bibfnamefont {J.}~\bibnamefont
  {Vu{\v{c}}kovi{\'{c}}}}, \bibinfo {author} {\bibfnamefont {A.}~\bibnamefont
  {Majumdar}},\ and\ \bibinfo {author} {\bibfnamefont {X.}~\bibnamefont {Xu}},\
  }\bibfield  {title} {\bibinfo {title} {{Monolayer semiconductor nanocavity
  lasers with ultralow thresholds}},\ }\href
  {https://doi.org/10.1038/nature14290} {\bibfield  {journal} {\bibinfo
  {journal} {Nature}\ }\textbf {\bibinfo {volume} {520}},\ \bibinfo {pages}
  {69} (\bibinfo {year} {2015})}\BibitemShut {NoStop}%
\bibitem [{\citenamefont {Ye}\ \emph {et~al.}(2015)\citenamefont {Ye},
  \citenamefont {Wong}, \citenamefont {Lu}, \citenamefont {Ni}, \citenamefont
  {Zhu}, \citenamefont {Chen}, \citenamefont {Wang},\ and\ \citenamefont
  {Zhang}}]{Ye2015}%
  \BibitemOpen
  \bibfield  {author} {\bibinfo {author} {\bibfnamefont {Y.}~\bibnamefont
  {Ye}}, \bibinfo {author} {\bibfnamefont {Z.~J.}\ \bibnamefont {Wong}},
  \bibinfo {author} {\bibfnamefont {X.}~\bibnamefont {Lu}}, \bibinfo {author}
  {\bibfnamefont {X.}~\bibnamefont {Ni}}, \bibinfo {author} {\bibfnamefont
  {H.}~\bibnamefont {Zhu}}, \bibinfo {author} {\bibfnamefont {X.}~\bibnamefont
  {Chen}}, \bibinfo {author} {\bibfnamefont {Y.}~\bibnamefont {Wang}},\ and\
  \bibinfo {author} {\bibfnamefont {X.}~\bibnamefont {Zhang}},\ }\bibfield
  {title} {\bibinfo {title} {{Monolayer excitonic laser}},\ }\href
  {https://doi.org/10.1038/nphoton.2015.197} {\bibfield  {journal} {\bibinfo
  {journal} {Nat. Photonics}\ }\textbf {\bibinfo {volume} {9}},\ \bibinfo
  {pages} {733} (\bibinfo {year} {2015})}\BibitemShut {NoStop}%
\bibitem [{\citenamefont {Li}\ \emph {et~al.}(2017)\citenamefont {Li},
  \citenamefont {Zhang}, \citenamefont {Huang}, \citenamefont {Sun},
  \citenamefont {Fan}, \citenamefont {Feng}, \citenamefont {Wang},\ and\
  \citenamefont {Ning}}]{Li2017}%
  \BibitemOpen
  \bibfield  {author} {\bibinfo {author} {\bibfnamefont {Y.}~\bibnamefont
  {Li}}, \bibinfo {author} {\bibfnamefont {J.}~\bibnamefont {Zhang}}, \bibinfo
  {author} {\bibfnamefont {D.}~\bibnamefont {Huang}}, \bibinfo {author}
  {\bibfnamefont {H.}~\bibnamefont {Sun}}, \bibinfo {author} {\bibfnamefont
  {F.}~\bibnamefont {Fan}}, \bibinfo {author} {\bibfnamefont {J.}~\bibnamefont
  {Feng}}, \bibinfo {author} {\bibfnamefont {Z.}~\bibnamefont {Wang}},\ and\
  \bibinfo {author} {\bibfnamefont {C.~Z.}\ \bibnamefont {Ning}},\ }\bibfield
  {title} {\bibinfo {title} {{Room-temperature continuous-wave lasing from
  monolayer molybdenum ditelluride integrated with a silicon nanobeam
  cavity}},\ }\href {https://doi.org/10.1038/nnano.2017.128} {\bibfield
  {journal} {\bibinfo  {journal} {Nat. Nanotechnol.}\ }\textbf {\bibinfo
  {volume} {12}},\ \bibinfo {pages} {987} (\bibinfo {year} {2017})}\BibitemShut
  {NoStop}%
\bibitem [{\citenamefont {Rosser}\ \emph {et~al.}(2022)\citenamefont {Rosser},
  \citenamefont {Gerace}, \citenamefont {Chen}, \citenamefont {Liu},
  \citenamefont {Whitehead}, \citenamefont {Ryou}, \citenamefont {Andreani},\
  and\ \citenamefont {Majumdar}}]{2010.05458}%
  \BibitemOpen
  \bibfield  {author} {\bibinfo {author} {\bibfnamefont {D.}~\bibnamefont
  {Rosser}}, \bibinfo {author} {\bibfnamefont {D.}~\bibnamefont {Gerace}},
  \bibinfo {author} {\bibfnamefont {Y.}~\bibnamefont {Chen}}, \bibinfo {author}
  {\bibfnamefont {Y.}~\bibnamefont {Liu}}, \bibinfo {author} {\bibfnamefont
  {J.}~\bibnamefont {Whitehead}}, \bibinfo {author} {\bibfnamefont
  {A.}~\bibnamefont {Ryou}}, \bibinfo {author} {\bibfnamefont {L.~C.}\
  \bibnamefont {Andreani}},\ and\ \bibinfo {author} {\bibfnamefont
  {A.}~\bibnamefont {Majumdar}},\ }\bibfield  {title} {\bibinfo {title}
  {Dispersive coupling between mose2 and an integrated zero-dimensional
  nanocavity},\ }\href {https://doi.org/10.1364/OME.443536} {\bibfield
  {journal} {\bibinfo  {journal} {Opt. Mater. Express}\ }\textbf {\bibinfo
  {volume} {12}},\ \bibinfo {pages} {59} (\bibinfo {year} {2022})}\BibitemShut
  {NoStop}%
\bibitem [{\citenamefont {Rhodes}\ \emph {et~al.}(2019)\citenamefont {Rhodes},
  \citenamefont {Chae}, \citenamefont {Ribeiro-Palau},\ and\ \citenamefont
  {Hone}}]{Rhodes2019}%
  \BibitemOpen
  \bibfield  {author} {\bibinfo {author} {\bibfnamefont {D.}~\bibnamefont
  {Rhodes}}, \bibinfo {author} {\bibfnamefont {S.~H.}\ \bibnamefont {Chae}},
  \bibinfo {author} {\bibfnamefont {R.}~\bibnamefont {Ribeiro-Palau}},\ and\
  \bibinfo {author} {\bibfnamefont {J.}~\bibnamefont {Hone}},\ }\bibfield
  {title} {\bibinfo {title} {{Disorder in van der Waals heterostructures of 2D
  materials}},\ }\href {https://doi.org/10.1038/s41563-019-0366-8} {\bibfield
  {journal} {\bibinfo  {journal} {Nat. Mater.}\ }\textbf {\bibinfo {volume}
  {18}},\ \bibinfo {pages} {541} (\bibinfo {year} {2019})}\BibitemShut
  {NoStop}%
\bibitem [{\citenamefont {Ju}\ \emph {et~al.}(2014)\citenamefont {Ju},
  \citenamefont {Velasco}, \citenamefont {Huang}, \citenamefont {Kahn},
  \citenamefont {Nosiglia}, \citenamefont {Tsai}, \citenamefont {Yang},
  \citenamefont {Taniguchi}, \citenamefont {Watanabe}, \citenamefont {Zhang},
  \citenamefont {Zhang}, \citenamefont {Crommie}, \citenamefont {Zettl},\ and\
  \citenamefont {Wang}}]{Ju2014}%
  \BibitemOpen
  \bibfield  {author} {\bibinfo {author} {\bibfnamefont {L.}~\bibnamefont
  {Ju}}, \bibinfo {author} {\bibfnamefont {J.}~\bibnamefont {Velasco}},
  \bibinfo {author} {\bibfnamefont {E.}~\bibnamefont {Huang}}, \bibinfo
  {author} {\bibfnamefont {S.}~\bibnamefont {Kahn}}, \bibinfo {author}
  {\bibfnamefont {C.}~\bibnamefont {Nosiglia}}, \bibinfo {author}
  {\bibfnamefont {H.-Z.}\ \bibnamefont {Tsai}}, \bibinfo {author}
  {\bibfnamefont {W.}~\bibnamefont {Yang}}, \bibinfo {author} {\bibfnamefont
  {T.}~\bibnamefont {Taniguchi}}, \bibinfo {author} {\bibfnamefont
  {K.}~\bibnamefont {Watanabe}}, \bibinfo {author} {\bibfnamefont
  {Y.}~\bibnamefont {Zhang}}, \bibinfo {author} {\bibfnamefont
  {G.}~\bibnamefont {Zhang}}, \bibinfo {author} {\bibfnamefont
  {M.}~\bibnamefont {Crommie}}, \bibinfo {author} {\bibfnamefont
  {A.}~\bibnamefont {Zettl}},\ and\ \bibinfo {author} {\bibfnamefont
  {F.}~\bibnamefont {Wang}},\ }\bibfield  {title} {\bibinfo {title}
  {Photoinduced doping in heterostructures of graphene and boron nitride},\
  }\href {https://doi.org/10.1038/nnano.2014.60} {\bibfield  {journal}
  {\bibinfo  {journal} {Nat. Nanotechnol.}\ }\textbf {\bibinfo {volume} {9}},\
  \bibinfo {pages} {348} (\bibinfo {year} {2014})}\BibitemShut {NoStop}%
\bibitem [{\citenamefont {Rosati}\ \emph {et~al.}(2020)\citenamefont {Rosati},
  \citenamefont {Perea-Causín}, \citenamefont {Brem},\ and\ \citenamefont
  {Malic}}]{C9NR07056G}%
  \BibitemOpen
  \bibfield  {author} {\bibinfo {author} {\bibfnamefont {R.}~\bibnamefont
  {Rosati}}, \bibinfo {author} {\bibfnamefont {R.}~\bibnamefont
  {Perea-Causín}}, \bibinfo {author} {\bibfnamefont {S.}~\bibnamefont
  {Brem}},\ and\ \bibinfo {author} {\bibfnamefont {E.}~\bibnamefont {Malic}},\
  }\bibfield  {title} {\bibinfo {title} {Negative effective excitonic diffusion
  in monolayer transition metal dichalcogenides},\ }\href
  {https://doi.org/10.1039/C9NR07056G} {\bibfield  {journal} {\bibinfo
  {journal} {Nanoscale}\ }\textbf {\bibinfo {volume} {12}},\ \bibinfo {pages}
  {356} (\bibinfo {year} {2020})}\BibitemShut {NoStop}%
\bibitem [{\citenamefont {Kato}\ and\ \citenamefont
  {Kaneko}(2016)}]{doi:10.1021/acsnano.6b05580}%
  \BibitemOpen
  \bibfield  {author} {\bibinfo {author} {\bibfnamefont {T.}~\bibnamefont
  {Kato}}\ and\ \bibinfo {author} {\bibfnamefont {T.}~\bibnamefont {Kaneko}},\
  }\bibfield  {title} {\bibinfo {title} {{Transport Dynamics of Neutral
  Excitons and Trions in Monolayer WS2}},\ }\href
  {https://doi.org/10.1021/acsnano.6b05580} {\bibfield  {journal} {\bibinfo
  {journal} {ACS Nano}\ }\textbf {\bibinfo {volume} {10}},\ \bibinfo {pages}
  {9687} (\bibinfo {year} {2016})}\BibitemShut {NoStop}%
\bibitem [{\citenamefont {Stobbe}\ \emph {et~al.}(2012)\citenamefont {Stobbe},
  \citenamefont {Kristensen}, \citenamefont {Mortensen}, \citenamefont {Hvam},
  \citenamefont {M\o{}rk},\ and\ \citenamefont {Lodahl}}]{PhysRevB.86.085304}%
  \BibitemOpen
  \bibfield  {author} {\bibinfo {author} {\bibfnamefont {S.}~\bibnamefont
  {Stobbe}}, \bibinfo {author} {\bibfnamefont {P.~T.}\ \bibnamefont
  {Kristensen}}, \bibinfo {author} {\bibfnamefont {J.~E.}\ \bibnamefont
  {Mortensen}}, \bibinfo {author} {\bibfnamefont {J.~M.}\ \bibnamefont {Hvam}},
  \bibinfo {author} {\bibfnamefont {J.}~\bibnamefont {M\o{}rk}},\ and\ \bibinfo
  {author} {\bibfnamefont {P.}~\bibnamefont {Lodahl}},\ }\bibfield  {title}
  {\bibinfo {title} {{Spontaneous emission from large quantum dots in
  nanostructures: Exciton-photon interaction beyond the dipole
  approximation}},\ }\href {https://doi.org/10.1103/PhysRevB.86.085304}
  {\bibfield  {journal} {\bibinfo  {journal} {Phys. Rev. B}\ }\textbf {\bibinfo
  {volume} {86}},\ \bibinfo {pages} {085304} (\bibinfo {year}
  {2012})}\BibitemShut {NoStop}%
\bibitem [{\citenamefont {Uddin}\ \emph {et~al.}(2020)\citenamefont {Uddin},
  \citenamefont {Kim}, \citenamefont {Lorenzon}, \citenamefont {Yeh},
  \citenamefont {Lien}, \citenamefont {Barnard}, \citenamefont {Htoon},
  \citenamefont {Weber-Bargioni},\ and\ \citenamefont
  {Javey}}]{doi:10.1021/acsnano.0c05305}%
  \BibitemOpen
  \bibfield  {author} {\bibinfo {author} {\bibfnamefont {S.~Z.}\ \bibnamefont
  {Uddin}}, \bibinfo {author} {\bibfnamefont {H.}~\bibnamefont {Kim}}, \bibinfo
  {author} {\bibfnamefont {M.}~\bibnamefont {Lorenzon}}, \bibinfo {author}
  {\bibfnamefont {M.}~\bibnamefont {Yeh}}, \bibinfo {author} {\bibfnamefont
  {D.-H.}\ \bibnamefont {Lien}}, \bibinfo {author} {\bibfnamefont {E.~S.}\
  \bibnamefont {Barnard}}, \bibinfo {author} {\bibfnamefont {H.}~\bibnamefont
  {Htoon}}, \bibinfo {author} {\bibfnamefont {A.}~\bibnamefont
  {Weber-Bargioni}},\ and\ \bibinfo {author} {\bibfnamefont {A.}~\bibnamefont
  {Javey}},\ }\bibfield  {title} {\bibinfo {title} {Neutral exciton diffusion
  in monolayer mos2},\ }\href {https://doi.org/10.1021/acsnano.0c05305}
  {\bibfield  {journal} {\bibinfo  {journal} {ACS Nano}\ }\textbf {\bibinfo
  {volume} {14}},\ \bibinfo {pages} {13433} (\bibinfo {year}
  {2020})}\BibitemShut {NoStop}%
\bibitem [{\citenamefont {Qian}\ \emph {et~al.}(2020)\citenamefont {Qian},
  \citenamefont {Xie}, \citenamefont {Yang},\ and\ \citenamefont
  {Xu}}]{doi:10.1002/qute.201900024}%
  \BibitemOpen
  \bibfield  {author} {\bibinfo {author} {\bibfnamefont {C.}~\bibnamefont
  {Qian}}, \bibinfo {author} {\bibfnamefont {X.}~\bibnamefont {Xie}}, \bibinfo
  {author} {\bibfnamefont {J.}~\bibnamefont {Yang}},\ and\ \bibinfo {author}
  {\bibfnamefont {X.}~\bibnamefont {Xu}},\ }\bibfield  {title} {\bibinfo
  {title} {{A Cratered Photonic Crystal Cavity Mode for Non-local
  Exciton–Photon Interactions}},\ }\href
  {https://doi.org/10.1002/qute.201900024} {\bibfield  {journal} {\bibinfo
  {journal} {Adv. Quantum Technol.}\ }\textbf {\bibinfo {volume} {3}},\
  \bibinfo {pages} {1900024} (\bibinfo {year} {2020})}\BibitemShut {NoStop}%
\bibitem [{\citenamefont {Yoshie}\ \emph {et~al.}(2004)\citenamefont {Yoshie},
  \citenamefont {Scherer}, \citenamefont {Hendrickson}, \citenamefont
  {Khitrova}, \citenamefont {Gibbs}, \citenamefont {Rupper}, \citenamefont
  {Ell}, \citenamefont {Shchekin},\ and\ \citenamefont {Deppe}}]{Yoshie2004}%
  \BibitemOpen
  \bibfield  {author} {\bibinfo {author} {\bibfnamefont {T.}~\bibnamefont
  {Yoshie}}, \bibinfo {author} {\bibfnamefont {A.}~\bibnamefont {Scherer}},
  \bibinfo {author} {\bibfnamefont {J.}~\bibnamefont {Hendrickson}}, \bibinfo
  {author} {\bibfnamefont {G.}~\bibnamefont {Khitrova}}, \bibinfo {author}
  {\bibfnamefont {H.~M.}\ \bibnamefont {Gibbs}}, \bibinfo {author}
  {\bibfnamefont {G.}~\bibnamefont {Rupper}}, \bibinfo {author} {\bibfnamefont
  {C.}~\bibnamefont {Ell}}, \bibinfo {author} {\bibfnamefont {O.~B.}\
  \bibnamefont {Shchekin}},\ and\ \bibinfo {author} {\bibfnamefont {D.~G.}\
  \bibnamefont {Deppe}},\ }\bibfield  {title} {\bibinfo {title} {{Vacuum Rabi
  splitting with a single quantum dot in a photonic crystal nanocavity}},\
  }\href {https://doi.org/10.1038/nature03119} {\bibfield  {journal} {\bibinfo
  {journal} {Nature}\ }\textbf {\bibinfo {volume} {432}},\ \bibinfo {pages}
  {200} (\bibinfo {year} {2004})}\BibitemShut {NoStop}%
\bibitem [{\citenamefont {Hennessy}\ \emph {et~al.}(2007)\citenamefont
  {Hennessy}, \citenamefont {Badolato}, \citenamefont {Winger}, \citenamefont
  {Gerace}, \citenamefont {Atat{\"u}re}, \citenamefont {Gulde}, \citenamefont
  {F{\"a}lt}, \citenamefont {Hu},\ and\ \citenamefont
  {Imamo{\u{g}}lu}}]{Hennessy2007}%
  \BibitemOpen
  \bibfield  {author} {\bibinfo {author} {\bibfnamefont {K.}~\bibnamefont
  {Hennessy}}, \bibinfo {author} {\bibfnamefont {A.}~\bibnamefont {Badolato}},
  \bibinfo {author} {\bibfnamefont {M.}~\bibnamefont {Winger}}, \bibinfo
  {author} {\bibfnamefont {D.}~\bibnamefont {Gerace}}, \bibinfo {author}
  {\bibfnamefont {M.}~\bibnamefont {Atat{\"u}re}}, \bibinfo {author}
  {\bibfnamefont {S.}~\bibnamefont {Gulde}}, \bibinfo {author} {\bibfnamefont
  {S.}~\bibnamefont {F{\"a}lt}}, \bibinfo {author} {\bibfnamefont {E.~L.}\
  \bibnamefont {Hu}},\ and\ \bibinfo {author} {\bibfnamefont {A.}~\bibnamefont
  {Imamo{\u{g}}lu}},\ }\bibfield  {title} {\bibinfo {title} {{Quantum nature of
  a strongly coupled single quantum dot--cavity system}},\ }\href
  {https://doi.org/10.1038/nature05586} {\bibfield  {journal} {\bibinfo
  {journal} {Nature}\ }\textbf {\bibinfo {volume} {445}},\ \bibinfo {pages}
  {896} (\bibinfo {year} {2007})}\BibitemShut {NoStop}%
\bibitem [{\citenamefont {Shakoor}\ \emph {et~al.}(2013)\citenamefont
  {Shakoor}, \citenamefont {Lo~Savio}, \citenamefont {Cardile}, \citenamefont
  {Portalupi}, \citenamefont {Gerace}, \citenamefont {Welna}, \citenamefont
  {Boninelli}, \citenamefont {Franzò}, \citenamefont {Priolo}, \citenamefont
  {Krauss}, \citenamefont {Galli},\ and\ \citenamefont
  {O'Faolain}}]{doi:10.1002/lpor.201200043}%
  \BibitemOpen
  \bibfield  {author} {\bibinfo {author} {\bibfnamefont {A.}~\bibnamefont
  {Shakoor}}, \bibinfo {author} {\bibfnamefont {R.}~\bibnamefont {Lo~Savio}},
  \bibinfo {author} {\bibfnamefont {P.}~\bibnamefont {Cardile}}, \bibinfo
  {author} {\bibfnamefont {S.~L.}\ \bibnamefont {Portalupi}}, \bibinfo {author}
  {\bibfnamefont {D.}~\bibnamefont {Gerace}}, \bibinfo {author} {\bibfnamefont
  {K.}~\bibnamefont {Welna}}, \bibinfo {author} {\bibfnamefont
  {S.}~\bibnamefont {Boninelli}}, \bibinfo {author} {\bibfnamefont
  {G.}~\bibnamefont {Franzò}}, \bibinfo {author} {\bibfnamefont
  {F.}~\bibnamefont {Priolo}}, \bibinfo {author} {\bibfnamefont {T.~F.}\
  \bibnamefont {Krauss}}, \bibinfo {author} {\bibfnamefont {M.}~\bibnamefont
  {Galli}},\ and\ \bibinfo {author} {\bibfnamefont {L.}~\bibnamefont
  {O'Faolain}},\ }\bibfield  {title} {\bibinfo {title} {Room temperature
  all-silicon photonic crystal nanocavity light emitting diode at sub-bandgap
  wavelengths},\ }\href
  {https://doi.org/https://doi.org/10.1002/lpor.201200043} {\bibfield
  {journal} {\bibinfo  {journal} {Laser Photonics Rev.}\ }\textbf
  {\bibinfo {volume} {7}},\ \bibinfo {pages} {114} (\bibinfo {year}
  {2013})}\BibitemShut {NoStop}%
\bibitem [{\citenamefont {Akahane}\ \emph {et~al.}(2003)\citenamefont
  {Akahane}, \citenamefont {Asano}, \citenamefont {Song},\ and\ \citenamefont
  {Noda}}]{Akahane2003}%
  \BibitemOpen
  \bibfield  {author} {\bibinfo {author} {\bibfnamefont {Y.}~\bibnamefont
  {Akahane}}, \bibinfo {author} {\bibfnamefont {T.}~\bibnamefont {Asano}},
  \bibinfo {author} {\bibfnamefont {B.-S.}\ \bibnamefont {Song}},\ and\
  \bibinfo {author} {\bibfnamefont {S.}~\bibnamefont {Noda}},\ }\bibfield
  {title} {\bibinfo {title} {{High-Q photonic nanocavity in a two-dimensional
  photonic crystal}},\ }\href {https://doi.org/10.1038/nature02063} {\bibfield
  {journal} {\bibinfo  {journal} {Nature}\ }\textbf {\bibinfo {volume} {425}},\
  \bibinfo {pages} {944} (\bibinfo {year} {2003})}\BibitemShut {NoStop}%
\bibitem [{\citenamefont {Huang}\ \emph {et~al.}(2020)\citenamefont {Huang},
  \citenamefont {Alharbi}, \citenamefont {Mayer}, \citenamefont {Cuniberto},
  \citenamefont {Taniguchi}, \citenamefont {Watanabe}, \citenamefont
  {Shabani},\ and\ \citenamefont {Shahrjerdi}}]{Huang2020}%
  \BibitemOpen
  \bibfield  {author} {\bibinfo {author} {\bibfnamefont {Z.}~\bibnamefont
  {Huang}}, \bibinfo {author} {\bibfnamefont {A.}~\bibnamefont {Alharbi}},
  \bibinfo {author} {\bibfnamefont {W.}~\bibnamefont {Mayer}}, \bibinfo
  {author} {\bibfnamefont {E.}~\bibnamefont {Cuniberto}}, \bibinfo {author}
  {\bibfnamefont {T.}~\bibnamefont {Taniguchi}}, \bibinfo {author}
  {\bibfnamefont {K.}~\bibnamefont {Watanabe}}, \bibinfo {author}
  {\bibfnamefont {J.}~\bibnamefont {Shabani}},\ and\ \bibinfo {author}
  {\bibfnamefont {D.}~\bibnamefont {Shahrjerdi}},\ }\bibfield  {title}
  {\bibinfo {title} {{Versatile construction of van der Waals heterostructures
  using a dual-function polymeric film}},\ }\href
  {https://doi.org/10.1038/s41467-020-16817-1} {\bibfield  {journal} {\bibinfo
  {journal} {Nat. Commun.}\ }\textbf {\bibinfo {volume} {11}},\ \bibinfo
  {pages} {3029} (\bibinfo {year} {2020})}\BibitemShut {NoStop}%
\bibitem [{\citenamefont {Pizzocchero}\ \emph {et~al.}(2016)\citenamefont
  {Pizzocchero}, \citenamefont {Gammelgaard}, \citenamefont {Jessen},
  \citenamefont {Caridad}, \citenamefont {Wang}, \citenamefont {Hone},
  \citenamefont {B{\o}ggild},\ and\ \citenamefont {Booth}}]{Pizzocchero2016}%
  \BibitemOpen
  \bibfield  {author} {\bibinfo {author} {\bibfnamefont {F.}~\bibnamefont
  {Pizzocchero}}, \bibinfo {author} {\bibfnamefont {L.}~\bibnamefont
  {Gammelgaard}}, \bibinfo {author} {\bibfnamefont {B.~S.}\ \bibnamefont
  {Jessen}}, \bibinfo {author} {\bibfnamefont {J.~M.}\ \bibnamefont {Caridad}},
  \bibinfo {author} {\bibfnamefont {L.}~\bibnamefont {Wang}}, \bibinfo {author}
  {\bibfnamefont {J.}~\bibnamefont {Hone}}, \bibinfo {author} {\bibfnamefont
  {P.}~\bibnamefont {B{\o}ggild}},\ and\ \bibinfo {author} {\bibfnamefont
  {T.~J.}\ \bibnamefont {Booth}},\ }\bibfield  {title} {\bibinfo {title} {{The
  hot pick-up technique for batch assembly of van der Waals
  heterostructures}},\ }\href {https://doi.org/10.1038/ncomms11894} {\bibfield
  {journal} {\bibinfo  {journal} {Nat. Commun.}\ }\textbf {\bibinfo {volume}
  {7}},\ \bibinfo {pages} {11894} (\bibinfo {year} {2016})}\BibitemShut
  {NoStop}%
\bibitem []{supplement}%
  \BibitemOpen
  See Supplemental Material for the detailed information on nanocavity design, fabrication methods and additional results, which includes Refs. [36-58] 
  \BibitemShut {NoStop}%
\bibitem [{\citenamefont {Grenadier}\ \emph {et~al.}(2013)\citenamefont
  {Grenadier}, \citenamefont {Li}, \citenamefont {Lin},\ and\ \citenamefont
  {Jiang}}]{doi:10.1116/1.4826363}%
  \BibitemOpen
  \bibfield  {author} {\bibinfo {author} {\bibfnamefont {S.}~\bibnamefont
  {Grenadier}}, \bibinfo {author} {\bibfnamefont {J.}~\bibnamefont {Li}},
  \bibinfo {author} {\bibfnamefont {J.}~\bibnamefont {Lin}},\ and\ \bibinfo
  {author} {\bibfnamefont {H.}~\bibnamefont {Jiang}},\ }\bibfield  {title}
  {\bibinfo {title} {Dry etching techniques for active devices based on
  hexagonal boron nitride epilayers},\ }\href
  {https://doi.org/10.1116/1.4826363} {\bibfield  {journal} {\bibinfo
  {journal} {J. Vac. Sci. Technol. A}\ }\textbf {\bibinfo {volume} {31}},\
  \bibinfo {pages} {061517} (\bibinfo {year} {2013})}\BibitemShut {NoStop}%
\bibitem [{\citenamefont {Caldwell}\ \emph {et~al.}(2014)\citenamefont
  {Caldwell}, \citenamefont {Kretinin}, \citenamefont {Chen}, \citenamefont
  {Giannini}, \citenamefont {Fogler}, \citenamefont {Francescato},
  \citenamefont {Ellis}, \citenamefont {Tischler}, \citenamefont {Woods},
  \citenamefont {Giles}, \citenamefont {Hong}, \citenamefont {Watanabe},
  \citenamefont {Taniguchi}, \citenamefont {Maier},\ and\ \citenamefont
  {Novoselov}}]{Caldwell2014}%
  \BibitemOpen
  \bibfield  {author} {\bibinfo {author} {\bibfnamefont {J.~D.}\ \bibnamefont
  {Caldwell}}, \bibinfo {author} {\bibfnamefont {A.~V.}\ \bibnamefont
  {Kretinin}}, \bibinfo {author} {\bibfnamefont {Y.}~\bibnamefont {Chen}},
  \bibinfo {author} {\bibfnamefont {V.}~\bibnamefont {Giannini}}, \bibinfo
  {author} {\bibfnamefont {M.~M.}\ \bibnamefont {Fogler}}, \bibinfo {author}
  {\bibfnamefont {Y.}~\bibnamefont {Francescato}}, \bibinfo {author}
  {\bibfnamefont {C.~T.}\ \bibnamefont {Ellis}}, \bibinfo {author}
  {\bibfnamefont {J.~G.}\ \bibnamefont {Tischler}}, \bibinfo {author}
  {\bibfnamefont {C.~R.}\ \bibnamefont {Woods}}, \bibinfo {author}
  {\bibfnamefont {A.~J.}\ \bibnamefont {Giles}}, \bibinfo {author}
  {\bibfnamefont {M.}~\bibnamefont {Hong}}, \bibinfo {author} {\bibfnamefont
  {K.}~\bibnamefont {Watanabe}}, \bibinfo {author} {\bibfnamefont
  {T.}~\bibnamefont {Taniguchi}}, \bibinfo {author} {\bibfnamefont {S.~A.}\
  \bibnamefont {Maier}},\ and\ \bibinfo {author} {\bibfnamefont {K.~S.}\
  \bibnamefont {Novoselov}},\ }\bibfield  {title} {\bibinfo {title}
  {Sub-diffractional volume-confined polaritons in the natural hyperbolic
  material hexagonal boron nitride},\ }\href
  {https://doi.org/10.1038/ncomms6221} {\bibfield  {journal} {\bibinfo
  {journal} {Nat. Commun.}\ }\textbf {\bibinfo {volume} {5}},\ \bibinfo {pages}
  {5221} (\bibinfo {year} {2014})}\BibitemShut {NoStop}%
\bibitem [{\citenamefont {Jain}\ \emph {et~al.}(2018)\citenamefont {Jain},
  \citenamefont {Bharadwaj}, \citenamefont {Heeg}, \citenamefont {Parzefall},
  \citenamefont {Taniguchi}, \citenamefont {Watanabe},\ and\ \citenamefont
  {Novotny}}]{Jain_2018}%
  \BibitemOpen
  \bibfield  {author} {\bibinfo {author} {\bibfnamefont {A.}~\bibnamefont
  {Jain}}, \bibinfo {author} {\bibfnamefont {P.}~\bibnamefont {Bharadwaj}},
  \bibinfo {author} {\bibfnamefont {S.}~\bibnamefont {Heeg}}, \bibinfo {author}
  {\bibfnamefont {M.}~\bibnamefont {Parzefall}}, \bibinfo {author}
  {\bibfnamefont {T.}~\bibnamefont {Taniguchi}}, \bibinfo {author}
  {\bibfnamefont {K.}~\bibnamefont {Watanabe}},\ and\ \bibinfo {author}
  {\bibfnamefont {L.}~\bibnamefont {Novotny}},\ }\bibfield  {title} {\bibinfo
  {title} {Minimizing residues and strain in 2d materials transferred from
  {PDMS}},\ }\href {https://doi.org/10.1088/1361-6528/aabd90} {\bibfield
  {journal} {\bibinfo  {journal} {Nanotechnology}\ }\textbf {\bibinfo {volume}
  {29}},\ \bibinfo {pages} {265203} (\bibinfo {year} {2018})}\BibitemShut
  {NoStop}%
\bibitem [{\citenamefont {Shree}\ \emph {et~al.}(2019)\citenamefont {Shree},
  \citenamefont {George}, \citenamefont {Lehnert}, \citenamefont {Neumann},
  \citenamefont {Benelajla}, \citenamefont {Robert}, \citenamefont {Marie},
  \citenamefont {Watanabe}, \citenamefont {Taniguchi}, \citenamefont {Kaiser},
  \citenamefont {Urbaszek},\ and\ \citenamefont {Turchanin}}]{Shree_2019}%
  \BibitemOpen
  \bibfield  {author} {\bibinfo {author} {\bibfnamefont {S.}~\bibnamefont
  {Shree}}, \bibinfo {author} {\bibfnamefont {A.}~\bibnamefont {George}},
  \bibinfo {author} {\bibfnamefont {T.}~\bibnamefont {Lehnert}}, \bibinfo
  {author} {\bibfnamefont {C.}~\bibnamefont {Neumann}}, \bibinfo {author}
  {\bibfnamefont {M.}~\bibnamefont {Benelajla}}, \bibinfo {author}
  {\bibfnamefont {C.}~\bibnamefont {Robert}}, \bibinfo {author} {\bibfnamefont
  {X.}~\bibnamefont {Marie}}, \bibinfo {author} {\bibfnamefont
  {K.}~\bibnamefont {Watanabe}}, \bibinfo {author} {\bibfnamefont
  {T.}~\bibnamefont {Taniguchi}}, \bibinfo {author} {\bibfnamefont
  {U.}~\bibnamefont {Kaiser}}, \bibinfo {author} {\bibfnamefont
  {B.}~\bibnamefont {Urbaszek}},\ and\ \bibinfo {author} {\bibfnamefont
  {A.}~\bibnamefont {Turchanin}},\ }\bibfield  {title} {\bibinfo {title} {High
  optical quality of {MoS}2 monolayers grown by chemical vapor deposition},\
  }\href {https://doi.org/10.1088/2053-1583/ab4f1f} {\bibfield  {journal}
  {\bibinfo  {journal} {2D Mater.}\ }\textbf {\bibinfo {volume} {7}},\ \bibinfo
  {pages} {015011} (\bibinfo {year} {2019})}\BibitemShut {NoStop}%
\bibitem [{\citenamefont {Lee}\ \emph {et~al.}(2019)\citenamefont {Lee},
  \citenamefont {Jeong}, \citenamefont {Jung},\ and\ \citenamefont
  {Yee}}]{doi:10.1002/pssb.201800417}%
  \BibitemOpen
  \bibfield  {author} {\bibinfo {author} {\bibfnamefont {S.-Y.}\ \bibnamefont
  {Lee}}, \bibinfo {author} {\bibfnamefont {T.-Y.}\ \bibnamefont {Jeong}},
  \bibinfo {author} {\bibfnamefont {S.}~\bibnamefont {Jung}},\ and\ \bibinfo
  {author} {\bibfnamefont {K.-J.}\ \bibnamefont {Yee}},\ }\bibfield  {title}
  {\bibinfo {title} {{Refractive Index Dispersion of Hexagonal Boron Nitride in
  the Visible and Near-Infrared}},\ }\href
  {https://doi.org/10.1002/pssb.201800417} {\bibfield  {journal} {\bibinfo
  {journal} {Phys. Status Solidi B}\ }\textbf {\bibinfo {volume} {256}},\
  \bibinfo {pages} {1800417} (\bibinfo {year} {2019})}\BibitemShut {NoStop}%
\bibitem [{\citenamefont {Rah}\ \emph {et~al.}(2019)\citenamefont {Rah},
  \citenamefont {Jin}, \citenamefont {Kim},\ and\ \citenamefont {Yu}}]{Rah:19}%
  \BibitemOpen
  \bibfield  {author} {\bibinfo {author} {\bibfnamefont {Y.}~\bibnamefont
  {Rah}}, \bibinfo {author} {\bibfnamefont {Y.}~\bibnamefont {Jin}}, \bibinfo
  {author} {\bibfnamefont {S.}~\bibnamefont {Kim}},\ and\ \bibinfo {author}
  {\bibfnamefont {K.}~\bibnamefont {Yu}},\ }\bibfield  {title} {\bibinfo
  {title} {{Optical analysis of the refractive index and birefringence of
  hexagonal boron nitride from the visible to near-infrared}},\ }\href
  {https://doi.org/10.1364/OL.44.003797} {\bibfield  {journal} {\bibinfo
  {journal} {Opt. Lett.}\ }\textbf {\bibinfo {volume} {44}},\ \bibinfo {pages}
  {3797} (\bibinfo {year} {2019})}\BibitemShut {NoStop}%
\bibitem [{\citenamefont {Zhou}\ \emph {et~al.}(2017)\citenamefont {Zhou},
  \citenamefont {Scuri}, \citenamefont {Wild}, \citenamefont {High},
  \citenamefont {Dibos}, \citenamefont {Jauregui}, \citenamefont {Shu},
  \citenamefont {De~Greve}, \citenamefont {Pistunova}, \citenamefont {Joe},
  \citenamefont {Taniguchi}, \citenamefont {Watanabe}, \citenamefont {Kim},
  \citenamefont {Lukin},\ and\ \citenamefont {Park}}]{Zhou2017}%
  \BibitemOpen
  \bibfield  {author} {\bibinfo {author} {\bibfnamefont {Y.}~\bibnamefont
  {Zhou}}, \bibinfo {author} {\bibfnamefont {G.}~\bibnamefont {Scuri}},
  \bibinfo {author} {\bibfnamefont {D.~S.}\ \bibnamefont {Wild}}, \bibinfo
  {author} {\bibfnamefont {A.~A.}\ \bibnamefont {High}}, \bibinfo {author}
  {\bibfnamefont {A.}~\bibnamefont {Dibos}}, \bibinfo {author} {\bibfnamefont
  {L.~A.}\ \bibnamefont {Jauregui}}, \bibinfo {author} {\bibfnamefont
  {C.}~\bibnamefont {Shu}}, \bibinfo {author} {\bibfnamefont {K.}~\bibnamefont
  {De~Greve}}, \bibinfo {author} {\bibfnamefont {K.}~\bibnamefont {Pistunova}},
  \bibinfo {author} {\bibfnamefont {A.~Y.}\ \bibnamefont {Joe}}, \bibinfo
  {author} {\bibfnamefont {T.}~\bibnamefont {Taniguchi}}, \bibinfo {author}
  {\bibfnamefont {K.}~\bibnamefont {Watanabe}}, \bibinfo {author}
  {\bibfnamefont {P.}~\bibnamefont {Kim}}, \bibinfo {author} {\bibfnamefont
  {M.~D.}\ \bibnamefont {Lukin}},\ and\ \bibinfo {author} {\bibfnamefont
  {H.}~\bibnamefont {Park}},\ }\bibfield  {title} {\bibinfo {title} {Probing
  dark excitons in atomically thin semiconductors via near-field coupling to
  surface plasmon polaritons},\ }\href {https://doi.org/10.1038/nnano.2017.106}
  {\bibfield  {journal} {\bibinfo  {journal} {Nat. Nanotechnol.}\ }\textbf
  {\bibinfo {volume} {12}},\ \bibinfo {pages} {856} (\bibinfo {year}
  {2017})}\BibitemShut {NoStop}%
\bibitem [{\citenamefont {Rivera}\ \emph {et~al.}(2018)\citenamefont {Rivera},
  \citenamefont {Yu}, \citenamefont {Seyler}, \citenamefont {Wilson},
  \citenamefont {Yao},\ and\ \citenamefont {Xu}}]{Rivera2018}%
  \BibitemOpen
  \bibfield  {author} {\bibinfo {author} {\bibfnamefont {P.}~\bibnamefont
  {Rivera}}, \bibinfo {author} {\bibfnamefont {H.}~\bibnamefont {Yu}}, \bibinfo
  {author} {\bibfnamefont {K.~L.}\ \bibnamefont {Seyler}}, \bibinfo {author}
  {\bibfnamefont {N.~P.}\ \bibnamefont {Wilson}}, \bibinfo {author}
  {\bibfnamefont {W.}~\bibnamefont {Yao}},\ and\ \bibinfo {author}
  {\bibfnamefont {X.}~\bibnamefont {Xu}},\ }\bibfield  {title} {\bibinfo
  {title} {{Interlayer valley excitons in heterobilayers of transition metal
  dichalcogenides}},\ }\href {https://doi.org/10.1038/s41565-018-0193-0}
  {\bibfield  {journal} {\bibinfo  {journal} {Nat. Nanotechnol.}\ }\textbf
  {\bibinfo {volume} {13}},\ \bibinfo {pages} {1004} (\bibinfo {year}
  {2018})}\BibitemShut {NoStop}%
\bibitem [{\citenamefont {Golla}\ \emph {et~al.}(2013)\citenamefont {Golla},
  \citenamefont {Chattrakun}, \citenamefont {Watanabe}, \citenamefont
  {Taniguchi}, \citenamefont {LeRoy},\ and\ \citenamefont
  {Sandhu}}]{doi:10.1063/1.4803041}%
  \BibitemOpen
  \bibfield  {author} {\bibinfo {author} {\bibfnamefont {D.}~\bibnamefont
  {Golla}}, \bibinfo {author} {\bibfnamefont {K.}~\bibnamefont {Chattrakun}},
  \bibinfo {author} {\bibfnamefont {K.}~\bibnamefont {Watanabe}}, \bibinfo
  {author} {\bibfnamefont {T.}~\bibnamefont {Taniguchi}}, \bibinfo {author}
  {\bibfnamefont {B.~J.}\ \bibnamefont {LeRoy}},\ and\ \bibinfo {author}
  {\bibfnamefont {A.}~\bibnamefont {Sandhu}},\ }\bibfield  {title} {\bibinfo
  {title} {Optical thickness determination of hexagonal boron nitride flakes},\
  }\href {https://doi.org/10.1063/1.4803041} {\bibfield  {journal} {\bibinfo
  {journal} {Appl. Phys. Lett.}\ }\textbf {\bibinfo {volume} {102}},\ \bibinfo
  {pages} {161906} (\bibinfo {year} {2013})}\BibitemShut {NoStop}%
\bibitem [{\citenamefont {Stanford}\ \emph {et~al.}(2018)\citenamefont
  {Stanford}, \citenamefont {Rack},\ and\ \citenamefont
  {Jariwala}}]{Stanford2018}%
  \BibitemOpen
  \bibfield  {author} {\bibinfo {author} {\bibfnamefont {M.~G.}\ \bibnamefont
  {Stanford}}, \bibinfo {author} {\bibfnamefont {P.~D.}\ \bibnamefont {Rack}},\
  and\ \bibinfo {author} {\bibfnamefont {D.}~\bibnamefont {Jariwala}},\
  }\bibfield  {title} {\bibinfo {title} {Emerging nanofabrication and quantum
  confinement techniques for 2d materials beyond graphene},\ }\href
  {https://doi.org/10.1038/s41699-018-0065-3} {\bibfield  {journal} {\bibinfo
  {journal} {npj 2D Mater. Appl.}\ }\textbf {\bibinfo {volume} {2}},\ \bibinfo
  {pages} {20} (\bibinfo {year} {2018})}\BibitemShut {NoStop}%
\bibitem [{\citenamefont {Lyons}\ \emph {et~al.}(2019)\citenamefont {Lyons},
  \citenamefont {Dufferwiel}, \citenamefont {Brooks}, \citenamefont {Withers},
  \citenamefont {Taniguchi}, \citenamefont {Watanabe}, \citenamefont
  {Novoselov}, \citenamefont {Burkard},\ and\ \citenamefont
  {Tartakovskii}}]{Lyons2019}%
  \BibitemOpen
  \bibfield  {author} {\bibinfo {author} {\bibfnamefont {T.~P.}\ \bibnamefont
  {Lyons}}, \bibinfo {author} {\bibfnamefont {S.}~\bibnamefont {Dufferwiel}},
  \bibinfo {author} {\bibfnamefont {M.}~\bibnamefont {Brooks}}, \bibinfo
  {author} {\bibfnamefont {F.}~\bibnamefont {Withers}}, \bibinfo {author}
  {\bibfnamefont {T.}~\bibnamefont {Taniguchi}}, \bibinfo {author}
  {\bibfnamefont {K.}~\bibnamefont {Watanabe}}, \bibinfo {author}
  {\bibfnamefont {K.~S.}\ \bibnamefont {Novoselov}}, \bibinfo {author}
  {\bibfnamefont {G.}~\bibnamefont {Burkard}},\ and\ \bibinfo {author}
  {\bibfnamefont {A.~I.}\ \bibnamefont {Tartakovskii}},\ }\bibfield  {title}
  {\bibinfo {title} {The valley zeeman effect in inter- and intra-valley trions
  in monolayer wse2},\ }\href {https://doi.org/10.1038/s41467-019-10228-7}
  {\bibfield  {journal} {\bibinfo  {journal} {Nat. Commun.}\ }\textbf {\bibinfo
  {volume} {10}},\ \bibinfo {pages} {2330} (\bibinfo {year}
  {2019})}\BibitemShut {NoStop}%
\bibitem [{\citenamefont {Mosor}\ \emph {et~al.}(2005)\citenamefont {Mosor},
  \citenamefont {Hendrickson}, \citenamefont {Richards}, \citenamefont {Sweet},
  \citenamefont {Khitrova}, \citenamefont {Gibbs}, \citenamefont {Yoshie},
  \citenamefont {Scherer}, \citenamefont {Shchekin},\ and\ \citenamefont
  {Deppe}}]{doi:10.1063/1.2076435}%
  \BibitemOpen
  \bibfield  {author} {\bibinfo {author} {\bibfnamefont {S.}~\bibnamefont
  {Mosor}}, \bibinfo {author} {\bibfnamefont {J.}~\bibnamefont {Hendrickson}},
  \bibinfo {author} {\bibfnamefont {B.~C.}\ \bibnamefont {Richards}}, \bibinfo
  {author} {\bibfnamefont {J.}~\bibnamefont {Sweet}}, \bibinfo {author}
  {\bibfnamefont {G.}~\bibnamefont {Khitrova}}, \bibinfo {author}
  {\bibfnamefont {H.~M.}\ \bibnamefont {Gibbs}}, \bibinfo {author}
  {\bibfnamefont {T.}~\bibnamefont {Yoshie}}, \bibinfo {author} {\bibfnamefont
  {A.}~\bibnamefont {Scherer}}, \bibinfo {author} {\bibfnamefont {O.~B.}\
  \bibnamefont {Shchekin}},\ and\ \bibinfo {author} {\bibfnamefont {D.~G.}\
  \bibnamefont {Deppe}},\ }\bibfield  {title} {\bibinfo {title} {Scanning a
  photonic crystal slab nanocavity by condensation of xenon},\ }\href
  {https://doi.org/10.1063/1.2076435} {\bibfield  {journal} {\bibinfo
  {journal} {Appl. Phys. Lett.}\ }\textbf {\bibinfo {volume} {87}},\ \bibinfo
  {pages} {141105} (\bibinfo {year} {2005})}\BibitemShut {NoStop}%
\bibitem [{\citenamefont {Elshaari}\ \emph {et~al.}(2016)\citenamefont
  {Elshaari}, \citenamefont {Zadeh}, \citenamefont {Jöns},\ and\ \citenamefont
  {Zwiller}}]{7463458}%
  \BibitemOpen
  \bibfield  {author} {\bibinfo {author} {\bibfnamefont {A.~W.}\ \bibnamefont
  {Elshaari}}, \bibinfo {author} {\bibfnamefont {I.~E.}\ \bibnamefont {Zadeh}},
  \bibinfo {author} {\bibfnamefont {K.~D.}\ \bibnamefont {Jöns}},\ and\
  \bibinfo {author} {\bibfnamefont {V.}~\bibnamefont {Zwiller}},\ }\bibfield
  {title} {\bibinfo {title} {{Thermo-Optic Characterization of Silicon Nitride
  Resonators for Cryogenic Photonic Circuits}},\ }\href
  {https://doi.org/10.1109/JPHOT.2016.2561622} {\bibfield  {journal} {\bibinfo
  {journal} {IEEE Photonics J.}\ }\textbf {\bibinfo {volume} {8}},\ \bibinfo
  {pages} {1} (\bibinfo {year} {2016})}\BibitemShut {NoStop}%
\bibitem [{\citenamefont {S.Takayoshi}\ \emph {et~al.}(2009)\citenamefont
  {S.Takayoshi}, \citenamefont {Kokuyama},\ and\ \citenamefont
  {Fukuyama}}]{STakayoshi2009TheBS}%
  \BibitemOpen
  \bibfield  {author} {\bibinfo {author} {\bibnamefont {S.Takayoshi}}, \bibinfo
  {author} {\bibfnamefont {W.}~\bibnamefont {Kokuyama}},\ and\ \bibinfo
  {author} {\bibfnamefont {H.}~\bibnamefont {Fukuyama}},\ }\bibfield  {title}
  {\bibinfo {title} {The boiling suppression of liquid nitrogen},\ }\href
  {https://doi.org/10.1016/j.cryogenics.2008.12.013} {\bibfield  {journal}
  {\bibinfo  {journal} {Cryogenics}\ }\textbf {\bibinfo {volume} {49}},\
  \bibinfo {pages} {221} (\bibinfo {year} {2009})}\BibitemShut {NoStop}%
\bibitem [{\citenamefont {Witkowski}\ \emph {et~al.}(2014)\citenamefont
  {Witkowski}, \citenamefont {Majkut},\ and\ \citenamefont
  {Rulik}}]{Witkowski2014}%
  \BibitemOpen
  \bibfield  {author} {\bibinfo {author} {\bibfnamefont {A.}~\bibnamefont
  {Witkowski}}, \bibinfo {author} {\bibfnamefont {M.}~\bibnamefont {Majkut}},\
  and\ \bibinfo {author} {\bibfnamefont {S.}~\bibnamefont {Rulik}},\ }\bibfield
   {title} {\bibinfo {title} {Analysis of pipeline transportation systems for
  carbon dioxide sequestration},\ }\href
  {https://doi.org/10.2478/aoter-2014-0008} {\bibfield  {journal} {\bibinfo
  {journal} {Arch. Thermodyn.}\ }\textbf {\bibinfo {volume} {35}},\ \bibinfo
  {pages} {117} (\bibinfo {year} {2014})}\BibitemShut {NoStop}%
\bibitem [{\citenamefont {Greben}\ \emph {et~al.}(2020)\citenamefont {Greben},
  \citenamefont {Arora}, \citenamefont {Harats},\ and\ \citenamefont
  {Bolotin}}]{doi:10.1021/acs.nanolett.9b05323}%
  \BibitemOpen
  \bibfield  {author} {\bibinfo {author} {\bibfnamefont {K.}~\bibnamefont
  {Greben}}, \bibinfo {author} {\bibfnamefont {S.}~\bibnamefont {Arora}},
  \bibinfo {author} {\bibfnamefont {M.~G.}\ \bibnamefont {Harats}},\ and\
  \bibinfo {author} {\bibfnamefont {K.~I.}\ \bibnamefont {Bolotin}},\
  }\bibfield  {title} {\bibinfo {title} {Intrinsic and extrinsic defect-related
  excitons in tmdcs},\ }\href {https://doi.org/10.1021/acs.nanolett.9b05323}
  {\bibfield  {journal} {\bibinfo  {journal} {Nano Lett.}\ }\textbf {\bibinfo
  {volume} {20}},\ \bibinfo {pages} {2544} (\bibinfo {year}
  {2020})}\BibitemShut {NoStop}%
\bibitem [{\citenamefont {He}\ \emph {et~al.}(2020)\citenamefont {He},
  \citenamefont {Rivera}, \citenamefont {Van~Tuan}, \citenamefont {Wilson},
  \citenamefont {Yang}, \citenamefont {Taniguchi}, \citenamefont {Watanabe},
  \citenamefont {Yan}, \citenamefont {Mandrus}, \citenamefont {Yu},
  \citenamefont {Dery}, \citenamefont {Yao},\ and\ \citenamefont
  {Xu}}]{He2020}%
  \BibitemOpen
  \bibfield  {author} {\bibinfo {author} {\bibfnamefont {M.}~\bibnamefont
  {He}}, \bibinfo {author} {\bibfnamefont {P.}~\bibnamefont {Rivera}}, \bibinfo
  {author} {\bibfnamefont {D.}~\bibnamefont {Van~Tuan}}, \bibinfo {author}
  {\bibfnamefont {N.~P.}\ \bibnamefont {Wilson}}, \bibinfo {author}
  {\bibfnamefont {M.}~\bibnamefont {Yang}}, \bibinfo {author} {\bibfnamefont
  {T.}~\bibnamefont {Taniguchi}}, \bibinfo {author} {\bibfnamefont
  {K.}~\bibnamefont {Watanabe}}, \bibinfo {author} {\bibfnamefont
  {J.}~\bibnamefont {Yan}}, \bibinfo {author} {\bibfnamefont {D.~G.}\
  \bibnamefont {Mandrus}}, \bibinfo {author} {\bibfnamefont {H.}~\bibnamefont
  {Yu}}, \bibinfo {author} {\bibfnamefont {H.}~\bibnamefont {Dery}}, \bibinfo
  {author} {\bibfnamefont {W.}~\bibnamefont {Yao}},\ and\ \bibinfo {author}
  {\bibfnamefont {X.}~\bibnamefont {Xu}},\ }\bibfield  {title} {\bibinfo
  {title} {Valley phonons and exciton complexes in a monolayer semiconductor},\
  }\href {https://doi.org/10.1038/s41467-020-14472-0} {\bibfield  {journal}
  {\bibinfo  {journal} {Nat. Commun.}\ }\textbf {\bibinfo {volume} {11}},\
  \bibinfo {pages} {618} (\bibinfo {year} {2020})}\BibitemShut {NoStop}%
\bibitem [{\citenamefont {Barbone}\ \emph {et~al.}(2018)\citenamefont
  {Barbone}, \citenamefont {Montblanch}, \citenamefont {Kara}, \citenamefont
  {Palacios-Berraquero}, \citenamefont {Cadore}, \citenamefont {De~Fazio},
  \citenamefont {Pingault}, \citenamefont {Mostaani}, \citenamefont {Li},
  \citenamefont {Chen}, \citenamefont {Watanabe}, \citenamefont {Taniguchi},
  \citenamefont {Tongay}, \citenamefont {Wang}, \citenamefont {Ferrari},\ and\
  \citenamefont {Atat{\"u}re}}]{Barbone2018}%
  \BibitemOpen
  \bibfield  {author} {\bibinfo {author} {\bibfnamefont {M.}~\bibnamefont
  {Barbone}}, \bibinfo {author} {\bibfnamefont {A.~R.-P.}\ \bibnamefont
  {Montblanch}}, \bibinfo {author} {\bibfnamefont {D.~M.}\ \bibnamefont
  {Kara}}, \bibinfo {author} {\bibfnamefont {C.}~\bibnamefont
  {Palacios-Berraquero}}, \bibinfo {author} {\bibfnamefont {A.~R.}\
  \bibnamefont {Cadore}}, \bibinfo {author} {\bibfnamefont {D.}~\bibnamefont
  {De~Fazio}}, \bibinfo {author} {\bibfnamefont {B.}~\bibnamefont {Pingault}},
  \bibinfo {author} {\bibfnamefont {E.}~\bibnamefont {Mostaani}}, \bibinfo
  {author} {\bibfnamefont {H.}~\bibnamefont {Li}}, \bibinfo {author}
  {\bibfnamefont {B.}~\bibnamefont {Chen}}, \bibinfo {author} {\bibfnamefont
  {K.}~\bibnamefont {Watanabe}}, \bibinfo {author} {\bibfnamefont
  {T.}~\bibnamefont {Taniguchi}}, \bibinfo {author} {\bibfnamefont
  {S.}~\bibnamefont {Tongay}}, \bibinfo {author} {\bibfnamefont
  {G.}~\bibnamefont {Wang}}, \bibinfo {author} {\bibfnamefont {A.~C.}\
  \bibnamefont {Ferrari}},\ and\ \bibinfo {author} {\bibfnamefont
  {M.}~\bibnamefont {Atat{\"u}re}},\ }\bibfield  {title} {\bibinfo {title}
  {{Charge-tuneable biexciton complexes in monolayer WSe2}},\ }\href
  {https://doi.org/10.1038/s41467-018-05632-4} {\bibfield  {journal} {\bibinfo
  {journal} {Nat. Commun.}\ }\textbf {\bibinfo {volume} {9}},\ \bibinfo {pages}
  {3721} (\bibinfo {year} {2018})}\BibitemShut {NoStop}%
\bibitem [{\citenamefont {Lee}\ \emph {et~al.}(2017)\citenamefont {Lee},
  \citenamefont {Wang}, \citenamefont {Xie}, \citenamefont {Mak},\ and\
  \citenamefont {Shan}}]{Lee2017}%
  \BibitemOpen
  \bibfield  {author} {\bibinfo {author} {\bibfnamefont {J.}~\bibnamefont
  {Lee}}, \bibinfo {author} {\bibfnamefont {Z.}~\bibnamefont {Wang}}, \bibinfo
  {author} {\bibfnamefont {H.}~\bibnamefont {Xie}}, \bibinfo {author}
  {\bibfnamefont {K.~F.}\ \bibnamefont {Mak}},\ and\ \bibinfo {author}
  {\bibfnamefont {J.}~\bibnamefont {Shan}},\ }\bibfield  {title} {\bibinfo
  {title} {{Valley magnetoelectricity in single-layer MoS}},\ }\href
  {https://doi.org/10.1038/nmat4931} {\bibfield  {journal} {\bibinfo  {journal}
  {Nat. Mater.}\ }\textbf {\bibinfo {volume} {16}},\ \bibinfo {pages} {887}
  (\bibinfo {year} {2017})}\BibitemShut {NoStop}%
\bibitem [{\citenamefont {Ge}\ \emph {et~al.}(2021)\citenamefont {Ge},
  \citenamefont {Zhang}, \citenamefont {Chen}, \citenamefont {Zhen},
  \citenamefont {Wang}, \citenamefont {Jiao}, \citenamefont {Zhang},\ and\
  \citenamefont {Zhu}}]{GE2021109338}%
  \BibitemOpen
  \bibfield  {author} {\bibinfo {author} {\bibfnamefont {D.}~\bibnamefont
  {Ge}}, \bibinfo {author} {\bibfnamefont {Y.}~\bibnamefont {Zhang}}, \bibinfo
  {author} {\bibfnamefont {H.}~\bibnamefont {Chen}}, \bibinfo {author}
  {\bibfnamefont {G.}~\bibnamefont {Zhen}}, \bibinfo {author} {\bibfnamefont
  {M.}~\bibnamefont {Wang}}, \bibinfo {author} {\bibfnamefont {J.}~\bibnamefont
  {Jiao}}, \bibinfo {author} {\bibfnamefont {L.}~\bibnamefont {Zhang}},\ and\
  \bibinfo {author} {\bibfnamefont {S.}~\bibnamefont {Zhu}},\ }\bibfield
  {title} {\bibinfo {title} {{Effect of patterned silicon nitride substrate on
  Raman scattering and stress of graphene}},\ }\href
  {https://doi.org/https://doi.org/10.1016/j.matdes.2020.109338} {\bibfield
  {journal} {\bibinfo  {journal} {Mater. Des.}\ }\textbf {\bibinfo {volume}
  {198}},\ \bibinfo {pages} {109338} (\bibinfo {year} {2021})}\BibitemShut
  {NoStop}%
\bibitem [{\citenamefont {Zhang}\ \emph {et~al.}(2020)\citenamefont {Zhang},
  \citenamefont {Ding}, \citenamefont {Zhou}, \citenamefont {Xiao},\ and\
  \citenamefont {Xu}}]{doi.org:10.1002/pssr.202000222}%
  \BibitemOpen
  \bibfield  {author} {\bibinfo {author} {\bibfnamefont {J.}~\bibnamefont
  {Zhang}}, \bibinfo {author} {\bibfnamefont {L.}~\bibnamefont {Ding}},
  \bibinfo {author} {\bibfnamefont {S.}~\bibnamefont {Zhou}}, \bibinfo {author}
  {\bibfnamefont {Y.~M.}\ \bibnamefont {Xiao}},\ and\ \bibinfo {author}
  {\bibfnamefont {W.}~\bibnamefont {Xu}},\ }\bibfield  {title} {\bibinfo
  {title} {Observation and optical control of saturable excitonic behaviors in
  monolayer mos2},\ }\href
  {https://doi.org/https://doi.org/10.1002/pssr.202000222} {\bibfield
  {journal} {\bibinfo  {journal} {Phys. Status Solidi RRL}\ }\textbf {\bibinfo
  {volume} {14}},\ \bibinfo {pages} {2000222} (\bibinfo {year}
  {2020})}\BibitemShut {NoStop}%
\bibitem [{\citenamefont {Radulaski}\ \emph {et~al.}(2017)\citenamefont
  {Radulaski}, \citenamefont {Fischer}, \citenamefont {Lagoudakis},
  \citenamefont {Zhang},\ and\ \citenamefont {Vu\ifmmode \check{c}\else
  \v{c}\fi{}kovi\ifmmode~\acute{c}\else \'{c}\fi{}}}]{PhysRevA.96.011801}%
  \BibitemOpen
  \bibfield  {author} {\bibinfo {author} {\bibfnamefont {M.}~\bibnamefont
  {Radulaski}}, \bibinfo {author} {\bibfnamefont {K.~A.}\ \bibnamefont
  {Fischer}}, \bibinfo {author} {\bibfnamefont {K.~G.}\ \bibnamefont
  {Lagoudakis}}, \bibinfo {author} {\bibfnamefont {J.~L.}\ \bibnamefont
  {Zhang}},\ and\ \bibinfo {author} {\bibfnamefont {J.}~\bibnamefont
  {Vu\ifmmode \check{c}\else \v{c}\fi{}kovi\ifmmode~\acute{c}\else
  \'{c}\fi{}}},\ }\bibfield  {title} {\bibinfo {title} {Photon blockade in
  two-emitter-cavity systems},\ }\href
  {https://doi.org/10.1103/PhysRevA.96.011801} {\bibfield  {journal} {\bibinfo
  {journal} {Phys. Rev. A}\ }\textbf {\bibinfo {volume} {96}},\ \bibinfo
  {pages} {011801} (\bibinfo {year} {2017})}\BibitemShut {NoStop}%
\bibitem [{\citenamefont {Kangawa}\ \emph {et~al.}(2002)\citenamefont
  {Kangawa}, \citenamefont {Ito}, \citenamefont {Taguchi}, \citenamefont
  {Shiraishi}, \citenamefont {Irisawa},\ and\ \citenamefont
  {Ohachi}}]{KANGAWA2002517}%
  \BibitemOpen
  \bibfield  {author} {\bibinfo {author} {\bibfnamefont {Y.}~\bibnamefont
  {Kangawa}}, \bibinfo {author} {\bibfnamefont {T.}~\bibnamefont {Ito}},
  \bibinfo {author} {\bibfnamefont {A.}~\bibnamefont {Taguchi}}, \bibinfo
  {author} {\bibfnamefont {K.}~\bibnamefont {Shiraishi}}, \bibinfo {author}
  {\bibfnamefont {T.}~\bibnamefont {Irisawa}},\ and\ \bibinfo {author}
  {\bibfnamefont {T.}~\bibnamefont {Ohachi}},\ }\bibfield  {title} {\bibinfo
  {title} {{Monte Carlo simulation for temperature dependence of Ga diffusion
  length on GaAs(0 0 1)}},\ }\href
  {https://doi.org/https://doi.org/10.1016/S0169-4332(01)00930-8} {\bibfield
  {journal} {\bibinfo  {journal} {Appl. Surf. Sci.}\ }\textbf {\bibinfo
  {volume} {190}},\ \bibinfo {pages} {517} (\bibinfo {year}
  {2002})}\BibitemShut {NoStop}%
  %
\bibitem [{\citenamefont {Jimenez-Mier}(1994)}]{JIMENEZMIER1994741}%
  \BibitemOpen
  \bibfield  {author} {\bibinfo {author} {\bibfnamefont {J.}~\bibnamefont
  {Jimenez-Mier}},\ }\bibfield  {title} {\bibinfo {title} {{Contribution of the
  instrument window function to the profile of autoionizing resonances}},\
  }\href {https://doi.org/10.1016/0022-4073(94)90128-7} {\bibfield  {journal}
  {\bibinfo  {journal} {J. Quant. Spectrosc. Radiat. Transfer}\ }\textbf
  {\bibinfo {volume} {51}},\ \bibinfo {pages} {741} (\bibinfo {year}
  {1994})}\BibitemShut {NoStop}%
\bibitem [{\citenamefont {Kim}\ \emph {et~al.}(2018)\citenamefont {Kim},
  \citenamefont {Fr{\"o}ch}, \citenamefont {Christian}, \citenamefont {Straw},
  \citenamefont {Bishop}, \citenamefont {Totonjian}, \citenamefont {Watanabe},
  \citenamefont {Taniguchi}, \citenamefont {Toth},\ and\ \citenamefont
  {Aharonovich}}]{Kim2018}%
  \BibitemOpen
  \bibfield  {author} {\bibinfo {author} {\bibfnamefont {S.}~\bibnamefont
  {Kim}}, \bibinfo {author} {\bibfnamefont {J.~E.}\ \bibnamefont {Fr{\"o}ch}},
  \bibinfo {author} {\bibfnamefont {J.}~\bibnamefont {Christian}}, \bibinfo
  {author} {\bibfnamefont {M.}~\bibnamefont {Straw}}, \bibinfo {author}
  {\bibfnamefont {J.}~\bibnamefont {Bishop}}, \bibinfo {author} {\bibfnamefont
  {D.}~\bibnamefont {Totonjian}}, \bibinfo {author} {\bibfnamefont
  {K.}~\bibnamefont {Watanabe}}, \bibinfo {author} {\bibfnamefont
  {T.}~\bibnamefont {Taniguchi}}, \bibinfo {author} {\bibfnamefont
  {M.}~\bibnamefont {Toth}},\ and\ \bibinfo {author} {\bibfnamefont
  {I.}~\bibnamefont {Aharonovich}},\ }\bibfield  {title} {\bibinfo {title}
  {{Photonic crystal cavities from hexagonal boron nitride}},\ }\href
  {https://doi.org/10.1038/s41467-018-05117-4} {\bibfield  {journal} {\bibinfo
  {journal} {Nat. Commun.}\ }\textbf {\bibinfo {volume} {9}},\ \bibinfo {pages}
  {2623} (\bibinfo {year} {2018})}\BibitemShut {NoStop}%
\bibitem [{\citenamefont {Fröch}\ \emph {et~al.}(2019)\citenamefont {Fröch},
  \citenamefont {Hwang}, \citenamefont {Kim}, \citenamefont {Aharonovich},\
  and\ \citenamefont {Toth}}]{doi:10.1002/adom.201801344}%
  \BibitemOpen
  \bibfield  {author} {\bibinfo {author} {\bibfnamefont {J.~E.}\ \bibnamefont
  {Fröch}}, \bibinfo {author} {\bibfnamefont {Y.}~\bibnamefont {Hwang}},
  \bibinfo {author} {\bibfnamefont {S.}~\bibnamefont {Kim}}, \bibinfo {author}
  {\bibfnamefont {I.}~\bibnamefont {Aharonovich}},\ and\ \bibinfo {author}
  {\bibfnamefont {M.}~\bibnamefont {Toth}},\ }\bibfield  {title} {\bibinfo
  {title} {{Photonic Nanostructures from Hexagonal Boron Nitride}},\ }\href
  {https://doi.org/10.1002/adom.201801344} {\bibfield  {journal} {\bibinfo
  {journal} {Adv. Opt. Mater.}\ }\textbf {\bibinfo {volume} {7}},\ \bibinfo
  {pages} {1801344} (\bibinfo {year} {2019})}\BibitemShut {NoStop}%
\bibitem [{\citenamefont {Fröch}\ \emph {et~al.}(2020)\citenamefont {Fröch},
  \citenamefont {Kim}, \citenamefont {Mendelson}, \citenamefont {Kianinia},
  \citenamefont {Toth},\ and\ \citenamefont
  {Aharonovich}}]{doi:10.1021/acsnano.0c01818}%
  \BibitemOpen
  \bibfield  {author} {\bibinfo {author} {\bibfnamefont {J.~E.}\ \bibnamefont
  {Fröch}}, \bibinfo {author} {\bibfnamefont {S.}~\bibnamefont {Kim}},
  \bibinfo {author} {\bibfnamefont {N.}~\bibnamefont {Mendelson}}, \bibinfo
  {author} {\bibfnamefont {M.}~\bibnamefont {Kianinia}}, \bibinfo {author}
  {\bibfnamefont {M.}~\bibnamefont {Toth}},\ and\ \bibinfo {author}
  {\bibfnamefont {I.}~\bibnamefont {Aharonovich}},\ }\bibfield  {title}
  {\bibinfo {title} {{Coupling Hexagonal Boron Nitride Quantum Emitters to
  Photonic Crystal Cavities}},\ }\href
  {https://doi.org/10.1021/acsnano.0c01818} {\bibfield  {journal} {\bibinfo
  {journal} {ACS Nano}\ }\textbf {\bibinfo {volume} {14}},\ \bibinfo {pages}
  {7085} (\bibinfo {year} {2020})}\BibitemShut {NoStop}%
\bibitem [{\citenamefont {Parak}\ \emph {et~al.}(2010)\citenamefont {Parak},
  \citenamefont {Manna},\ and\ \citenamefont
  {Nann}}]{10.1002/9783527628155.nanotech004}%
  \BibitemOpen
  \bibfield  {author} {\bibinfo {author} {\bibfnamefont {W.~J.}\ \bibnamefont
  {Parak}}, \bibinfo {author} {\bibfnamefont {L.}~\bibnamefont {Manna}},\ and\
  \bibinfo {author} {\bibfnamefont {T.}~\bibnamefont {Nann}},\ }\bibinfo
  {title} {Fundamental principles of quantum dots},\ in\ \href
  {https://doi.org/https://doi.org/10.1002/9783527628155.nanotech004} {\emph
  {\bibinfo {booktitle} {Nanotechnology}}}\ (\bibinfo  {publisher} {American
  Cancer Society},\ \bibinfo {year} {2010})\ Chap.~\bibinfo {chapter} {4}, pp.\
  \bibinfo {pages} {73--96}\BibitemShut {NoStop}%
\bibitem [{\citenamefont {Kwak}(2019)}]{KWAK2019102202}%
  \BibitemOpen
  \bibfield  {author} {\bibinfo {author} {\bibfnamefont {J.~Y.}\ \bibnamefont
  {Kwak}},\ }\bibfield  {title} {\bibinfo {title} {Absorption coefficient
  estimation of thin mos2 film using attenuation of silicon substrate raman
  signal},\ }\href {https://doi.org/https://doi.org/10.1016/j.rinp.2019.102202}
  {\bibfield  {journal} {\bibinfo  {journal} {Results in Physics}\ }\textbf
  {\bibinfo {volume} {13}},\ \bibinfo {pages} {102202} (\bibinfo {year}
  {2019})}\BibitemShut {NoStop}%
\bibitem [{\citenamefont {Andersen}\ \emph {et~al.}(2011)\citenamefont
  {Andersen}, \citenamefont {Stobbe}, \citenamefont {S{\o}rensen},\ and\
  \citenamefont {Lodahl}}]{Andersen2011}%
  \BibitemOpen
  \bibfield  {author} {\bibinfo {author} {\bibfnamefont {M.~L.}\ \bibnamefont
  {Andersen}}, \bibinfo {author} {\bibfnamefont {S.}~\bibnamefont {Stobbe}},
  \bibinfo {author} {\bibfnamefont {A.~S.}\ \bibnamefont {S{\o}rensen}},\ and\
  \bibinfo {author} {\bibfnamefont {P.}~\bibnamefont {Lodahl}},\ }\bibfield
  {title} {\bibinfo {title} {{Strongly modified plasmon--matter interaction
  with mesoscopic quantum emitters}},\ }\href
  {https://doi.org/10.1038/nphys1870} {\bibfield  {journal} {\bibinfo
  {journal} {Nat. Phys.}\ }\textbf {\bibinfo {volume} {7}},\ \bibinfo {pages}
  {215} (\bibinfo {year} {2011})}\BibitemShut {NoStop}%
\bibitem [{\citenamefont {Dias}\ \emph {et~al.}(2018)\citenamefont {Dias},
  \citenamefont {Iranzo}, \citenamefont {Gon\ifmmode~\mbox{\c{c}}\else
  \c{c}\fi{}alves}, \citenamefont {Hajati}, \citenamefont {Bludov},
  \citenamefont {Jauho}, \citenamefont {Mortensen}, \citenamefont {Koppens},\
  and\ \citenamefont {Peres}}]{PhysRevB.97.245405}%
  \BibitemOpen
  \bibfield  {author} {\bibinfo {author} {\bibfnamefont {E.~J.~C.}\
  \bibnamefont {Dias}}, \bibinfo {author} {\bibfnamefont {D.~A.}\ \bibnamefont
  {Iranzo}}, \bibinfo {author} {\bibfnamefont {P.~A.~D.}\ \bibnamefont
  {Gon\ifmmode~\mbox{\c{c}}\else \c{c}\fi{}alves}}, \bibinfo {author}
  {\bibfnamefont {Y.}~\bibnamefont {Hajati}}, \bibinfo {author} {\bibfnamefont
  {Y.~V.}\ \bibnamefont {Bludov}}, \bibinfo {author} {\bibfnamefont {A.-P.}\
  \bibnamefont {Jauho}}, \bibinfo {author} {\bibfnamefont {N.~A.}\ \bibnamefont
  {Mortensen}}, \bibinfo {author} {\bibfnamefont {F.~H.~L.}\ \bibnamefont
  {Koppens}},\ and\ \bibinfo {author} {\bibfnamefont {N.~M.~R.}\ \bibnamefont
  {Peres}},\ }\bibfield  {title} {\bibinfo {title} {Probing nonlocal effects in
  metals with graphene plasmons},\ }\href
  {https://doi.org/10.1103/PhysRevB.97.245405} {\bibfield  {journal} {\bibinfo
  {journal} {Phys. Rev. B}\ }\textbf {\bibinfo {volume} {97}},\ \bibinfo
  {pages} {245405} (\bibinfo {year} {2018})}\BibitemShut {NoStop}%
\bibitem [{\citenamefont {Alcaraz~Iranzo}\ \emph {et~al.}(2018)\citenamefont
  {Alcaraz~Iranzo}, \citenamefont {Nanot}, \citenamefont {Dias}, \citenamefont
  {Epstein}, \citenamefont {Peng}, \citenamefont {Efetov}, \citenamefont
  {Lundeberg}, \citenamefont {Parret}, \citenamefont {Osmond}, \citenamefont
  {Hong}, \citenamefont {Kong}, \citenamefont {Englund}, \citenamefont
  {Peres},\ and\ \citenamefont {Koppens}}]{Iranzo291}%
  \BibitemOpen
  \bibfield  {author} {\bibinfo {author} {\bibfnamefont {D.}~\bibnamefont
  {Alcaraz~Iranzo}}, \bibinfo {author} {\bibfnamefont {S.}~\bibnamefont
  {Nanot}}, \bibinfo {author} {\bibfnamefont {E.~J.~C.}\ \bibnamefont {Dias}},
  \bibinfo {author} {\bibfnamefont {I.}~\bibnamefont {Epstein}}, \bibinfo
  {author} {\bibfnamefont {C.}~\bibnamefont {Peng}}, \bibinfo {author}
  {\bibfnamefont {D.~K.}\ \bibnamefont {Efetov}}, \bibinfo {author}
  {\bibfnamefont {M.~B.}\ \bibnamefont {Lundeberg}}, \bibinfo {author}
  {\bibfnamefont {R.}~\bibnamefont {Parret}}, \bibinfo {author} {\bibfnamefont
  {J.}~\bibnamefont {Osmond}}, \bibinfo {author} {\bibfnamefont {J.-Y.}\
  \bibnamefont {Hong}}, \bibinfo {author} {\bibfnamefont {J.}~\bibnamefont
  {Kong}}, \bibinfo {author} {\bibfnamefont {D.~R.}\ \bibnamefont {Englund}},
  \bibinfo {author} {\bibfnamefont {N.~M.~R.}\ \bibnamefont {Peres}},\ and\
  \bibinfo {author} {\bibfnamefont {F.~H.~L.}\ \bibnamefont {Koppens}},\
  }\bibfield  {title} {\bibinfo {title} {Probing the ultimate plasmon
  confinement limits with a van der waals heterostructure},\ }\href
  {https://doi.org/10.1126/science.aar8438} {\bibfield  {journal} {\bibinfo
  {journal} {Science}\ }\textbf {\bibinfo {volume} {360}},\ \bibinfo {pages}
  {291} (\bibinfo {year} {2018})}\BibitemShut {NoStop}%
\bibitem [{\citenamefont {Qian}\ \emph {et~al.}(2019)\citenamefont {Qian},
  \citenamefont {Xie}, \citenamefont {Yang}, \citenamefont {Peng},
  \citenamefont {Wu}, \citenamefont {Song}, \citenamefont {Sun}, \citenamefont
  {Dang}, \citenamefont {Yu}, \citenamefont {Steer}, \citenamefont {Thayne},
  \citenamefont {Jin}, \citenamefont {Gu},\ and\ \citenamefont
  {Xu}}]{PhysRevLett.122.087401}%
  \BibitemOpen
  \bibfield  {author} {\bibinfo {author} {\bibfnamefont {C.}~\bibnamefont
  {Qian}}, \bibinfo {author} {\bibfnamefont {X.}~\bibnamefont {Xie}}, \bibinfo
  {author} {\bibfnamefont {J.}~\bibnamefont {Yang}}, \bibinfo {author}
  {\bibfnamefont {K.}~\bibnamefont {Peng}}, \bibinfo {author} {\bibfnamefont
  {S.}~\bibnamefont {Wu}}, \bibinfo {author} {\bibfnamefont {F.}~\bibnamefont
  {Song}}, \bibinfo {author} {\bibfnamefont {S.}~\bibnamefont {Sun}}, \bibinfo
  {author} {\bibfnamefont {J.}~\bibnamefont {Dang}}, \bibinfo {author}
  {\bibfnamefont {Y.}~\bibnamefont {Yu}}, \bibinfo {author} {\bibfnamefont
  {M.~J.}\ \bibnamefont {Steer}}, \bibinfo {author} {\bibfnamefont {I.~G.}\
  \bibnamefont {Thayne}}, \bibinfo {author} {\bibfnamefont {K.}~\bibnamefont
  {Jin}}, \bibinfo {author} {\bibfnamefont {C.}~\bibnamefont {Gu}},\ and\
  \bibinfo {author} {\bibfnamefont {X.}~\bibnamefont {Xu}},\ }\bibfield
  {title} {\bibinfo {title} {{Enhanced Strong Interaction between Nanocavities
  and $p$-shell Excitons Beyond the Dipole Approximation}},\ }\href
  {https://doi.org/10.1103/PhysRevLett.122.087401} {\bibfield  {journal}
  {\bibinfo  {journal} {Phys. Rev. Lett.}\ }\textbf {\bibinfo {volume} {122}},\
  \bibinfo {pages} {087401} (\bibinfo {year} {2019})}\BibitemShut {NoStop}%
\bibitem [{\citenamefont {Stier}\ \emph {et~al.}(2016)\citenamefont {Stier},
  \citenamefont {Wilson}, \citenamefont {Clark}, \citenamefont {Xu},\ and\
  \citenamefont {Crooker}}]{doi:10.1021/acs.nanolett.6b03276}%
  \BibitemOpen
  \bibfield  {author} {\bibinfo {author} {\bibfnamefont {A.~V.}\ \bibnamefont
  {Stier}}, \bibinfo {author} {\bibfnamefont {N.~P.}\ \bibnamefont {Wilson}},
  \bibinfo {author} {\bibfnamefont {G.}~\bibnamefont {Clark}}, \bibinfo
  {author} {\bibfnamefont {X.}~\bibnamefont {Xu}},\ and\ \bibinfo {author}
  {\bibfnamefont {S.~A.}\ \bibnamefont {Crooker}},\ }\bibfield  {title}
  {\bibinfo {title} {{Probing the Influence of Dielectric Environment on
  Excitons in Monolayer WSe2: Insight from High Magnetic Fields}},\ }\href
  {https://doi.org/10.1021/acs.nanolett.6b03276} {\bibfield  {journal}
  {\bibinfo  {journal} {Nano Lett.}\ }\textbf {\bibinfo {volume} {16}},\
  \bibinfo {pages} {7054} (\bibinfo {year} {2016})}\BibitemShut {NoStop}%
\bibitem [{\citenamefont {Wang}\ \emph {et~al.}(2016)\citenamefont {Wang},
  \citenamefont {Zhang}, \citenamefont {Chan}, \citenamefont {Manolatou},
  \citenamefont {Tiwari},\ and\ \citenamefont {Rana}}]{Wang2016}%
  \BibitemOpen
  \bibfield  {author} {\bibinfo {author} {\bibfnamefont {H.}~\bibnamefont
  {Wang}}, \bibinfo {author} {\bibfnamefont {C.}~\bibnamefont {Zhang}},
  \bibinfo {author} {\bibfnamefont {W.}~\bibnamefont {Chan}}, \bibinfo {author}
  {\bibfnamefont {C.}~\bibnamefont {Manolatou}}, \bibinfo {author}
  {\bibfnamefont {S.}~\bibnamefont {Tiwari}},\ and\ \bibinfo {author}
  {\bibfnamefont {F.}~\bibnamefont {Rana}},\ }\bibfield  {title} {\bibinfo
  {title} {{Radiative lifetimes of excitons and trions in monolayers of the
  metal dichalcogenide ${\mathrm{MoS}}_{2}$}},\ }\href
  {https://doi.org/10.1103/PhysRevB.93.045407} {\bibfield  {journal} {\bibinfo
  {journal} {Phys. Rev. B}\ }\textbf {\bibinfo {volume} {93}},\ \bibinfo
  {pages} {045407} (\bibinfo {year} {2016})}\BibitemShut {NoStop}%
\bibitem [{\citenamefont {Zinov’ev}\ \emph {et~al.}(1983)\citenamefont
  {Zinov’ev}, \citenamefont {Ivanov}, \citenamefont {Lang}, \citenamefont
  {Pavlov}, \citenamefont {Prokaznikov},\ and\ \citenamefont
  {Yaroshetskii}}]{Zinovev1983}%
  \BibitemOpen
  \bibfield  {author} {\bibinfo {author} {\bibfnamefont {N.}~\bibnamefont
  {Zinov’ev}}, \bibinfo {author} {\bibfnamefont {L.}~\bibnamefont {Ivanov}},
  \bibinfo {author} {\bibfnamefont {I.}~\bibnamefont {Lang}}, \bibinfo {author}
  {\bibfnamefont {S.}~\bibnamefont {Pavlov}}, \bibinfo {author} {\bibfnamefont
  {A.}~\bibnamefont {Prokaznikov}},\ and\ \bibinfo {author} {\bibfnamefont
  {I.}~\bibnamefont {Yaroshetskii}},\ }\bibfield  {title} {\bibinfo {title}
  {Exciton diffusion and the mechanism of exciton momentum scattering in
  semiconductors},\ }\href
  {http://www.jetp.ac.ru/cgi-bin/e/index/e/57/6/p1254?a=list} {\bibfield
  {journal} {\bibinfo  {journal} {Sov. Phys. JETP}\ }\textbf {\bibinfo {volume}
  {57}},\ \bibinfo {pages} {1254} (\bibinfo {year} {1983})}\BibitemShut
  {NoStop}%
\bibitem [{\citenamefont {Chai}\ \emph {et~al.}(2017)\citenamefont {Chai},
  \citenamefont {Su}, \citenamefont {Yan}, \citenamefont {Ozkan}, \citenamefont
  {Lake},\ and\ \citenamefont {Ozkan}}]{Chai2017}%
  \BibitemOpen
  \bibfield  {author} {\bibinfo {author} {\bibfnamefont {Y.}~\bibnamefont
  {Chai}}, \bibinfo {author} {\bibfnamefont {S.}~\bibnamefont {Su}}, \bibinfo
  {author} {\bibfnamefont {D.}~\bibnamefont {Yan}}, \bibinfo {author}
  {\bibfnamefont {M.}~\bibnamefont {Ozkan}}, \bibinfo {author} {\bibfnamefont
  {R.}~\bibnamefont {Lake}},\ and\ \bibinfo {author} {\bibfnamefont {C.~S.}\
  \bibnamefont {Ozkan}},\ }\bibfield  {title} {\bibinfo {title} {{Strain Gated
  Bilayer Molybdenum Disulfide Field Effect Transistor with Edge Contacts}},\
  }\href {https://doi.org/10.1038/srep41593} {\bibfield  {journal} {\bibinfo
  {journal} {Sci. Rep.}\ }\textbf {\bibinfo {volume} {7}},\ \bibinfo {pages}
  {41593} (\bibinfo {year} {2017})}\BibitemShut {NoStop}%
\bibitem [{\citenamefont {John}\ \emph {et~al.}(2020)\citenamefont {John},
  \citenamefont {Thenapparambil},\ and\ \citenamefont
  {Thalakulam}}]{John_2020}%
  \BibitemOpen
  \bibfield  {author} {\bibinfo {author} {\bibfnamefont {A.~P.}\ \bibnamefont
  {John}}, \bibinfo {author} {\bibfnamefont {A.}~\bibnamefont
  {Thenapparambil}},\ and\ \bibinfo {author} {\bibfnamefont {M.}~\bibnamefont
  {Thalakulam}},\ }\bibfield  {title} {\bibinfo {title} {Strain-engineering the
  schottky barrier and electrical transport on {MoS}2},\ }\href
  {https://doi.org/10.1088/1361-6528/ab83b7} {\bibfield  {journal} {\bibinfo
  {journal} {Nanotechnology}\ }\textbf {\bibinfo {volume} {31}},\ \bibinfo
  {pages} {275703} (\bibinfo {year} {2020})}\BibitemShut {NoStop}%
\bibitem [{\citenamefont {Klein}\ \emph {et~al.}(2019)\citenamefont {Klein},
  \citenamefont {Lorke}, \citenamefont {Florian}, \citenamefont {Sigger},
  \citenamefont {Sigl}, \citenamefont {Rey}, \citenamefont {Wierzbowski},
  \citenamefont {Cerne}, \citenamefont {M{\"u}ller}, \citenamefont
  {Mitterreiter}, \citenamefont {Zimmermann}, \citenamefont {Taniguchi},
  \citenamefont {Watanabe}, \citenamefont {Wurstbauer}, \citenamefont
  {Kaniber}, \citenamefont {Knap}, \citenamefont {Schmidt}, \citenamefont
  {Finley},\ and\ \citenamefont {Holleitner}}]{Klein2019}%
  \BibitemOpen
  \bibfield  {author} {\bibinfo {author} {\bibfnamefont {J.}~\bibnamefont
  {Klein}}, \bibinfo {author} {\bibfnamefont {M.}~\bibnamefont {Lorke}},
  \bibinfo {author} {\bibfnamefont {M.}~\bibnamefont {Florian}}, \bibinfo
  {author} {\bibfnamefont {F.}~\bibnamefont {Sigger}}, \bibinfo {author}
  {\bibfnamefont {L.}~\bibnamefont {Sigl}}, \bibinfo {author} {\bibfnamefont
  {S.}~\bibnamefont {Rey}}, \bibinfo {author} {\bibfnamefont {J.}~\bibnamefont
  {Wierzbowski}}, \bibinfo {author} {\bibfnamefont {J.}~\bibnamefont {Cerne}},
  \bibinfo {author} {\bibfnamefont {K.}~\bibnamefont {M{\"u}ller}}, \bibinfo
  {author} {\bibfnamefont {E.}~\bibnamefont {Mitterreiter}}, \bibinfo {author}
  {\bibfnamefont {P.}~\bibnamefont {Zimmermann}}, \bibinfo {author}
  {\bibfnamefont {T.}~\bibnamefont {Taniguchi}}, \bibinfo {author}
  {\bibfnamefont {K.}~\bibnamefont {Watanabe}}, \bibinfo {author}
  {\bibfnamefont {U.}~\bibnamefont {Wurstbauer}}, \bibinfo {author}
  {\bibfnamefont {M.}~\bibnamefont {Kaniber}}, \bibinfo {author} {\bibfnamefont
  {M.}~\bibnamefont {Knap}}, \bibinfo {author} {\bibfnamefont {R.}~\bibnamefont
  {Schmidt}}, \bibinfo {author} {\bibfnamefont {J.~J.}\ \bibnamefont
  {Finley}},\ and\ \bibinfo {author} {\bibfnamefont {A.~W.}\ \bibnamefont
  {Holleitner}},\ }\bibfield  {title} {\bibinfo {title} {{Site-selectively
  generated photon emitters in monolayer MoS2 via local helium ion
  irradiation}},\ }\href {https://doi.org/10.1038/s41467-019-10632-z}
  {\bibfield  {journal} {\bibinfo  {journal} {Nat. Commun.}\ }\textbf {\bibinfo
  {volume} {10}},\ \bibinfo {pages} {2755} (\bibinfo {year}
  {2019})}\BibitemShut {NoStop}%
\bibitem [{\citenamefont {Seyler}\ \emph {et~al.}(2019)\citenamefont {Seyler},
  \citenamefont {Rivera}, \citenamefont {Yu}, \citenamefont {Wilson},
  \citenamefont {Ray}, \citenamefont {Mandrus}, \citenamefont {Yan},
  \citenamefont {Yao},\ and\ \citenamefont {Xu}}]{Seyler2019}%
  \BibitemOpen
  \bibfield  {author} {\bibinfo {author} {\bibfnamefont {K.~L.}\ \bibnamefont
  {Seyler}}, \bibinfo {author} {\bibfnamefont {P.}~\bibnamefont {Rivera}},
  \bibinfo {author} {\bibfnamefont {H.}~\bibnamefont {Yu}}, \bibinfo {author}
  {\bibfnamefont {N.~P.}\ \bibnamefont {Wilson}}, \bibinfo {author}
  {\bibfnamefont {E.~L.}\ \bibnamefont {Ray}}, \bibinfo {author} {\bibfnamefont
  {D.~G.}\ \bibnamefont {Mandrus}}, \bibinfo {author} {\bibfnamefont
  {J.}~\bibnamefont {Yan}}, \bibinfo {author} {\bibfnamefont {W.}~\bibnamefont
  {Yao}},\ and\ \bibinfo {author} {\bibfnamefont {X.}~\bibnamefont {Xu}},\
  }\bibfield  {title} {\bibinfo {title} {{Signatures of moir{\'e}-trapped
  valley excitons in MoSe2/WSe2 heterobilayers}},\ }\href
  {https://doi.org/10.1038/s41586-019-0957-1} {\bibfield  {journal} {\bibinfo
  {journal} {Nature}\ }\textbf {\bibinfo {volume} {567}},\ \bibinfo {pages}
  {66} (\bibinfo {year} {2019})}\BibitemShut {NoStop}%
\bibitem [{\citenamefont {Baek}\ \emph {et~al.}(2020)\citenamefont {Baek},
  \citenamefont {Brotons-Gisbert}, \citenamefont {Koong}, \citenamefont
  {Campbell}, \citenamefont {Rambach}, \citenamefont {Watanabe}, \citenamefont
  {Taniguchi},\ and\ \citenamefont {Gerardot}}]{Baek2020}%
  \BibitemOpen
  \bibfield  {author} {\bibinfo {author} {\bibfnamefont {H.}~\bibnamefont
  {Baek}}, \bibinfo {author} {\bibfnamefont {M.}~\bibnamefont
  {Brotons-Gisbert}}, \bibinfo {author} {\bibfnamefont {Z.~X.}\ \bibnamefont
  {Koong}}, \bibinfo {author} {\bibfnamefont {A.}~\bibnamefont {Campbell}},
  \bibinfo {author} {\bibfnamefont {M.}~\bibnamefont {Rambach}}, \bibinfo
  {author} {\bibfnamefont {K.}~\bibnamefont {Watanabe}}, \bibinfo {author}
  {\bibfnamefont {T.}~\bibnamefont {Taniguchi}},\ and\ \bibinfo {author}
  {\bibfnamefont {B.~D.}\ \bibnamefont {Gerardot}},\ }\bibfield  {title}
  {\bibinfo {title} {Highly energy-tunable quantum light from moir{\'e}-trapped
  excitons},\ }\href {https://doi.org/10.1126/sciadv.aba8526} {\bibfield
  {journal} {\bibinfo  {journal} {Sci. Adv.}\ }\textbf {\bibinfo {volume}
  {6}},\ \bibinfo {pages} {eaba8526} (\bibinfo {year} {2020})}\BibitemShut
  {NoStop}%
\end{thebibliography}
\end{document}